\documentclass [prb,preprint,superscriptaddress,floatfix,amsmath,amssymb] {revtex4-1}
\usepackage{amsmath}
\usepackage{graphicx}
\usepackage{hyperref}
\usepackage{color}
\usepackage{amssymb}
\usepackage{bm}

\definecolor{darkgreen}{RGB}{0,170,0}

\newcommand{\beq} {\begin{equation}}
\newcommand{\eeq} {\end{equation}}
\newcommand{\bea} {\begin{eqnarray}}
\newcommand{\eea} {\end{eqnarray}}
\newcommand{\be} {\begin{equation}}
\newcommand{\ee} {\end{equation}}

\begin{document}
\title {Interplay between superconductivity and non-Fermi liquid  at a quantum-critical point in a metal.\\
III: The $\gamma$ model and its phase diagram across  $\gamma = 1$.}
\author{Yi-Ming Wu}
\affiliation{School of Physics and Astronomy and William I. Fine Theoretical Physics Institute,
University of Minnesota, Minneapolis, MN 55455, USA}
\author{Artem Abanov}
\affiliation{Department of Physics, Texas A\&M University, College Station,  USA}
\author{Andrey V. Chubukov}
\affiliation{School of Physics and Astronomy and William I. Fine Theoretical Physics Institute,
University of Minnesota, Minneapolis, MN 55455, USA}
\date{\today}
\begin{abstract}
 In this paper we continue our analysis of the interplay between the pairing and the non-Fermi liquid behavior
  in a metal for a set of quantum-critical models with an effective dynamical electron-electron interaction
 $V(\Omega_m) \propto 1/|\Omega_m|^\gamma$ (the $\gamma$-model).   We analyze both the original model and its extension, in which we introduce an extra parameter $N$ to account for non-equal
   interactions in the particle-hole and particle-particle channel.
 In two previous papers,\cite{paper_1,paper_2} we considered the case $0 < \gamma <1$ and argued that (i) at $T=0$, there exists an infinite discrete set of topologically different gap functions, $\Delta_n (\omega_m)$, all with the same  spatial symmetry, and (ii) each $\Delta_n$
  evolves with
   temperature and terminates at a particular $T_{p,n}$.  In this paper, we analyze how the system behavior changes between  $\gamma <1$ and $\gamma >1$, both at $T=0$ and a finite $T$.
     The limit $\gamma \to 1$ is singular
      due to
        infra-red  divergence of $\int d \omega_m V(\Omega_m)$, and the system behavior is highly sensitive to how this limit is taken.  We
  show that for $N =1$,  the divergencies  in the gap equation cancel out, and
   $\Delta_n (\omega_m)$   gradually evolve through $\gamma=1$ both at $T=0$ and a finite $T$.
     For $N \neq 1$,  divergent terms do not cancel,
    and
     a
     qualitatively new behavior emerges for $\gamma >1$. Namely, the
      form of $\Delta_n (\omega_m)$ changes qualitatively,
      and the spectrum of condensation energies, $E_{c,n}$ becomes  continuous at $T=0$.
       We
       introduce
        different extension of the model,
       which is free from singularities for $\gamma >1$.

\end{abstract}
\maketitle

\section{ Introduction.}

In this paper we continue our analysis of the competition between non-Fermi liquid (NFL) physics and superconductivity (SC) near a quantum-critical point (QCP) in a metal
  with
   four-fermion interaction, mediated by a critical soft boson.
      We consider a  class of models, for which soft bosons are slow modes compared to dressed electrons.
      In this situation, the low-energy physics at a QCP is governed by an effective dynamical interaction $V(\Omega_m) = {\bar g}^\gamma/|\Omega_m|^\gamma$
       integrated along the Fermi surface (the $\gamma$-model).
     This interaction
 is singular, and gives rise to two
 opposite
 tendencies: NFL
  behavior  in the normal state, with fermionic self-energy
$\Sigma (\omega_m) \propto \omega^{1-\gamma}_m$, and
 an attraction in at least one pairing channel.
  The two tendencies compete with each other as
  a NFL self-energy reduces the magnitude of the pairing kernel, while the feedback from the pairing reduces fermionic self-energy.

In the first paper in the series,
   Ref. \cite{paper_1}, we listed quantum-critical systems, whose low-energy physics is described by the $\gamma$-model with different $\gamma$ and presented  an extensive list of references to earlier publications on this subject.
   In this and in the subsequent paper, Ref. \cite{paper_2}, hereafter referred to as Paper I and Paper II,  respectively,
    we   analyzed the behavior of the $\gamma$-model for $0 <\gamma <1$ at
$T=0$ (Paper I) and at a finite $T$ (Paper II).    We found that the
 system does become unstable towards pairing.
  However, in qualitative distinction with BCS/Eliashberg theory of superconductivity, in which there is a single
  solution of the gap equation,
   $\Delta (\omega_m)$,
   here we found an infinite discrete set of solutions $\Delta_n (\omega_m)$.
   All
   solutions have the same spatial
   symmetry, but are topologically distinct as $\Delta_n (\omega_m)$ changes sign $n$ times as a function of Matsubara frequency (each such point is a center of a dynamical vortex).
   The gap functions $\Delta_n (\omega_m)$
     with finite $n$
     tend to finite values at zero frequency, but the magnitude of $\Delta_n (0)$ decreases with $n$ and at large enough $n$ scales as
   $\Delta_n (0) \propto e^{-A n}$, where $A$ is a function of $\gamma$.
    In the limit $n \to \infty$, $\Delta_\infty$ is
    the solution of the linearized gap equation.
      We found the exact form of $\Delta_\infty (\omega_m)$.
    It
      oscillates as a function of $\log (|\omega_m|/{\bar g})$
        down to the lowest frequencies and up to $\omega_{max}$, which is generally of order ${\bar g}$, except for the smallest $\gamma$, where  $\omega_{max}\sim {\bar g} (1/\gamma)^{1/\gamma}$. At $\omega > \omega_{max}$, $\Delta_{\infty} (\omega_m)$ decays as
     $1/|\omega_m|^\gamma$.  A function $\Delta_n (\omega_m)$ with a finite $n$ saturates
     at $\Delta_n (0)$
      below
       $\omega_m \sim \Delta_n (0)$, and at larger $\omega$
     retains the functional
     form of $\Delta_{\infty} (\omega_m)$.  At a finite $T$, each $\Delta_n (\omega_m)$ evolves with $T$ and terminates at its own $T_{p,n} \sim \Delta_n (0)$.

 In this paper we extend the analysis to larger $\gamma$.  We will be particularly interested in the
  evolution of the system behavior between $\gamma \leq 1$ and $\gamma \geq 1$.
   At $T=0$, the frequency integral $\int d \Omega_m V(\Omega_m) \propto \int d \Omega_m /|\Omega_m|^\gamma$ diverges at small $\Omega_{m}$
    for $\gamma \geq 1$, and from a general perspective one could expect that this divergence introduces  qualitative changes in the system behavior.  Indeed,  the
   pairing vertex and the fermionic self-energy at $T=0$ do become singular for $\gamma \geq 1$.  We show, however, that singular terms cancel out in the equation for the gap function $\Delta (\omega_m)$. As a consequence,  for  both $\gamma \leq 1$ and $\gamma \geq 1$  the  full non-linear gap equation  has an  infinite number of solutions $\Delta_n (\omega_m)$ at $T=0$ and  each $\Delta_n$ terminates at its own $T_{p,n}$. All functions $\Delta_n (\omega_m)$ evolve smoothly through $\gamma =1$.
   The corresponding  condensation energies, $E_{c,n}$ form a discrete set in which $E_{c,0}$ is the largest.

We next analyze
 a more general model
 with different interaction strength in particle-particle and  particle-hole channels.  A natural way to account for this is to multiply the interaction in the pairing channel by a factor $1/N$ leaving
     the interaction in the particle-hole channel intact~\cite{paper_1,paper_2,Wang2016,Chubukov_2020a,Abanov_19,*Wu_19_1}
 Another way to split the strength of the two interactions is to extend the original $\gamma-$ model  to matrix $SU(N)$ model\cite{raghu_15,Wang_H_17,Wang_H_18}.

  The factor $N$ plays the role of the eigenvalue in the linearized matrix gap equation, and understanding the system behavior for $N \neq 1$ is also essential for the interpretation of the flow of the eigenvalues and the eigenfunctions in the numerical analysis of the gap equation even for $N \to 1$.
     We show that for $\gamma \to 1$, the system  behavior becomes very sensitive to small deviations from $N=1$, because the $T=0$ divergencies in the self-energy and the pairing vertex do not cancel for  any $N \neq 1$.
      As a consequence,
      the limits $\gamma \to 1$ and $N \to 1$ do not commute, and the structure of the gap function strongly  depends on the ratio $(N-1)/(1-\gamma)$.  We show that the behavior at $\gamma \to 1$
       and $N <1$  is qualitatively  different from that for $N \equiv 1$.
      Namely, for $N <1$ the set of condensation energies  becomes a continuous one at $\gamma =1$: $E_{c,n}$ for all finite $n$ become the same as $E_{c,0}$ and $E_{c,\infty}$ form a continuous one-parameter gapless spectrum, similar to how a continuous phonon spectrum emerges in a continuum limit.   This opens up a channel of massless 'longitudinal" gap fluctuations, in a truly qualitative distinction from BCS-type physics. We  show that this behavior holds at $T=0$ for $\gamma >1$, and that the structure of $\Delta_n (\omega_m)$ at a finite $T$  also becomes qualitatively different from that for $\gamma <1$ and gives rise to highly unconventional form of the density of eigenvalues for $N <1$, as we show both analytically and numerically.

We also discuss another extension of the theory, which does not introduce singular contributions that were responsible for qualitatively different behavior at $N =1$ and $N \neq 1$. This extension also allows one to vary the relative strength of the interactions in the particle-hole and particle-particle channels (although in a less obvious way), and  tune between NFL and SC states for $\gamma >1$, similar to how it was done before for $\gamma <1$ in Refs. \cite{Wang2016,Chubukov_2020a,Abanov_19,*Wu_19_1,paper_1,paper_2,raghu_15,Wang_H_17,Wang_H_18}.

 The structure of the paper is the following.  In Sec. \ref{sec:model} we briefly review the $\gamma$ model and present the equations  for the pairing vertex, the self-energy, and the gap function, which we will use later in the paper.  In Sec.~\ref{sec:cancellation} we show that the singularities, imposed by the divergence of $\int d \Omega_m V(\omega_m)$ for $\gamma \geq 1$, cancel out in the gap equation for $N=1$. We argue that the full non-linear gap equation at $T=0$ has an infinite set of solutions $\Delta_n (\omega_m)$ both for $\gamma \leq 1$ and $\gamma \geq 1$,  and show that the solutions vary smoothly through $\gamma =1$.
  In Sec. \ref{sec:N} we extend the $\gamma$ model to $N \neq 1$ and show that for a generic $N$ the system
 behavior changes qualitatively between $\gamma <1$ and $\gamma \geq 1$. We discuss the double limit $\gamma \to 1$, $N \to 1$ at $T=0$, show how the set of the condensation energies, $E_{c,n}$, becomes continuous at $\gamma >1$, and discuss the new structure of $\Delta_n (\omega_m)$ at $\gamma >1$ and a finite $T$.
 In Sec. \ref{sec:tilde N} we discuss another extension of the $\gamma$ model, which does not introduce the divergencies.

  In Paper IV, the next in the series, we consider the case $N=1$, $1< \gamma<2$ in more detail, and argue that as $\gamma$ increases, the
    dynamical vortices  emerge one by one   and form an array in the upper frequency half-plane. The number of vortices tends to infinity for $\gamma \to 2$.

\section{$\gamma$-model, Eliashberg equations}
\label{sec:model}

The $\gamma$-model was introduced in  Paper I and in earlier publications as a low-energy model for the interaction between soft bosons and electrons~\cite{paper_1,paper_2,acf,acs,moon_2,max,senthil,scal,efetov,
max_last,raghu_15,Wang2016,Kotliar2018,Abanov_19,Wu_19_1,Chubukov_2020a},
 and we refer the reader to these works for the justification of the model and its relation to various quantum-critical systems.  The model describes low-energy fermions with an effective dynamical interaction $V(\Omega_m) = {\bar g}^\gamma/|\Omega_m|^\gamma$, averaged over momenta on the Fermi surface with a proper weight.  The case $\gamma \approx 1$ corresponds to, e.g., pairing by a weakly damped soft
  optical phonon with static  susceptibility peaked some finite momentum $Q_0$.\cite{Chubukov_2020b}
The coupled equations for the fermionic self-energy $\Sigma (\omega_m)$ and the pairing vertex $\Phi (\omega_m)$ in the most attractive pairing channel are similar to Eliashberg equations for the case of a dispersionless phonon, and we will use the term ``Eliashberg equations" for our case.

At a finite $T$ the coupled Eliashberg equations for  $\Phi (\omega_m)$ and $\Sigma (\omega_m)$ are,
 in Matsubara formalism
    \bea \label{eq:gapeq}
    \Phi (\omega_m) &=&
     {\bar g}^\gamma \pi T \sum_{m' \neq m} \frac{\Phi (\omega_{m'})}{\sqrt{{\tilde \Sigma}^2 (\omega_{m'}) +\Phi^2 (\omega_{m'})}}
    ~\frac{1}{|\omega_m - \omega_{m'}|^\gamma}, \nonumber \\
     {\tilde \Sigma} (\omega_m) &=& \omega_m
   +  {\bar g}^\gamma \pi T \sum_{m' \neq m}  \frac{{\tilde \Sigma}(\omega_{m'})}{\sqrt{{\tilde \Sigma}^2 (\omega_{m'})  +\Phi^2 (\omega_{m'})}}
    ~\frac{1}{|\omega_m - \omega_{m'}|^\gamma}
\eea
 where ${\tilde \Sigma}(\omega_{m}) = \omega_m + \Sigma (\omega_m)$. In these notations, $\Sigma (\omega_{m})$ is a  real function, odd in frequency.

 The  SC gap function $\Delta (\omega_m)$ is defined as
\beq
 \Delta (\omega_m) = \omega_m  \frac{\Phi (\omega_m)}{{\tilde \Sigma} (\omega_m)} = \frac{\Phi (\omega_m)}{1 + \Sigma (\omega_m)/\omega_m}
  \label{ss_1}
  \eeq
   The equation for $\Delta (\omega_{m})$ is readily obtained from (\ref{eq:gapeq}):
   \beq
   \Delta (\omega_m) = {\bar g}^\gamma \pi T \sum_{m' \neq m} \frac{\Delta (\omega_{m'}) - \Delta (\omega_m) \frac{\omega_{m'}}{\omega_m}}{\sqrt{(\omega_{m'})^2 +\Delta^2 (\omega_{m'})}}
    ~\frac{1}{|\omega_m - \omega_{m'}|^\gamma}.
     \label{ss_11}
  \eeq
   This equation contains a single function $\Delta (\omega_{m})$, but at the cost that $\Delta (\omega_m)$ appears also in the r.h.s.  Both $\Phi (\omega_m)$ and $\Delta (\omega_m)$ are defined up to an overall $U(1)$ phase factor, which we set to zero for definiteness.   Eqs. (\ref{eq:gapeq}) and (\ref{ss_11}) exclude the self-action term with $m'=m$. This term cancels out by Anderson theorem~\cite{agd}, because scattering with zero frequency transfer mimics the effect of scattering by non-magnetic impurities.

Below we will analyze the full non-linear  equations  and the linearized equations, for infinitesimally small $\Phi (\omega_m)$ and $\Delta (\omega_m)$.  The latter determine, e.g., critical temperatures $T_{p,n}$.
  The linearized gap equation is
  \beq
   \Delta (\omega_m) = {\bar g}^\gamma \pi T \sum_{m'} \frac{\Delta (\omega_{m'}) - \Delta (\omega_m) \frac{\omega_{m'}}{\omega_m}}{|\omega_{m'}|}\frac{1}{|\omega_m - \omega_{m'}|^\gamma}.
     \label{ss_11_l}
  \eeq
 The linearized equation for the pairing vertex $\Phi (\omega_m)$ is
 \beq
    \Phi (\omega_m) =
     {\bar g}^\gamma \pi T \sum_{m' \neq m} \frac{\Phi (\omega_{m'})}{| \omega_{m'} + \Sigma_{\text{norm}} (\omega_{m'})|}    ~\frac{1}{|\omega_m - \omega_{m'}|^\gamma}
 \label{eq:gapeq_l}
 \eeq
 where $\Sigma_{\text{norm}} (\omega_{m})$ is the self-energy of the normal state,
  \beq
  \Sigma_{\text{norm}} (\omega_m) =  {\bar g}^\gamma (2\pi T)^{1-\gamma} \sum_{m'=1}^m \frac{1}{|m'|^{\gamma}} = {\bar g}^\gamma (2\pi T)^{1-\gamma} H_{m, \gamma}
\label{ss_111_a}
\eeq
and $H_{m, \gamma}$ is the Harmonic number.  This expression holds for $\omega_m \neq \pm \pi T$. For the two lowest  Matsubara frequencies, $\Sigma_{\text{norm}} (\pm \pi T) =0$.  We emphasize that
 $\Sigma_{\text{norm}} (\omega_m)$ in (\ref{ss_111_a}) is not the full normal state self-energy, as the summation in (\ref{eq:gapeq_l})
 excludes the term $m'=m$.

At $T=0$,
   \bea
   \Delta (\omega_m) &=& \frac{{\bar g}^\gamma}{2} \int d \omega'_m \frac{\Delta (\omega'_{m}) - \Delta (\omega_m) \frac{\omega'_{m}}{\omega_m}}{\sqrt{(\omega'_{m})^2 + \Delta^2 (\omega'_m)}}
    ~\frac{1}{|\omega_m - \omega'_{m}|^\gamma}, \nonumber \\
    \Phi (\omega_m) &=&
     \frac{{\bar g}^\gamma}{2}  \int d \omega'_m  \frac{\Phi (\omega'_{m})}{\sqrt{{\tilde \Sigma}^2 (\omega'_m) + \Phi^2 (\omega'_m)}} ~\frac{1}{|\omega_m - \omega'_{m}|^\gamma}
 \label{eq:gapeq_0}
  \eea
 The linearized equations are
   \bea
   \Delta (\omega_m) &=& \frac{{\bar g}^\gamma}{2} \int d \omega'_m \frac{\Delta (\omega'_{m}) - \Delta (\omega_m) \frac{\omega'_{m}}{\omega_m}}{|\omega'_{m}|}
    ~\frac{1}{|\omega_m - \omega'_{m}|^\gamma}, \nonumber \\
   \Phi (\omega_m) &=&
     \frac{{\bar g}^\gamma}{2}  \int d \omega'_m  \frac{\Phi (\omega'_{m})}{|\omega'_m + \Sigma_{\text{norm}} (\omega'_m)|} ~\frac{1}{|\omega_m - \omega'_{m}|^\gamma}
     \label{ss_11_l0}
  \eea
 where
\beq
\Sigma_{\text{norm}} (\omega_m) = \omega^\gamma_0 |\omega_m|^{1-\gamma} {\text{sgn}}(\omega_m)
\label{s0}
\eeq
 and
\beq
\omega_0 = {\bar g}/(1-\gamma)^{1/\gamma}.
\label{w0}
\eeq
 At small $\gamma$, $\omega_0 = {\bar g} e$. At $\gamma \to 1$, $\omega^\gamma_0$ diverges as $1/(1-\gamma)$.

\section{Transformation from $\gamma \leq 1$ to  $\gamma \geq 1$}
\label{sec:cancellation}

\begin{figure}
  \includegraphics[width=8cm]{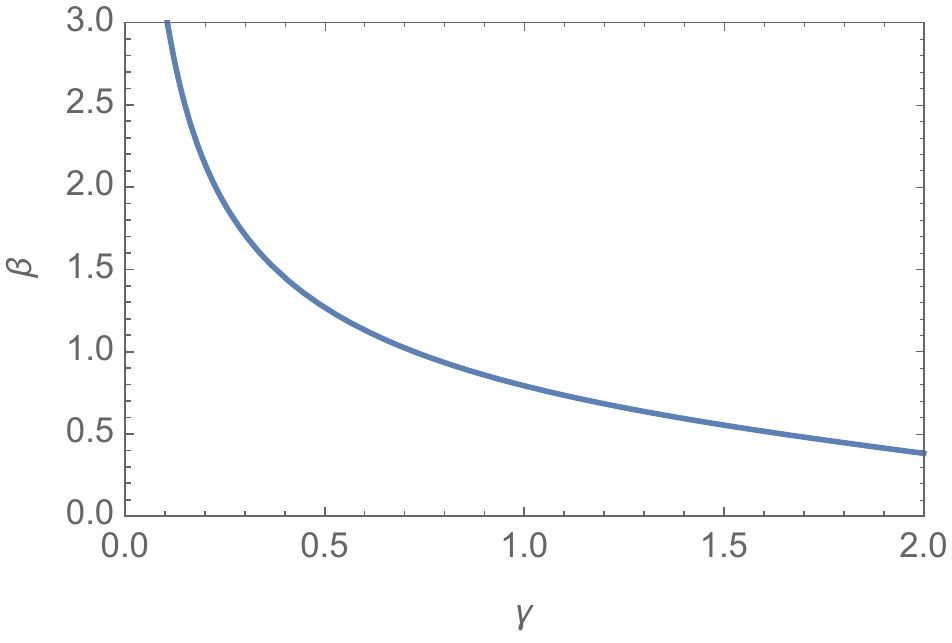}
  \caption{The solution of $\epsilon_{i\beta} =1$, with $\epsilon_{i\beta}$ given by Eq. (\ref{su_15_2}), as a function of the exponent $\gamma$.  }\label{fig:beta}
 \end{figure}

\subsection{A generic $\gamma <1$}

 We found in Papers I and II that
 \begin{itemize}
 \item
  The non-linear gap equation has an infinite discrete set of solutions $\Delta_n (\omega_m)$, $n =0, 1, 2...$.  All $\Delta_n (\omega_m)$ with finite $n$ tend to finite $\Delta_n (0)$ at zero frequency and decay as $1/|\omega_m|^\gamma$ at large frequencies.  The function $\Delta_n (\omega_m)$ changes sign $n$ times.  At large $n$, $\Delta_n (0) \propto e^{-A n}$, where $A = O(1)$ is a function of $\gamma $.
  \item
 The end point of the set,  $\Delta_{\infty} (\omega_m)$, is the solution of the linearized gap equation.
     At small $\omega_m \ll \omega_0$,
     \beq
    \Delta_{\infty} (\omega_m) = C |\omega_m|^{\gamma/2} \cos\left({\beta \log{\left(\frac{|\omega_m|}{\omega_0}\right)^\gamma} + \phi}\right)
     \label{3_23}
     \eeq
      where $\beta$ depends on $\gamma$ (see Eq. (\ref{su_15_2}) below and Fig.\ref{fig:beta}),
     and $\phi$ is some $\gamma$-dependent number.  Eq. (\ref{3_23}) is readily obtained if one
     neglects $\omega_m$ compared to $\Sigma_{\text{norm}} (\omega_m) \propto \omega^{1-\gamma}_m$ in Eq. (\ref{ss_11_l0}) for the pairing vertex. The corrections to log-oscillating form  hold in powers of
    $z = (|\omega_m|/\omega_0)^\gamma = \omega_m/\Sigma_{\text{norm}} (\omega_m)$.
  \item
  We found the exact form of $\Delta_{\infty} (z)$ for all $z$. It oscillates up to $z = O(1)$ and decays as $1/z$ at larger $z$.  A gap function $\Delta_n (z)$ with a finite $n$  also decays as $1/z$ for $z >1$,
   oscillates $n$ times at smaller $z$, and saturates at the lowest frequencies at a finite $\Delta_n (0)$.
 \item
   At a finite $T$, each $\Delta_n (\omega_m)$ develops at the onset temperature $T_{p,n} \sim \Delta_n (0)$. At large $n$,  $T_{p,n} \propto e^{-An}$.  The magnitude of $\Delta_n (\omega_m)$ increases with decreasing $T$, and at $T=0$ it coincides with the $n$-th solution of the non-linear gap equation.
  \end{itemize}

\subsection{$\gamma \approx  1$}

\subsubsection{Linearized gap equation,  $T=0$}

\begin{figure}
    \includegraphics[width=8cm]{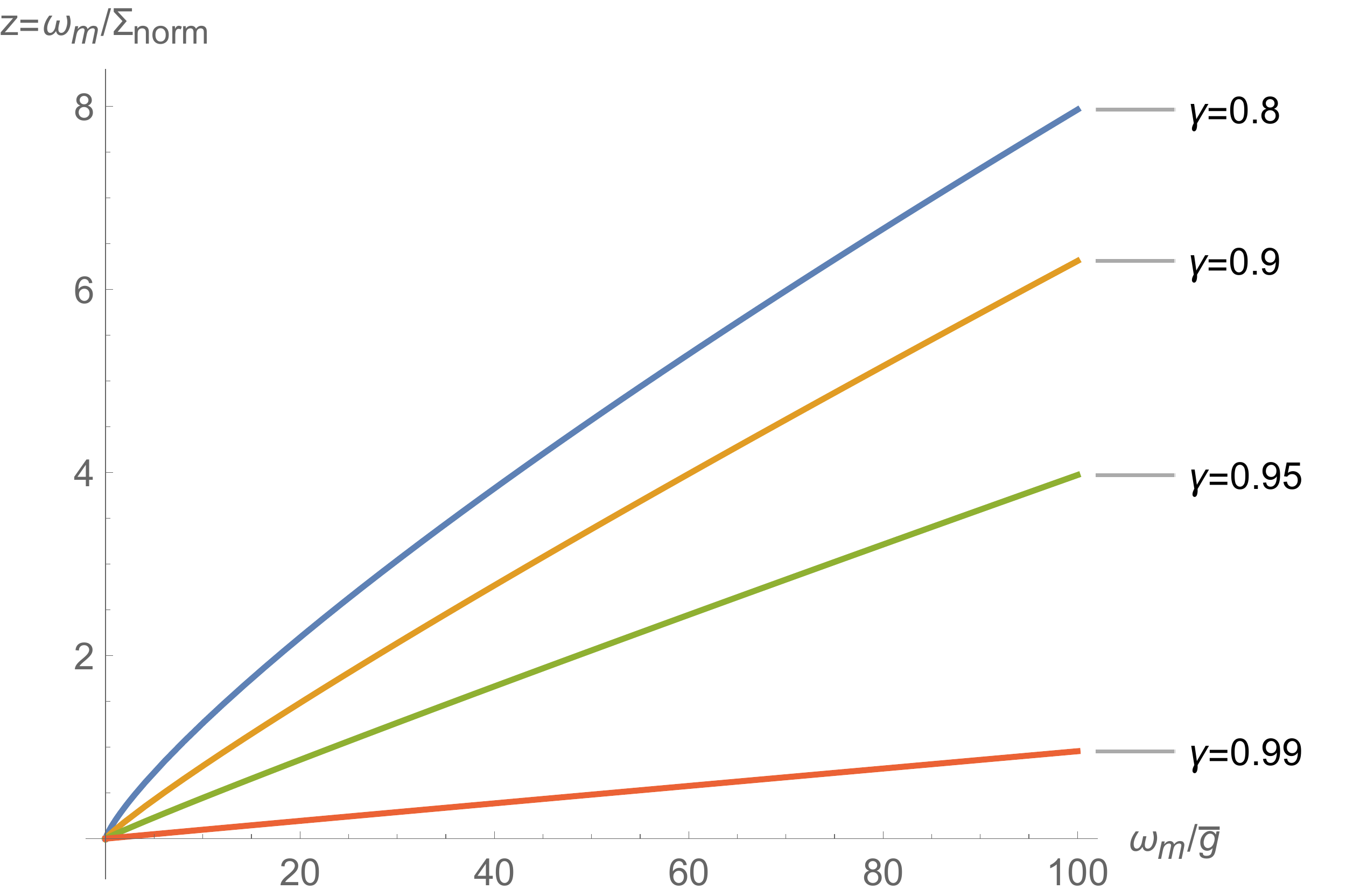}
    \caption{The dimensionless parameter $z = \omega_m/\Sigma_{\text{norm}}$  as a function of $\omega_m$ for various $\gamma$. As $\gamma \to 1$, the slope of $z(\omega_m)$ decreases, and
     $z$ becomes $O(1)$ at  progressively larger $\omega_m$. }\label{fig:normal_state_selfenergy}
  \end{figure}

 We now analyze what happens when $\gamma$ increases  and approaches $1$.  We begin with the linearized gap equation for $\Delta_{\infty} (\omega_m)$. At low frequencies, the solution is Eq. (\ref{3_23}).
  The pre-logarithmic factor $\beta$ there is the root of $\epsilon_{i\beta} =1$, where
    \beq
    \epsilon_{i\beta}
      =\frac{1-\gamma}{2}\frac{|\Gamma (\gamma /2(1 +2i\beta) )|^{2}}{\Gamma (\gamma )}\left(1+\frac{\cosh (\pi \gamma \beta)}{\cos (\pi \gamma /2)} \right)
    \label{su_15_2}
\eeq
The solution  exists for all $\gamma <1$, and $\beta$ approaches  a finite value $ 0.792 $ when $\gamma \to 1$
 (Fig. \ref{fig:beta}).
However, other quantities do become singular at $\gamma \to 1$.  We see from
  (\ref{s0}) and (\ref{w0}) that  the normal state self-energy diverges because
  $\omega_0 = {\bar g}/(1- \gamma)^{1/\gamma} \to \infty$.
   Accordingly, $z  = \omega_m/\Sigma_{\text{norm}} (\omega_m)$
   remains small at frequencies of order ${\bar g}$ and becomes $O(1)$ only at a much larger $\omega_m \sim \omega_0$. We show this in Fig.\ref{fig:normal_state_selfenergy}.
   Taken at a face value, this would imply that at $\gamma =1$, the corrections from the expansion in $\omega_m/\Sigma_{\text{norm}} (\omega_m)$ become totally irrelevant, and
  log-oscillations  of $\Delta_{\infty} (z)$ extend to all frequencies.
   This would  have a profound effect on the behavior of all other $\Delta_n (z)$ and on $T_{p,n}$, as
    it is set by a frequency at which log-oscillations end.

  We show that this is not the case, and  $\Delta_{\infty} (\omega_m)$ evolves smoothly through
   $\gamma =1$. Namely $\Delta_{\infty} (\omega_m)$
  displays log-oscillations only up to $\omega_m = O({\bar g})$, even  at $\gamma \to 1$, and decays as $1/z$
   at larger frequencies.
   We  show that this happens because the expansion in $\omega_m/\Sigma_{\text{norm}} (\omega_m)$ in the limit $\gamma \rightarrow 1$  actually holds in
     \beq
     y = \frac{z}{1-\gamma} = \left(\frac{|\omega_m|}{\bar g}\right)^\gamma
     \label{y_1}
     \eeq
so that the singularity in $z$ is canceled in this limit.

     To demonstrate this, we analyze the structure of the corrections to the log-oscillating form of
    $\Delta_{\infty} (z)$.   As we discussed in Paper I, there are two types of corrections  from the expansion in $\omega_m/\Sigma_{\text{norm}} (\omega_m)$:  local corrections, which come from fermions with  frequencies of order $\omega_m$, and non-local corrections, which come from  fermions with frequencies of order ${\bar g}$:
     $\Delta_{\infty} (\omega_m) = \Delta_{\infty, L} (\omega_m) + \Delta_{\infty,NL} (\omega_m)$
     The expansion in powers of
     $\omega_m/\Sigma_{\text{norm}} (\omega_m)$ comes from the local corrections, and  we analyze now the structure of these  corrections  for $\gamma \to 1$.

    The series of local corrections  can be obtained analytically in the order-by-order expansion.
     For a generic $\gamma <1$, this expansion holds in powers of $z$ with prefactors of order one.
     Specifically,
      \beq
    \label{eq:x<}
 \Delta_{\infty, L} (z) \propto  \frac{\sqrt{z}}{1+z} {\text Re} \left[ e^{(i \beta \log{z} + i\phi)}
  \sum_{m=0}^{\infty }
  C_m~ z^{m}\right]
  \eeq
   Here $C_m$, subject to  $C_0 =1$,
    are  complex coefficients given by
 \beq
  C_{m>0} = I_m  \displaystyle\prod_{{m'}=1}^{m} \frac{1}{I_{m'}-1},
  \label{dd_10}
  \eeq
  where
 \begin{widetext}
 \bea
&&I_{m'}  =  \frac{(1-\gamma)}{2} \frac{\Gamma((m'+1/2)\gamma +i{\beta} \gamma) \Gamma((1/2-m')\gamma -i{\beta} \gamma)}{\Gamma(\gamma)}  +  \nonumber \\
&&\frac{\Gamma(2-\gamma)}{2} \left(\frac{\Gamma((m'+1/2)\gamma +i{\beta} \gamma)}{\Gamma(1-(1/2-m')\gamma +i{\beta} \gamma)} + \frac{\Gamma((1/2-m')\gamma -i{\beta} \gamma)}{\Gamma(1-(m'+1/2)\gamma -i{\beta} \gamma)}\right),
\label{dd_7a}
\eea
\end{widetext}
and $\Gamma (...)$ are Gamma-functions.   The phase $\phi$ is a free parameter in $\Delta_{\infty, L} (z)$. Its value is set by the requirement that the total $\Delta_{\infty} (z) = \Delta_{\infty, L} (z) + \Delta_{\infty,NL} (z)$ decay as $1/z$ at large $z$.

\begin{figure}
  \includegraphics[width=10cm]{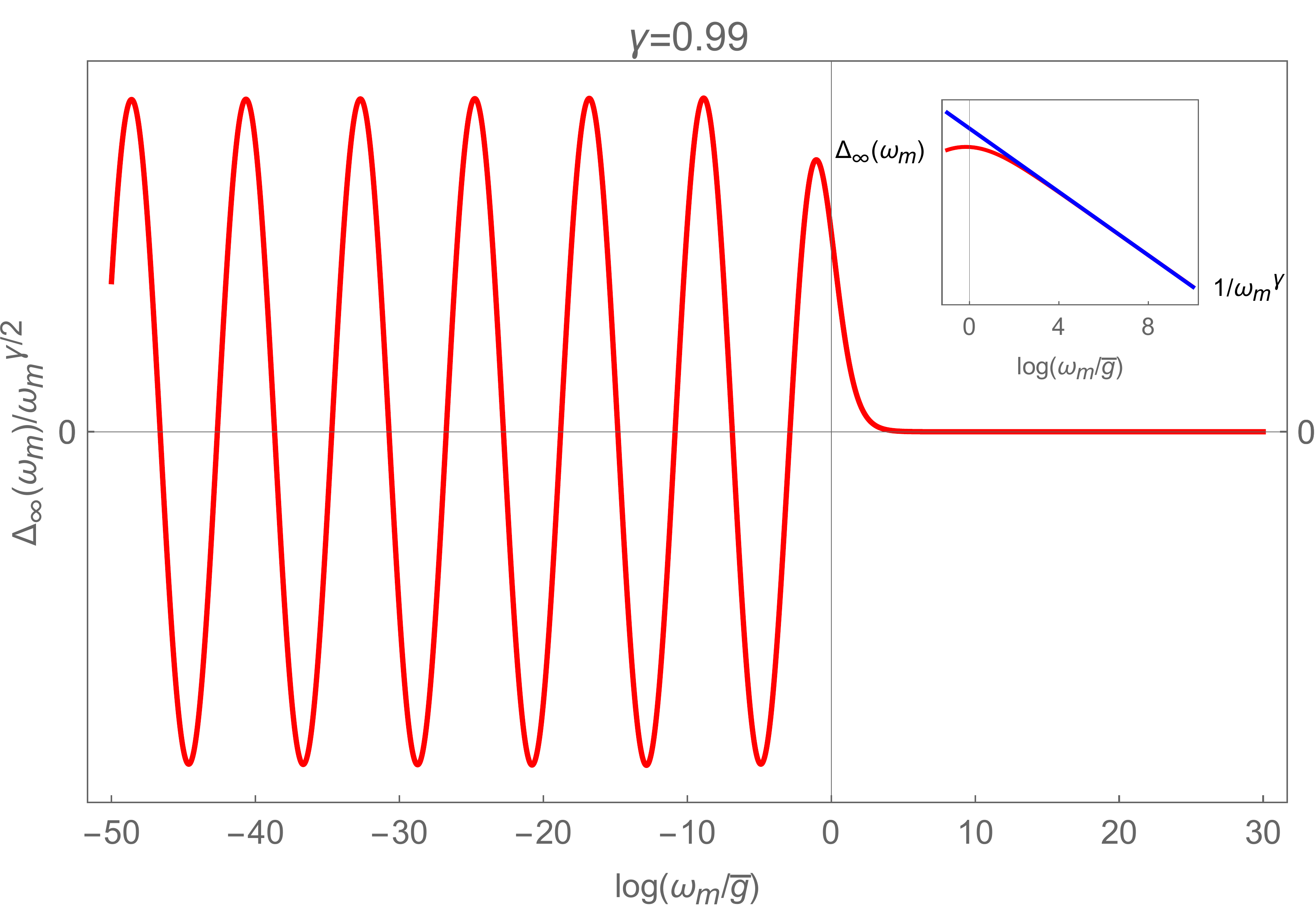}
  \caption{The exact $\Delta_{\infty} (\omega_m)$ for $\gamma \to 1$.  Log-oscillations of $\Delta_\infty(\omega_m)$ exist up to $\omega_m\sim \bar g$, like for smaller $\gamma$. At larger frequencies
   $\Delta_\infty(\omega_m)$ decays as $1/\omega_m^\gamma$ (the  upright inset).}\label{fig:Delta_inf}
\end{figure}

For $\gamma \approx 1$,  all $I_{m'}$ tend to 1, and the coefficients $C_m$ become singular.  Expanding $I_{m'}$ in
 (\ref{dd_7a}) near $\gamma =1$, we obtain $I_{m'} = 1 + (1-\gamma) {\bar I}_{m'}$, where
 \begin{eqnarray}
&& {\bar I}_{m'} = \frac{((-1)^{m'} -1) \pi}{2\cosh{(\pi \beta)}} + \nonumber\\
&&\frac{1}{2} \left(\Psi{(1/2 + i \beta)} + \Psi{(1/2 - i \beta)} - \Psi{(1/2 +m'  + i \beta)}- \Psi{(1/2-m' - i \beta)}\right) \nonumber \\
&&= \frac{((-1)^{m'} -1) \pi}{2\cosh{(\pi \beta)}} - \sum_{p=0}^{m'-1} \frac{1}{1/2+ i\beta +p}
\label{dd_23}
\end{eqnarray}
 where $\Psi (...)$ is a di-Gamma function.
  Substituting into (\ref{dd_10}), we find that the coefficients $C_m$ scale as
  $C_m~ \propto  1/(1-\gamma)^m$. Substituting these $C_m$  into (\ref{eq:x<}) we find that the expansion actually holds in  $ z/(1-\gamma) = (|\omega_m|/{\bar g})^\gamma$, which  is non-critical at $\gamma =1$.
    The corrections to log-oscillating behavior then become relevant  at a finite characteristic frequency $\omega_m \sim {\bar g}$.
    The same behavior can be detected by plotting the exact solution for $\Delta_{\infty} (\omega_m)$ for $\gamma \to 1$. We present the plot in Fig.  \ref{fig:Delta_inf}. We see  that
 $\Delta_{\infty} (\omega_m)$ indeed oscillates up to $\omega_m \sim {\bar g}$ and then decays as $1/|\omega_m|$.
 We emphasize again that the largest scale for the oscillations is a finite ${\bar g}$, despite that the
   expansion in frequencies in the exact solution  formally holds in powers of $z \propto (1-\gamma)$.  We discuss this issue in Appendix ~\ref{app:exact}.

\begin{figure}
  \includegraphics[width=10cm]{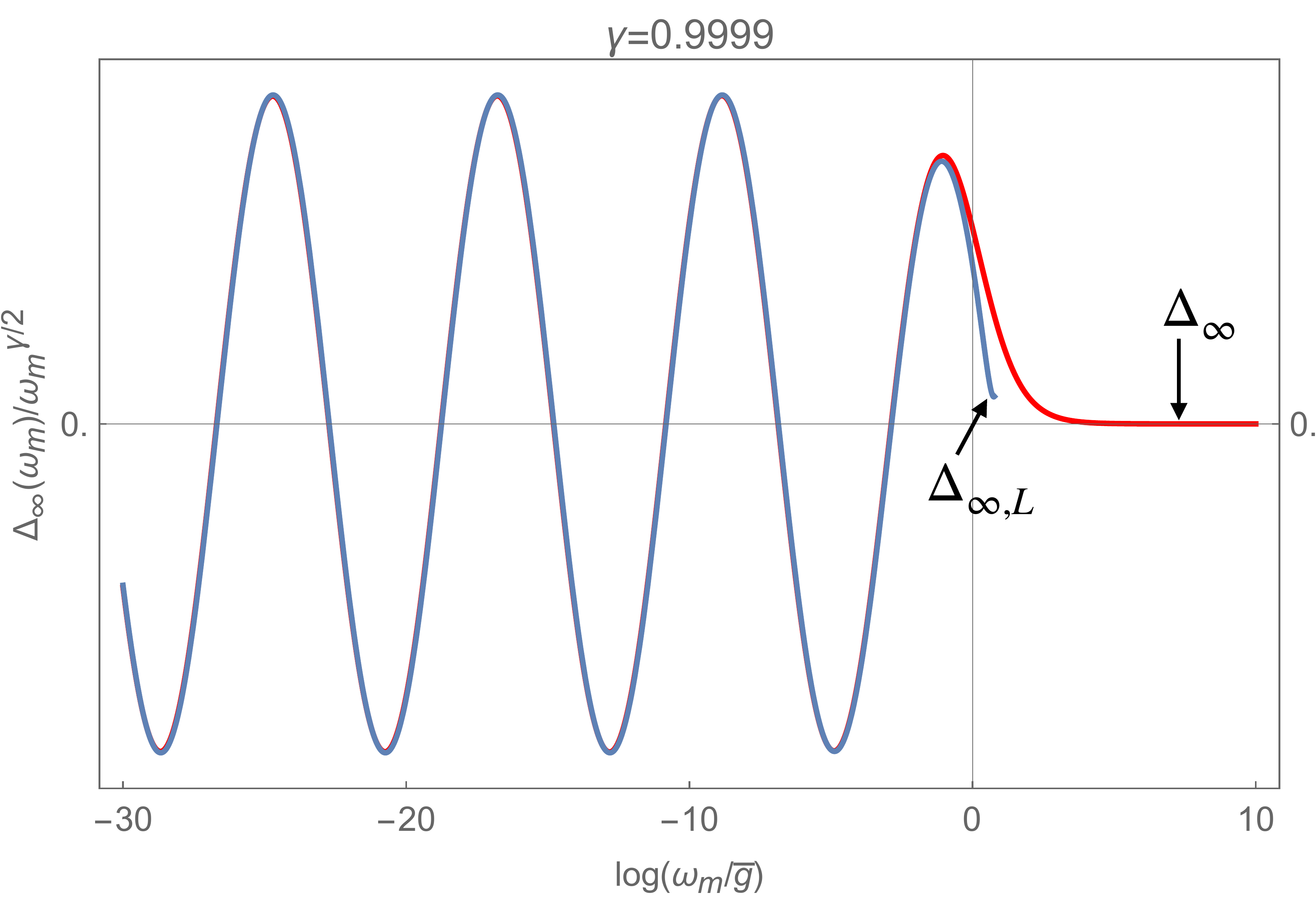}
  \caption{The 'local" part of the gap function,  $\Delta_{\infty, L} (\omega_m)$, along with the exact
  $\Delta_{\infty} (\omega_m)$ at $\gamma\to1$. Over the whole frequency range where
    $\Delta_{\infty} (\omega_m)$ oscillates, it is  almost exactly reproduced by  $\Delta_{\infty, L} (\omega_m)$.}
  \label{fig:compare}
\end{figure}

 We also note that at $m \gg 1$,  $z^m C_m \propto  (-1)^m ((|\omega_m|/{\bar g})/\log m)^m$.  Because of $\log m$ in the denominator,  the series in (\ref{eq:x<}) converge absolutely, i.e.,  one can obtain
   $\Delta_{\infty, L} (\omega_m)$ for {\it any} $|\omega_m|/{\bar g}$  by summing up enough  terms in the perturbation series, although in practice it can be done only up to some $\omega/{\bar g} \geq 1$.  We plot the result of the summation of $1000$ terms in Fig.\ref{fig:compare} along with the exact
    $\Delta_{\infty} (\omega_m)$ for
     $\gamma = 0.9999$.  We see that over the whole frequency range where $\Delta_{\infty} (\omega_m)$ oscillates, it practically coincides
      with $\Delta_{\infty,L} (\omega_m)$.  To reproduce the $1/|\omega_m|$ behavior at larger frequencies
        we would need to include the non-local part $\Delta_{\infty,NL} (\omega_m)$.

 \subsubsection{Nonlinear gap equation, $T=0$}

   We now look at the evolution of $\Delta_n (\omega_m)$ with some finite $n$. At $T=0$, $\Delta_{n} (\omega_m)$ tends to a finite value at $\omega_m \to 0$, and we first check whether $\Delta_n (0)$ remain continuous through $\gamma =1$.

   The gap function $\Delta_n (\omega_m)$ is the solution of the non-linear gap equation (\ref{ss_11_l0}).
    For $\gamma <1$,  one can safely move the term with $\Delta (\omega_m)$ to the l.h.s. of the gap equation and re-express it as
 \bea
&& \Delta_n (\omega_m)  \left[1 + \frac{{\bar g}^\gamma}{2 \omega_m} \int_0^\infty \frac{d \omega'_m ~ \omega'_m}{\sqrt{\Delta^2_n (\omega'_m)+ (\omega'_m)^2}} \left(\frac{1}{|\omega_m - \omega'_m|^\gamma} -\frac{1}{ |\omega_m + \omega'_m|^\gamma}\right)\right] = \nonumber \\
&&\frac{{\bar g}^\gamma}{2} \int_0^\infty \frac{d \omega'_m ~ \Delta_n(\omega'_m)}{\sqrt{\Delta^2_n (\omega'_m)+ (\omega'_m)^2}} \left(\frac{1}{|\omega_m - \omega'_m|^\gamma} +\frac{1}{|\omega_m + \omega'_m|^\gamma}\right)
\label{new_1}
\eea
 Each integral is non-singular in the infra-red limit, provided that $\Delta_n (0)$ is finite.  Then relevant
  $\omega'_m$ are finite, and at small $\omega$, one can expand  in the integrands as
\beq
\frac{1}{|\omega'_m \mp  \omega_m|^\gamma} \approx \frac{1}{|\omega'_m|^\gamma} \left(1 \pm \gamma \frac{\omega_m}{\omega'_m}\right)
 \label{new_2}
 \eeq
 Substituting the expansion into (\ref{new_1}) and taking the limit $\omega_m \to 0$, we obtain the condition on
 $\Delta_n (0)$:
 \beq
 \Delta_n (0) \left[1 + {\bar g}^\gamma \gamma \int_0^\infty \frac{d \omega'_m }{|\omega'_m|^\gamma \sqrt{\Delta^2_n (\omega'_m)^2+ (\omega'_m)^2}}\right] = {\bar g}^\gamma \int_0^\infty \frac{d \omega'_m  \Delta_n (\omega'_m)}{|\omega'_m|^\gamma \sqrt{\Delta^2_n (\omega'_m)+ (\omega'_m)^2}}
 \label{new_3}
 \eeq
 At $\gamma \to 1$, each integral diverges as $1/(1-\gamma)$, but the divergent terms cancel each other.
  As the result, $\Delta_n (0)$ remain finite at $\gamma \to 1$.  To see this more explicitly, consider the solution with $n=0$.  A sign-preserving  $\Delta_0 (\omega_m)$ remains roughly equal to $\Delta_0 (0)$ up to $\omega_m \sim \Delta_0 (0)$, at which both integrals in (\ref{new_3}) already converge.
   Approximating then $\Delta_0 (\omega'_m)$ by $\Delta_0 (0)$, we obtain from (\ref{new_3})
    \beq
 1 =  {\bar g}^\gamma (1-\gamma) \int_0^\infty \frac{d \omega'_m }{|\omega'_m|^\gamma \sqrt{(\Delta_0 (0))^2 + (\omega'_m)^2}}=(1-\gamma )\left(\frac{\bar{g}}{\Delta_{0}(0)} \right)^{\gamma }\frac{\Gamma \left(\frac{1}{2}-\frac{\gamma}{2} \right)\Gamma \left(\frac{\gamma}{2} \right)}{2\sqrt{\pi }}.
 \label{new_4}
 \eeq
This yields
\begin{equation}\label{eq:new4}
 \Delta_{0}(0)=\bar{g}\left[\frac{(1-\gamma ) \Gamma \left(\frac{1}{2}-\frac{\gamma}{2} \right)\Gamma \left(\frac{\gamma}{2} \right)}{2 \sqrt{\pi }} \right]^{1/\gamma }\approx  \bar{g}\left(1+(1-\gamma )\log 2 \right)
\end{equation}
   We see that $\Delta_0 (0) = {\bar g}$ at $\gamma \to 1$ from below.  At smaller $\gamma$, $\Delta_0 (0)$ increases in the same way as $T_{c,0}$, and the ratio $2\Delta_0 (0)/T_{c,0}$ remains of order one.  This is consistent with the more detailed study of $2\Delta_0 (0)/T_{c,0}$ ratio  in Ref.\cite{Wu_19}

 For $\gamma >1$,  $\int d x/|x|^\gamma$ diverges, and one cannot separate the two terms in the r.h.s. of (\ref{ss_11_l0}). However, we  can now use the identity
 \beq
 \int_{-\infty}^\infty d\omega'_m \frac{1 - \frac{\omega'_m}{\omega_m}}{|\omega_m-\omega'_m|^\gamma} =0
 \label{new_5}
 \eeq
 This identity holds for $\gamma >1$, but not for $\gamma \leq 1$.  Using (\ref{new_5}), we
  re-express the equation on $\Delta_n (\omega_m)$ as
 \bea
 &&\Delta_n(\omega_m) = \frac{{\bar g}^\gamma}{2} \int_{-\infty}^\infty \frac{d \omega'_m ~ \left(\Delta_n(\omega'_m)-\Delta_n (\omega_m)\right)}{\sqrt{\Delta^2_n (\omega'_m)+ (\omega'_m)^2}|\omega_m - \omega'_m|^\gamma} \nonumber \\
&&- \frac{{\bar g}^\gamma}{2} \int_{-\infty}^\infty d \omega'_m \frac{1 - \frac{\omega'_m}{\omega_m}}{|\omega_m - \omega'_m|^\gamma} \left[\sqrt{\Delta^2_n (\omega'_m)+ (\omega'_m)^2} - \Delta_n (\omega'_m)\right]
\label{new_6}
\eea
Each integral in (\ref{new_6}) is now regular.  In the limit $\omega_m \to 0$, one can again use (\ref{new_2}) and obtain
 \bea
\Delta_n(0) =
&& {\bar g}^\gamma \int_{0}^\infty \frac{d \omega'_m ~ \left(\Delta_n(\omega'_m)-\Delta_n (0)\right)}{\sqrt{\Delta^2_n (\omega'_m)+ (\omega'_m)^2}|\omega'_m|^\gamma} \nonumber \\
&&  + {\bar g}^\gamma (\gamma -1) \int_{0}^\infty d \omega'_m  \frac{\sqrt{\Delta^2_n (\omega'_m)+ (\omega'_m)^2} - \Delta_n (\omega'_m)}{\sqrt{\Delta^2_n (\omega'_m)+ (\omega'_m)^2}|\omega'_m|^\gamma}
\label{new_7}
\eea
Like we did for $\gamma <1$,  we set $n=0$ and approximate $\Delta_0 (\omega_m)$ by $\Delta_0 (0)$. Substituting into (\ref{new_7}), we obtain
 \bea
 \Delta_{0}(0) &=& {\bar g}^\gamma (\gamma -1) \int_{0}^\infty d \omega'_m  \frac{\sqrt{\Delta^2_0 (0) + (\omega'_m)^2} - \Delta_0(0)}{\sqrt{\Delta^2_0 (0) + (\omega'_m)^2}|\omega'_m|^\gamma} \nonumber \\
 &&
=\Delta_{0}(0)\left(\frac{\bar{g}}{\Delta_{0}(0)} \right)^{\gamma } \left[(1-\gamma) \frac{ \Gamma \left(\frac{1}{2}-\frac{\gamma }{2}\right) \Gamma \left(\frac{\gamma }{2}\right)}{2 \sqrt{\pi }}\right]
\label{new_8}
\eea
This gives exactly the same $\Delta_{0}(0)$ as \eqref{eq:new4}. This proves that $\Delta_0 (0)$ evolves continuously through $\gamma =1$.

 The verification that the same holds for $\Delta_n$ with a finite $n >0$  requires more efforts as one has to solve the actual non-linear gap equation for $\gamma <1$ and $\gamma >1$ and check whether the solutions match at $\gamma =1$.  This is technically quite challenging, but from physics perspective one should indeed expect
  $\Delta_n (\omega_m)$ to vary continuously through $\gamma =1$.

  \subsubsection{Linearized  gap equation, finite $T$}
\label{sec:e}

We next analyze how the onset temperatures for the pairing, $T_{p,n}$ change around $\gamma =1$.
 For a generic $\gamma <1$, we found in paper II that at large $n$, $T_{p,n} \propto e^{-\pi n/
 (\gamma \beta)}$.   We now show that this relation holds also for $\gamma \geq 1$, but the derivation requires more efforts than for $\gamma <1$.

  The computations  are more transparent when done for the  pairing vertex $\Phi (\omega_m)$,
   expressed via the normal state ${\tilde \Sigma}_{\text{norm}} (\omega_m)$. The gap function $\Delta (\omega_m) = \Phi (\omega_m) \omega_m/{\tilde \Sigma}_{\text{norm}} (\omega_m)$.
     We have from (\ref{eq:gapeq_l})
 \bea \label{eq:gapeq_a_11}
    \Phi (\omega_m) &=&
     {\bar g}^\gamma \pi T \sum_{m' \neq m} \frac{\Phi (\omega_{m'})}{|{\tilde \Sigma}_{\text{norm}} (\omega_{m'})|}
    ~\frac{1}{|\omega_m - \omega_{m'}|^\gamma}, \nonumber \\
     {\tilde \Sigma}_{\text{norm}}  (\omega_m) &=& \omega_m
   +  {\bar g}^\gamma \pi T \sum_{m' \neq m}
    ~\frac{{\text{sign}} (\omega_{m'}) }{|\omega_m - \omega_{m'}|^\gamma}
\eea
Evaluating ${\tilde \Sigma}_{\text{norm}} (\omega_m) \equiv  {\tilde \Sigma}_{\text{norm}} (m)$,  we obtain
\beq
{\tilde \Sigma}_{\text{norm}}(m) = \pi T \left(2m+1 + K A(m) {\text{sign}} (2m+1) \right)
\label{ee_7}
\eeq
 where
 \bea
 A (m) &=&  2 \sum_{1}^{m} \frac{1}{n^\gamma} , ~~m >0 \nonumber \\
 A (m) &=& A (-m-1), ~~ m<-1 \nonumber\\
 A (0) &=& A (-1) =0
 \eea
 and
\beq
K = \left(\frac{{\bar g}}{2\pi T}\right)^\gamma
\eeq
 The expression for
 $A(m)$
  is the same for $\gamma <1$ and $\gamma >1$.  The distinction is in that
 for $\gamma <1$, $A(m) \propto m^{1-\gamma}$, and for
 $\gamma >1$,
 $A(m)$
 tends to finite value at $m \to \infty$: $A(m \to \infty) = 2\zeta (\gamma)$, where $\zeta (\gamma)$ is a Zeta-function.
  Substituting the self-energy into the equation for $\Phi (\omega_m) = \Phi (m)$ and eliminating the term
   with $m=0$, we obtain
   \bea
 \Phi (m>0) &=&  \sum_{n=1,n\neq m}^\infty \frac{\Phi (n)}{A (n) + \frac{2n+1}{K}} \frac{1}{|n-m|^\gamma} + \sum_{n=1}^\infty \frac{\Phi (n)}{A (n) + \frac{2n+1}{K}}\frac{1}{(n+m+1)^\gamma} - \nonumber \\
  && \frac{K}{K-1}  \sum_{n=1}^\infty \frac{\Phi (n)}{A (n) + \frac{2n+1}{K}} \left(\frac{1}{n^\gamma} + \frac{1}{(n+1)^\gamma}\right) \left(\frac{1}{m^\gamma} + \frac{1}{(m+1)^\gamma}\right)
   \label{ch_31}
 \eea
 At small $T$, when $K \gg 1$,  we obtain from (\ref{ch_31}):
   \bea
 \Phi (m>0) &=&  \sum_{n=1,n\neq m}^\infty \frac{\Phi (n)}{A (n)} \frac{1}{|n-m|^\gamma} + \sum_{n=1}^\infty \frac{\Phi (n)}{A (n)}\frac{1}{(n+m+1)^\gamma} - \nonumber \\
  &&  \sum_{n=1}^\infty \frac{\Phi (n)}{A (n)} \left(\frac{1}{n^\gamma} + \frac{1}{(n+1)^\gamma}\right) \left(\frac{1}{m^\gamma} + \frac{1}{(m+1)^\gamma}\right)
   \label{ch_31_a}
 \eea
 For $m\gg 1$,  Eq. (\ref{ch_31_a}) reduces to
  \beq
 \Phi (m >0) =  \sum_{n=1,n\neq m}^\infty \frac{\Phi (n)}{A (n)} \frac{1}{|n-m|^\gamma} +
 \sum_{n=1}^\infty \frac{\Phi (n)}{A (n)} \frac{1}{(n+m)^\gamma}  - \frac{2}{m^\gamma}  \sum_{n=1}^\infty \frac{\Phi (n)}{A (n)} \left(\frac{1}{n^\gamma} + \frac{1}{(n+1)^\gamma}\right)
\label{ch_31_aa}
 \eeq
 One can easily verify that relevant $n$ in the sums are of order $m$, are also large.
 It is tempting to replace the sum by the integral, with the lower limit of order $T$.  However, this can be done only for $\gamma <1$, when the integral does not diverge.  Keeping $\gamma <1$,  replacing the summation by integration, and restoring Matsubara frequencies $\omega_m$ instead of Matsubara numbers,
  we obtain
  \beq
 \Phi (\omega_m) = \frac{1-\gamma}{2}  \int_{-\infty}^{\infty} d \omega'_m  \frac{\Phi (\omega'_{m})}{|\omega'_m|^{1-\gamma} |\omega_m - \omega'_{m}|^\gamma} - \frac{2(1-\gamma)(2\pi T)^\gamma}{|\omega_m|^\gamma} \int^\infty_{O(T)} \frac{\Phi (\omega'_{m})}{\omega'_{m}}
     \label{ss_111_l0}
  \eeq
At $\omega_m \gg  T$, the last term is irrelevant, and $\Phi (\omega_m)$ has the same form as at $T=0$:
 $\Phi (\omega_m) \propto |\omega_m|^{-\gamma/2} \cos{(\beta \log{(|\omega_m|/{\bar g})^\gamma} +\phi)}$.
  The phase $\phi$ is set by matching this form and $\Phi (\omega_m) \propto |\omega_m|^{-\gamma}$ at
  $\omega_m \sim {\bar g}$.  At $\omega_m \sim T$, i.e., at Matsubara numbers $m = O(1)$, the last term  cannot be neglected. However, it vanishes for certain $T$, then log-oscillating $\Phi (\omega_m)$ is the
   solution of the full Eq. (\ref{ss_111_l0}).  Substituting log-oscillating form into the last term we find that it vanishes when
   \beq
   \beta \gamma \log {\frac{T}{\bar g}} = n \pi + {\text{const}},  ~~n= 0,1,2,...
   \label{ee}
   \eeq
   This yields the set of $T_{p,n} \propto e^{-\pi n/(\beta \gamma)}$.  Because we assumed that $K \gg 1$, i.e., $2\pi T \ll {\bar g}$,  Eq. (\ref{ee}) is, strictly speaking,  valid for $n \gg 1$.

  For $\gamma >1$, one cannot convert the summation in (\ref{ch_31_aa}) into integration as the integral will be divergent. Instead, we use the fact that $A(\infty) = 2 \xi (\gamma)$ is now finite and do the following trick: \\

\begin{figure}
  \includegraphics[width=6cm]{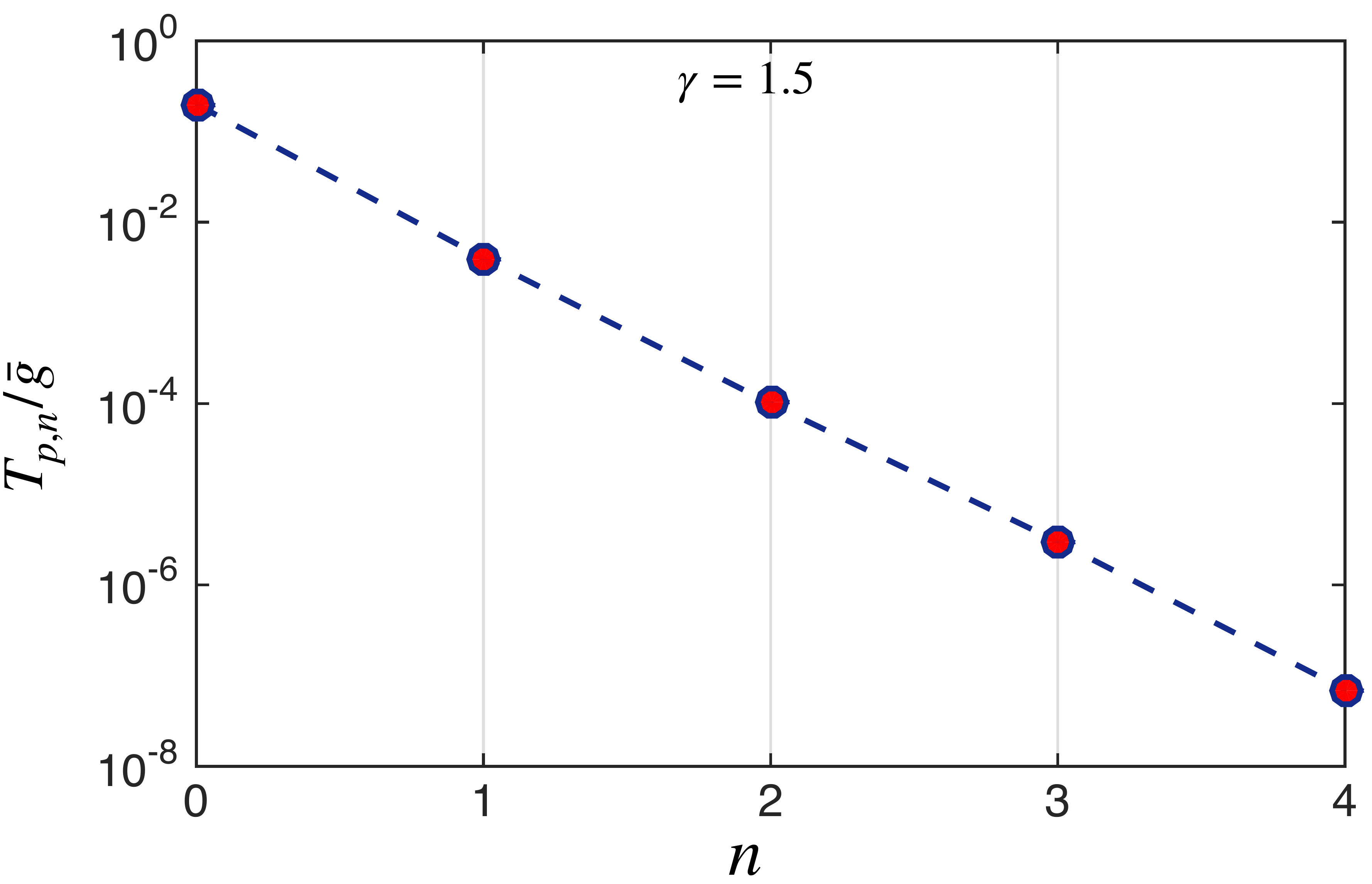}
  \caption{$T_{p,n}$ as a function of $n$ for $\gamma=1.5$. The onset temperature depends
  on $n$ exponentially, as $T_{p,n} \propto e^{-n\pi/(\beta \gamma)}$, like for $\gamma <1$.
   The slope of $\log(T_{p,n}/\bar g)$ vs $n$ is -3.716, in good agreement with the analytical
    result 3.785.}\label{fig:Tpng15}
\end{figure}

 (i)  rewrite the normal state  ${\tilde \Sigma}_{\text{norm}} (m)$  as
 \beq
{\tilde \Sigma}_{\text{norm}} (m)  = \pi T \left(2m+1 + K A(\infty) - K(A(\infty)-A(m))\right) {\text{sign}} (2m+1)
 \label{ee_1}
\eeq
(ii)  introduce  ${\bar{\tilde \Sigma}}_{\text{norm}} (m)$ via
\beq
 {\bar {\tilde \Sigma}}_{\text{norm}} (m) = {\tilde \Sigma}_{\text{norm}}(m) \left(1 - \frac{\pi T K A(\infty)}{{\tilde \Sigma}_{\text{norm}}(m) }\right) = -\pi T K \left(A (\infty) - A(m)\right)
 \label{ee_2}
 \eeq
(ii) introduce simultaneously
 \beq
 {\bar \Phi} (m) = \Phi (m) \left(1 - \frac{\pi T K A(\infty)}{{\tilde \Sigma}_{\text{norm}}(m) }\right)
 \label{ee_3}
 \eeq
Because $\Phi (m)/{\tilde \Sigma}_{\text{norm}} (m) = {\bar \Phi} (m)/{\bar {\tilde \Sigma}}_{\text{norm}} (m)$,  Eq. (\ref{ch_31_aa}) becomes
   \bea
 {\bar \Phi} (m>0) &=&  \sum_{n=1,n\neq m}^\infty \left(\frac{{\bar \Phi} (m)}{A (\infty)- A(m)}- \frac{{\bar \Phi} (n)}{A (\infty)- A(n)}\right) \frac{1}{|n-m|^\gamma} \nonumber\\
 && + \sum_{n=1}^\infty
  \left(\frac{{\bar \Phi} (m)}{A (\infty)- A(m)}- \frac{{\bar \Phi} (n)}{A (\infty)- A(n)}\right)
  \frac{1}{(n+m)^\gamma} - \nonumber \\
  &&  \frac{2}{m^\gamma}  \left(\sum_{n=1}^\infty \frac{{\bar \Phi} (n)}{A (\infty)- A(n)} \left(\frac{1}{n^\gamma} + \frac{1}{(n+1)^\gamma}\right) + \frac{{\bar \Phi} (m)}{A (\infty)- A(m)}\right)
\label{ee_4}
 \eea
Converting the summation over $n$ into integration, we see that the integral is now free from divergencies.
 Using that at large $m$, $A(\infty) - A(m) = m^{1-\gamma}/(\gamma-1)$ and replacing $m$ by $\omega_m$, we obtain
  \bea
 {\bar \Phi} (\omega_m) &=&  \frac{\gamma-1}{2}
 \int_{-\infty}^\infty  d \omega'_m \left({\bar \Phi} (\omega_m)|\omega_m|^{\gamma-1} - {\bar \Phi} (\omega'_m)|\omega'_m|^{\gamma-1}\right) \frac{1}{|\omega_m-\omega'_m|^\gamma} - \nonumber \\
  &&2 \left(\frac{2\pi T}{|\omega_m|}\right)^\gamma
  \left(\int_{O(T)}^\infty \frac{{\bar \Phi} (\omega'_m)}{\omega'_m} + \pi T \frac{{\bar \Phi} (\omega_m)}{\omega_m}\right)
\label{ee_5}
 \eea
At $\omega_m \gg 2\pi T$, the last term can be neglected, and we obtain
 \beq
 {\bar \Phi} (\omega_m) =  \frac{\gamma-1}{2}
 \int_{-\infty}^\infty  d \omega'_m \left({\bar \Phi} (\omega_m)|\omega_m|^{\gamma-1} - {\bar \Phi} (\omega'_m)|\omega'_m|^{\gamma-1}\right) \frac{1}{|\omega_m-\omega'_m|^\gamma}
\label{ee_6}
 \eeq
 The solution of this equation is the same log-oscillating function
 $\Phi (\omega_m) \propto |\omega_m|^{-\gamma/2} \cos{(\beta \log{(|\omega_m|/{\bar g})^\gamma} +\phi)}$
  as for $\gamma <1$, and $\beta$ is again determined by $\epsilon_{i\beta} =1$, where $\epsilon_{i\beta}$ is given by Eq. (\ref{su_15_2}).  Like for $\gamma <1$, the set of $T_{p,n}$, where Eq. (\ref{ee_5}) is valid, is determined by the condition that the last term in (\ref{ee_5}) vanishes.  Substituting log-oscillating form of $\Phi(\omega_m)$, we find the same condition on $T_{p,n}$ as in Eq. (\ref{ee_5}):
   $T_{p,n} \propto e^{-\pi n/(\gamma \beta)}$.  In Fig.\ref{fig:Tpng15}  we show numerical result for  $T_{p,n}$ for $\gamma =1.5$ as a function of $n$.   We see that its dependence on $n$ is exponential, like for $\gamma <1$.

\begin{figure}
  \includegraphics[width=8cm]{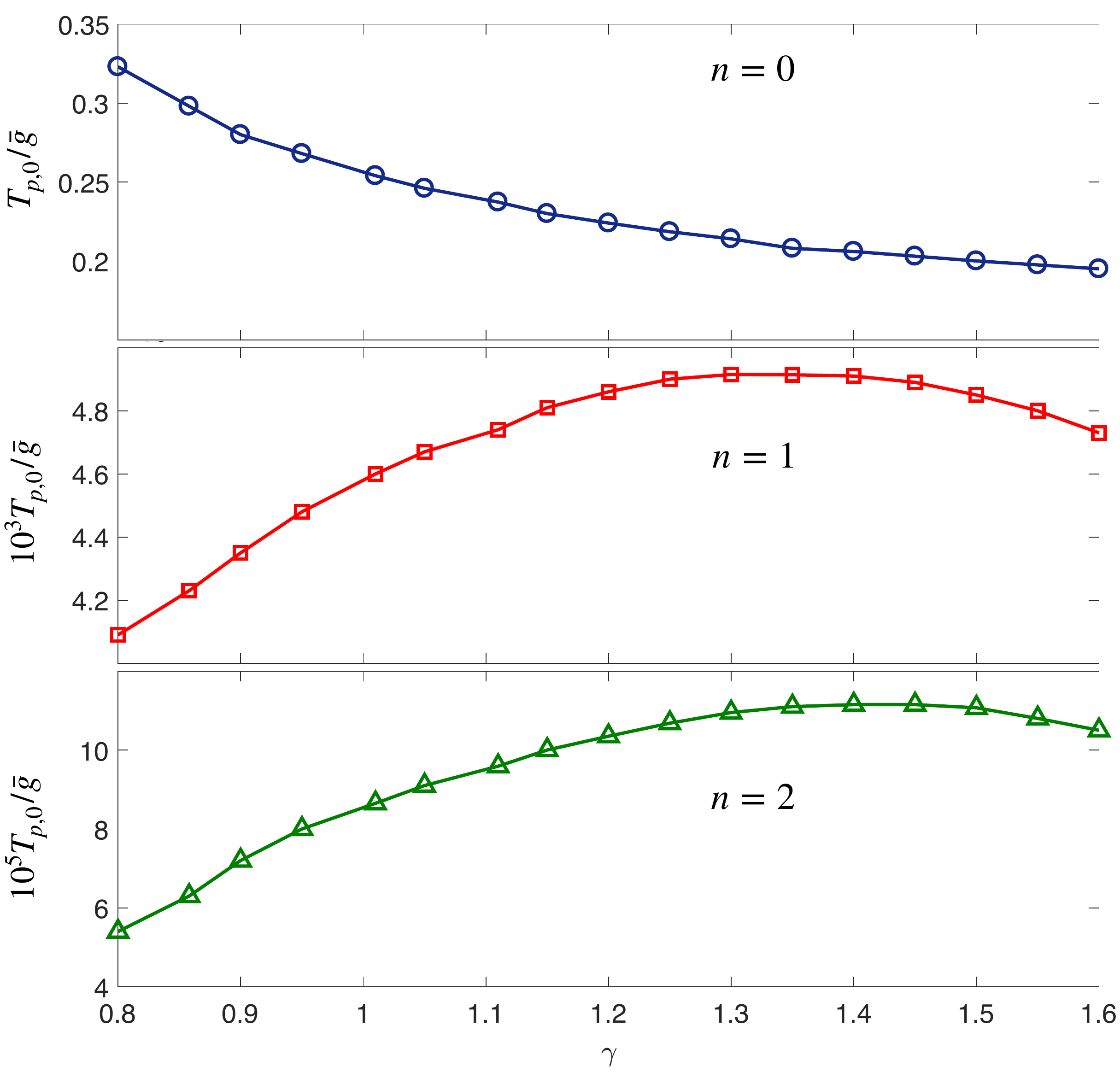}
  \caption{ Variations of the onset temperatures for the pairing, $T_{p,n}$, through $\gamma = 1$  for $n=0,1,2$.}\label{fig:Tpn}
\end{figure}

 The computation of the prefactor for $T_{p,n}$ requires more efforts, and we didn't find it analytically.
 In Fig. \ref{fig:Tpn} we show numerical results for  the onset temperatures $T_{p,n}$ for $\gamma$ around one and  $n = 0, 1, 2$.  We see that all $T_{p,n}$  evolve smoothly through $\gamma =1$.

\subsubsection{Nonlinear  gap equation, finite $T$}
\begin{figure}
  \includegraphics[width=8cm]{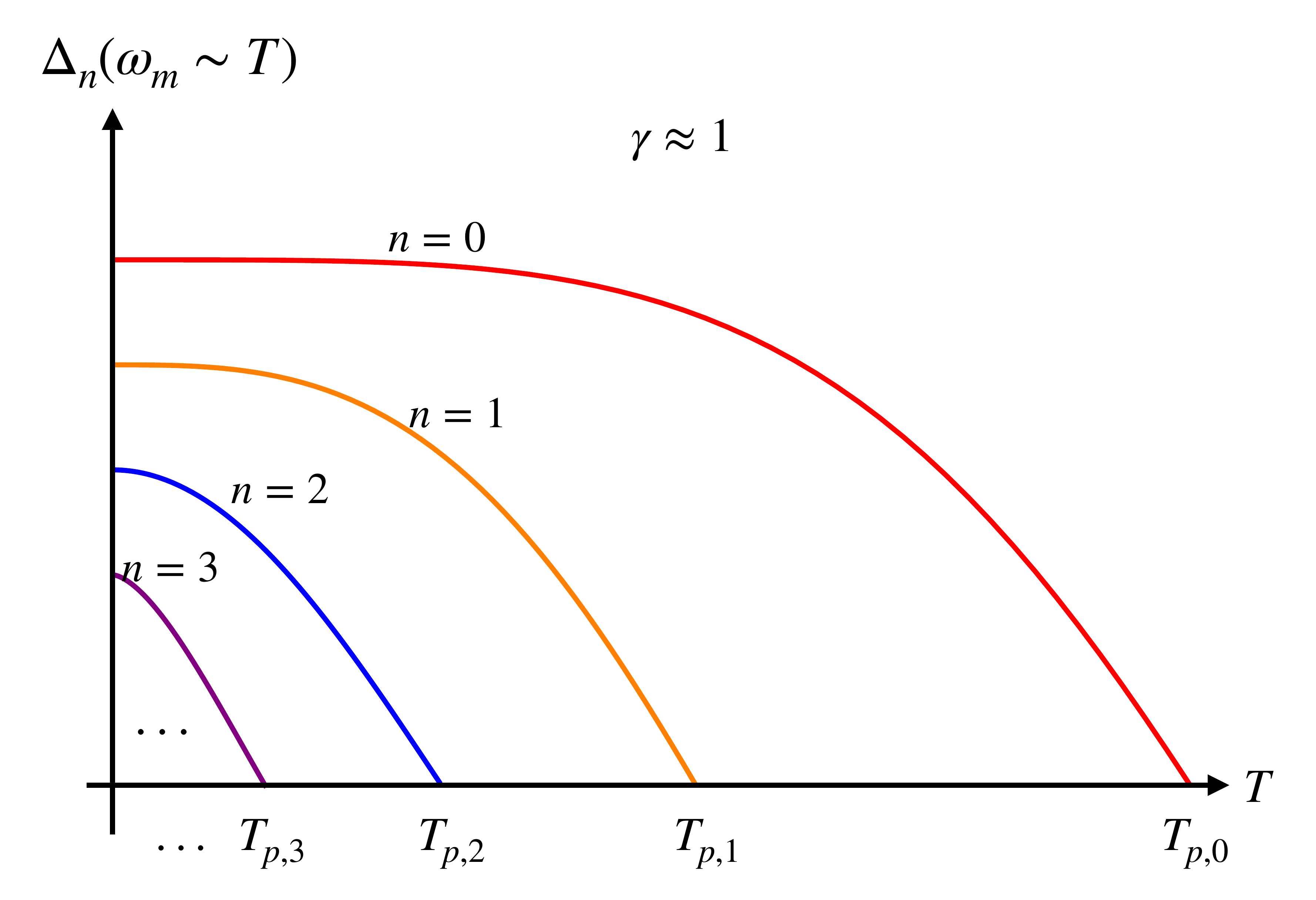}
  \caption{The sketch of the behavior of  $\Delta_n(\omega_m\sim T)$. The gap functions with different $n$ emerge
   at different onset temperatures $T_{p,n}$ and at $T =0$ have different overall magnitudes. The behavior of $\Delta_n(\omega_m\sim T)$ in the extended $\gamma$-model with $N \neq 1$ is different, see Fig. \ref{fig:nonlinear}.}
   \label{fig:nonlinear_a}
\end{figure}

We did not attempt to solve the non-linear gap equation at a finite $T < T_{p,n}$.  Given that
 $\Delta_n (\omega_m)$ with different $n$ are topologically distinct, and that there is a set of
  $\Delta_n (\omega_m)$ at $T=0$, we conjecture that the amplitude of $\Delta_n (\omega_m)$, which emerges at $T_{p,n}$, increases as $T$ decreases, and at $T =0$ it coincides with the $n$th solution of the non-linear gap equation.  We illustrate this in
Fig. \ref{fig:nonlinear_a}.

 \section{Extension to $N \neq 1$}
\label{sec:N}

We now extend the model and introduce a parameter $N$, which controls the relative strength of the interactions in the particle-hole and particle-particle channels.  Like we said in the Introduction, we treat $N$ as a continuous variable.   With this
extension,
 \bea \label{eq:gapeq_a}
    \Phi (\omega_m) &=&
     \frac{{\bar g}^\gamma}{N}  \pi T \sum_{m' \neq m} \frac{\Phi (\omega'_{m})}{\sqrt{{\tilde \Sigma}^2 (\omega'_{m}) +\Phi^2 (\omega'_{m})}}
    ~\frac{1}{|\omega_m - \omega'_{m}|^\gamma}, \nonumber \\
     {\tilde \Sigma} (\omega_m) &=& \omega_m
   +  {\bar g}^\gamma \pi T \sum_{m' \neq m}  \frac{{\tilde \Sigma}(\omega'_m)}{\sqrt{{\tilde \Sigma}^2 (\omega'_{m})  +\Phi^2 (\omega'_{m})}}
    ~\frac{1}{|\omega_m - \omega'_{m}|^\gamma}
\eea
and
\beq
   \Delta (\omega_m) = \frac{{\bar g}^\gamma}{N} \pi T \sum_{m' \neq m} \frac{\Delta (\omega_{m'}) - N \Delta (\omega_m) \frac{\omega_{m'}}{\omega_m}}{\sqrt{(\omega_{m'})^2 +\Delta^2 (\omega_{m'})}}
    ~\frac{1}{|\omega_m - \omega_{m'}|^\gamma}.
     \label{ss_11_b}
  \eeq
 Note that here, like in earlier papers, we extend  Eliashberg equations to $N \neq 1$ after
 cancelling out the divergent contribution from thermal fluctuations (the $m'=m$ term in the sum over Matsubara frequencies). An alternative approach, suggested in Ref. \cite{Wang_H_18}, is to extend to $N \neq 1$ without first subtracting the $m'=m$ terms in Eq. (\ref{ss_11_b}).
  In this case, one has to regularize the divergencies in the r.h.s of these equations and also in the gap equation. In general, the contribution from thermal fluctuations has to be computed differently from other terms in the frequency sum because one cannot factorize the momentum integration based on the separation between fast electrons and slow bosons. We refer a reader to Refs\cite{Wang_H_18,Abanov_19,*Wu_19_1,Chubukov_2020a},
   where  this issue has been addressed in detail.

We now consider how the solutions of the gap equation, $\Delta_n (\omega_m)$, evolve near $\gamma =1$.
  For this we consider separately the cases $\gamma <1$ and $\gamma >1$.

\subsection{A generic $\gamma <1$}

We first briefly summarize the results for a generic $\gamma <1$  (Parts I and II)
 and then move to $\gamma \to 1$.

\subsubsection{Linearized gap equation, $T=0$.}

The linearized gap equation at $T=0$ is
 \beq
    \label{eq:gapeq_d_a}
    \Delta_{\infty} (\omega_m) = \frac{{\bar g}^\gamma}{2N}  \int d \omega'_m \frac{\Delta_{\infty} (\omega'_{m}) - N \Delta_{\infty} (\omega_m) \frac{\omega'_{m}}{\omega_m}}{|\omega'_{m}|}
    ~\frac{1}{|\omega_m - \omega'_{m}|^\gamma},
    \eeq
 or, equivalently,
 \beq
 D_\infty(\omega_m) \omega_m \left(1 + \lambda \left(\frac{\bar g}{|\omega_m|}\right)^\gamma \right) = \frac{{\bar g}^\gamma}{2N} \int d \omega'_m \frac{D_\infty(\omega'_m)-D_\infty (\omega_m)}{|\omega_m-\omega'_m|^\gamma} {\text{sign}} \omega'_m,
 \label{3_11}
 \eeq
  where $D(\omega_m) = \Delta (\omega_m)/\omega_m$ and $\lambda = (1-1/N)/(1-\gamma)$.

\begin{figure}
  \includegraphics[width=6cm]{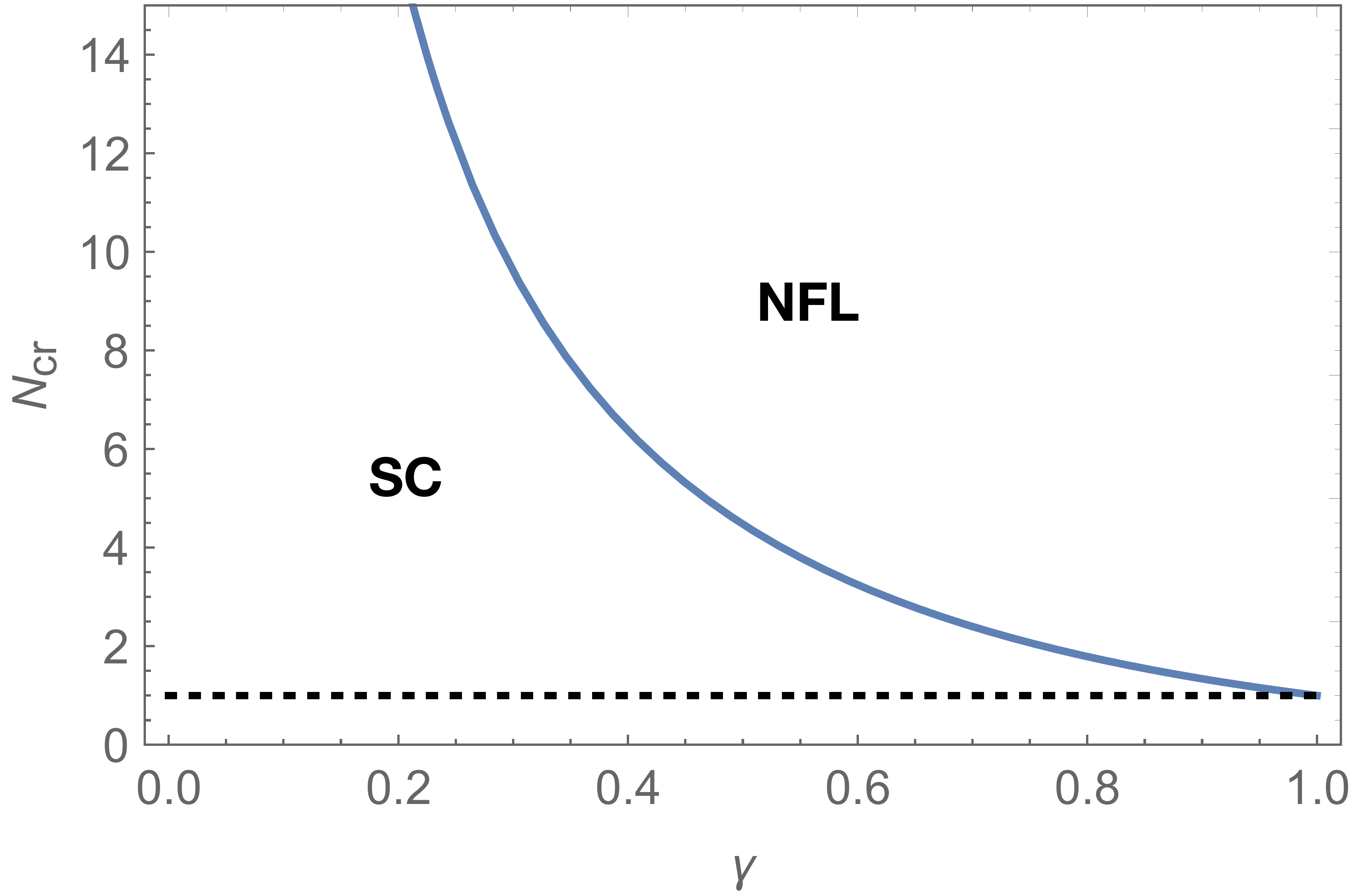}
  \caption{$N_{cr}$ from Eq. (\ref{su_15_1}) as a function of $\gamma$. This critical $N$  separates a SC state at $N < N_{cr}$ and a NFL normal state at $N > N_{cr}$.  At $\gamma \to 1$, $N_{cr} \to 1$.}
  \label{fig:Ncr}
\end{figure}

  A non-zero solution, $\Delta_{\infty} (\omega_m)$,  exists for $N < N_{cr}$, where
\beq
N_{cr}=\frac{1-\gamma}{2}\frac{\Gamma^{2} (\gamma /2)}{\Gamma (\gamma )}\left(1+\frac{1}{\cos (\pi \gamma /2)} \right),
\label{su_15_1}
\eeq
We plot $N_{cr}$ vs $\gamma$ in Fig.\ref{fig:Ncr}.
 For all $\gamma <1$, $N_{cr} >1$.
Similar to the case $N=1$, $\Delta_{\infty} (\omega_m)$ undergoes log-oscillations at $\omega_m < {\bar g}$:   $\Delta_{\infty} (\omega_m) \propto |\omega_m|^{\gamma/2} \cos (\beta_N \log{(|\omega_m|/{\bar g})^\gamma} + \phi)$, where $\beta_N$ is the solution of $\epsilon_{i\beta_N} = N$ and $\epsilon_{i\beta}$ is given by (\ref{su_15_2}).
 A non-zero $\beta_N$ exists for $N < N_{cr}$. For $N \lesssim N_{cr}$, $\beta_N \propto (N_{cr}-N)^{1/2}$.

\subsubsection{Linearized gap equation, $T \neq 0$.}

At a finite $T$, the solution of the linearized gap equation exists for a set of critical temperatures, $T_{p,n}$, like for $N=1$.  An eigenfunction $\Delta_n (\omega_m)$ changes sign $n$ times as a function of discrete Matsubara frequency $\omega_m = \pi T (2m +1)$.  All critical lines $T_{p,n} (N)$ for $n >0$ terminate at $T=0$ at $N = N_{cr}$, while
 $T_{p,0}$ scales as $1/N^{1/\gamma}$ for large $N$ (Ref. \cite{Wang2016}).
 At $N = O(1)$, $T_{p,n} \propto e^{-An}$ for large $n$.

\subsection{The limit $\gamma \to 1$}

\subsubsection{Linearized gap equation, $T=0$.}

\begin{figure}
  \includegraphics[width=7cm]{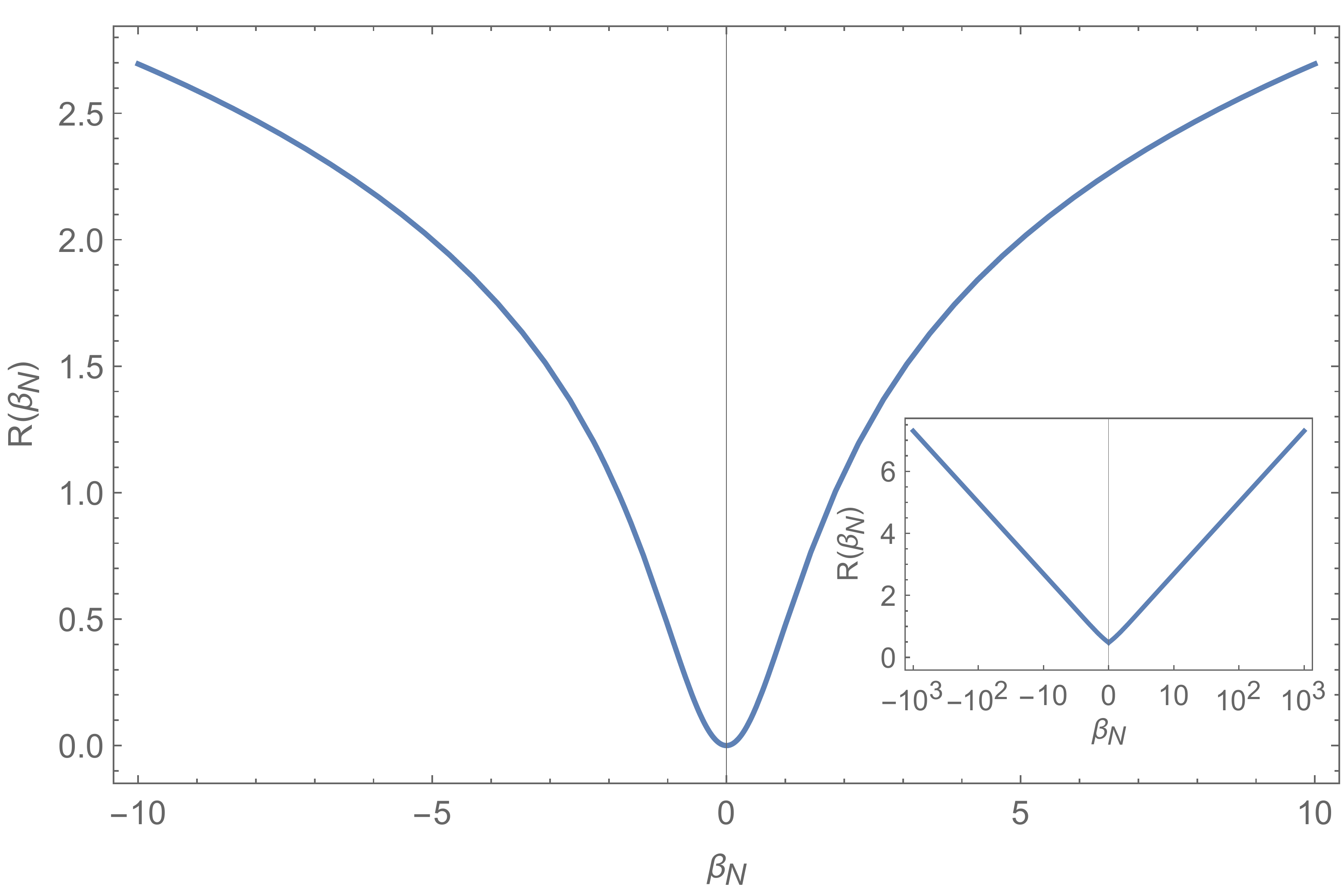}
  \caption{
$R(\beta_N)$ from Eq. (\ref{an_1})  as a function of $\beta_N$. At large $\beta_N$, $R(\beta_N) \approx \log{|\beta_N|}$.}
  \label{fig:Rbeta}
\end{figure}

In the limit $\gamma \to 1$, $N_{cr} =1 + (\pi /2+\log (4)) (1-\gamma) + O(1-\gamma)^2$ tends to $1$, i.e., relevant $N < N_{cr}$ become $N \leq 1$.  Simultaneously, the function $\epsilon_{i\beta}$ becomes flat:
$$
\epsilon_{i\beta_N }\approx N_{cr}-(1-\gamma )R(\beta_N )  =1 + \left(\frac{\pi}{2}+\log (4) - R(\beta_N)\right),
$$
where
\beq
 R(\beta_N) =  \frac{1}{2} \left(\Psi{(1/2 + i \beta_N)}+ \Psi{(1/2- i \beta_N)}\right) - \Psi (1/2) - \frac{\pi}{2} \left(1-\frac{1}{\cosh{\pi \beta_N}}\right)
 \label{an_1}
 \eeq
  We plot $R(\beta_N)$ in Fig. \ref{fig:Rbeta}. At large $\beta_N$, $R(\beta_N) \approx \log{|\beta_N|}$.
   Because $\epsilon_{i\beta_N }$ becomes flat,
   $\beta_N$ remains finite for $N = 1$, but exponentially grows for any $N <1$ and becomes infinite
     at $\gamma = 1$.   This implies that the system behavior
  at $N=1$ and $N <1$ changes discontinuously at $\gamma = 1$.   To understand this change, it is instructive
   to consider the double limit when both $\gamma$ and $N$  tend to one, and $\beta_N$ is a
    continuous function of the ratio $(1-N)/(1-\gamma)$, or, equivalently, of $(N_{cr} -N)/(1-\gamma) = R(\beta_N)$.
For $N = N_{cr}$, $\beta_{N_{cr}} =0$, for $N=1$, $\beta_{N=1} =\beta$ tends to $0.792$, and for
$1-N \gg (1-\gamma)$, $\beta_N \approx 0.561 /N^{1/(1-\gamma )} \gg 1$.  The case $N <1$ and $\gamma \to 1$ corresponds to the limit $\beta_N \rightarrow \infty$.  We emphasize that a continuous evolution is only possible
  if we keep  $N$ as a continuous parameter.

   \begin{figure}
  \includegraphics[width=.9\columnwidth]{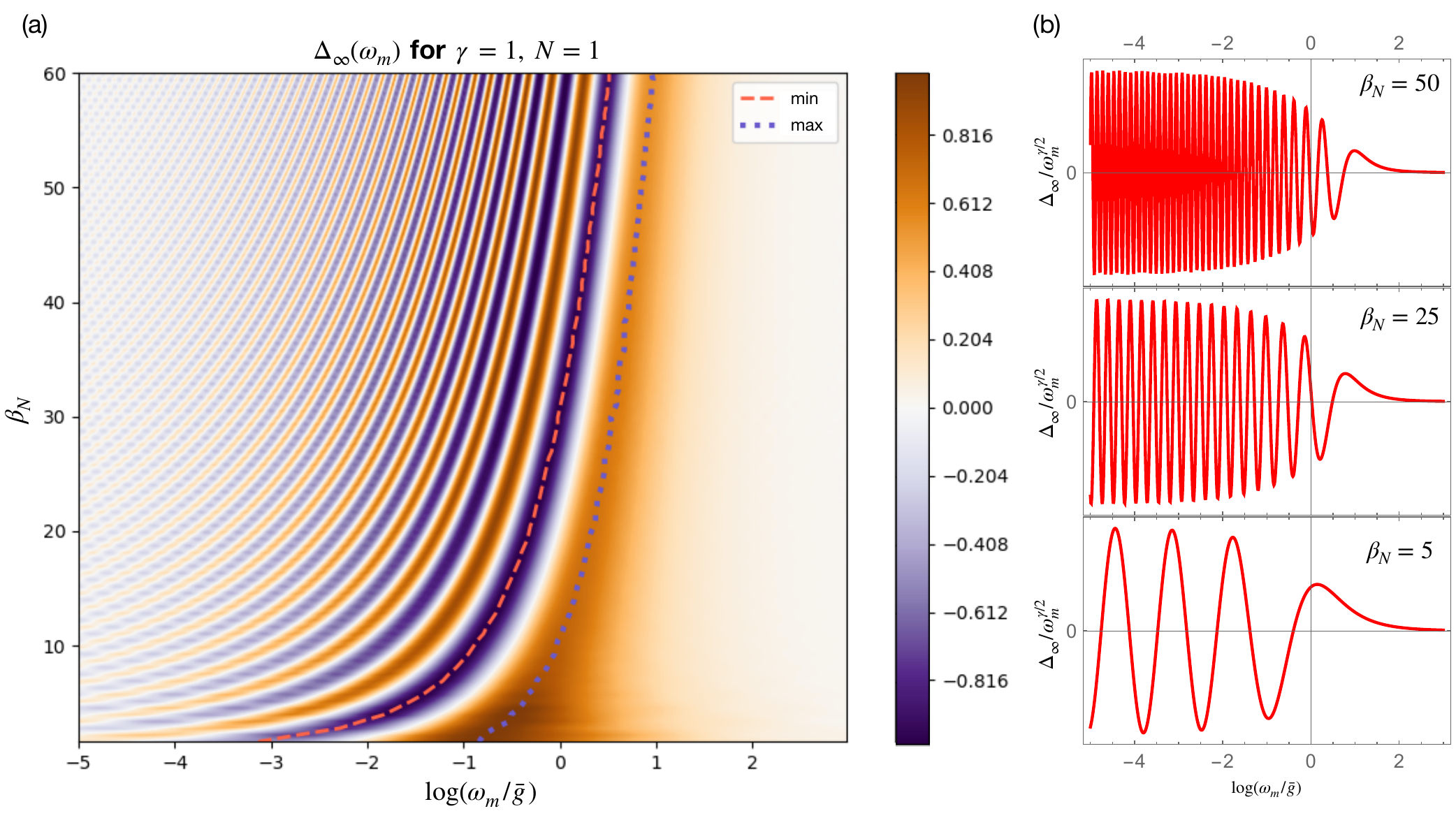}
  \caption{The exact $\Delta_{\infty} (\omega_m)$ for $\gamma \to 1$, $N \to 1$ and
   different $\beta_{N}$, which depends on the ratio $(1-N)/(1-\gamma)$.
    Right panel:  $\Delta_\infty (\omega_m )/|\omega_m|^{1/2}$ as a function of $\log{(|\omega_m|/{\bar g})}$.  As $\beta_N$ increases, oscillations of $\Delta_\infty (\omega_m )$ extend to larger frequencies.  Left panel: color plot of $\Delta (\omega_m )$, normalized to max $(|\Delta_\infty (\omega_m )|)$. The lines show the positions of the first maximum of the oscillations (blue dotted line) and the first minimum (red dashed line). }\label{fig:ExactG1}
\end{figure}
\begin{figure}
	\includegraphics[width=7cm]{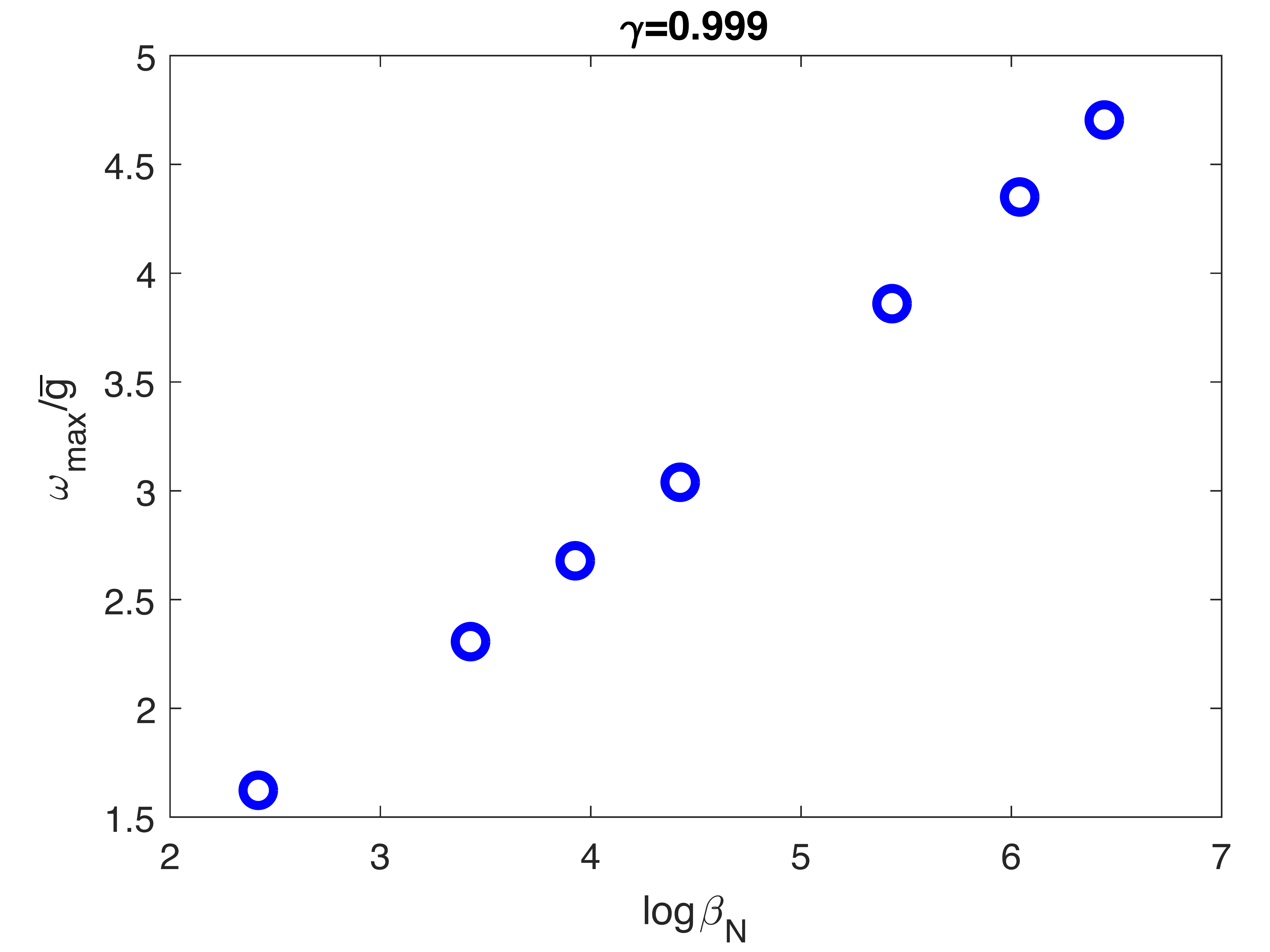}
	\caption{Numerical results  for the dependence of  the largest frequency for oscillations of $\Delta_{\infty} (\omega_m)$,  $\omega_{max}$,  on $\beta_N$.  The data show that $\omega_{max} \propto \log\beta_N$.}\label{fig:omegamax_logbeta}
\end{figure}

The exact solution for $\Delta_{\infty} (\omega_m)$ can be obtained for any $\beta_N$.   We plot $\Delta_{\infty} (\omega_m)$ for different $\beta_{N}$ in Fig. \ref{fig:ExactG1}. To demonstrate the behavior over a large range of frequencies, we  use $\log(\omega_m/{\bar g})$ as a variable. We see  that for $\beta_N = O(1)$, $\Delta_{\infty} (\omega_m)$  oscillates on the logarithmic scale for $\omega_m \leq {\bar g}$  and decreases as $1/|\omega_m|$ at larger frequencies. This agrees with our earlier result for $N \equiv 1$.  However, as $\beta_N$ increases,
 new non-logarithmic oscillations develop at $\omega_m \geq 1$ and extend up $\omega_{max}$.  Numerical results strongly indicate that for large enough $\beta_N$, $\omega_{max} \sim {\bar g} \log {\beta_N}$, see Fig.\ref{fig:omegamax_logbeta}.
  This is expected on general grounds because ${\bar g} \log {\beta_N} \sim {\bar g} (1-N)/(1-\gamma)$, and the latter is the scale at which divergencies in the gap equation are cut when $N \leq 1$.
 We also see from Fig.\ref{fig:ExactG1}(b)
   the overall magnitude of $\Delta_{\infty} (\omega_m)$ decreases with ${\omega_m}$, while
    the period of oscillations increases.

\begin{figure*}
  \includegraphics[width=16cm]{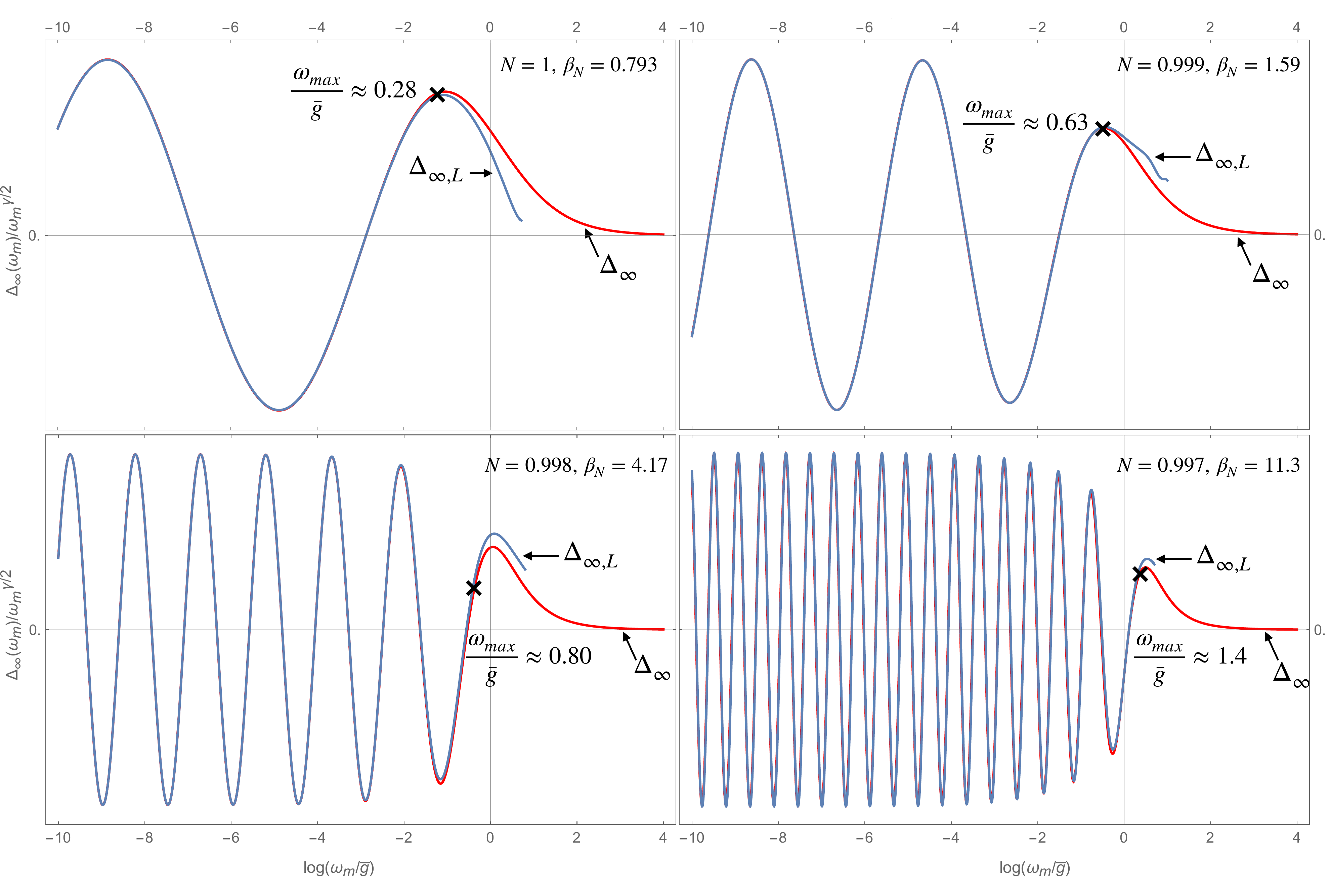}
  \caption{Comparisons between $\Delta_{\infty}(\omega_m)$ and $\Delta_{\infty,L}(\omega_m)$  at $\gamma\to 1$ for different $\beta_N$.  Both are plotted as functions of $y = \log{(\omega_m/{\bar g})}$.  We adjusted a free phase factor  in $\Delta_{\infty, L} (\omega_m)$) to match $\Delta_{\infty} (\omega_m)$ at small $\omega_m$.  The two functions nearly coincide up to $y_{max}$, over the full frequency range where  $\Delta_{\infty,L} (\omega_m)$ oscillates.   The scale $y_{max}$  increases with increasing $\beta_N$.}\label{fig:largebeta}
\end{figure*}

 To rationalize this observation we again compute the local series $\Delta_{\infty, L} (\omega_m)$. We have
 \beq
 \Delta_L (\omega_m) \propto  |\omega_m|^{1/2} Re \left[e^{i(\beta_N \log{|\omega_m|/{\bar g}}+ \phi)} \sum_{m=0}^\infty {\tilde C}^N_m \left(\frac{|\omega_m|}{\bar g} \right)^m \right]
\label{dd_23_aa}
 \eeq
 where
 ${\tilde C}^N_m =\displaystyle\prod_{m'=1}^{m} \frac{1}{{\bar I}^N_{m'}}$, and
 \beq
 {\bar I}^N_{m'} =  \frac{((-1)^{m'} -1) \pi}{2\cosh{(\pi \beta_N)}} - \sum_{p=0}^{m'-1} \frac{1}{1/2+ i\beta_N +p}
\label{dd_23a}
 \eeq
  The series again converge absolutely, i.e., $\Delta_L (\omega_m)$ can be obtained for any $\omega_m$ by summing up enough terms in the series.
  In Fig. \ref{fig:largebeta} we show both the exact $\Delta_{\infty} (\omega_m)$ and $\Delta_{\infty,L} (\omega_m)$.  We see that they nearly coincide over the full range where $\Delta_{\infty} (\omega_m)$ oscillates.  We can also expand the series in (\ref{dd_23_aa}) in $1/\beta_N$ and obtain the analytical expansion in $\omega_m/{\bar g}$ for the overall factor of
  $\Delta_{\infty,L} (\omega)$ and the period of oscillations. To leading order in $\beta_N$ we find after straightforward but lengthy calculation:
     \beq
 \Delta_{\infty,L} (\omega_m) \propto |\omega_m|^{1/2} f_1 \left(\frac{|\omega_m|}{\bar g} \right)
 \cos{\left(\beta_N \left(\log {|\omega_m|/{\bar g}} + f_2 \left(\frac{|\omega_m|}{\bar g} \right) + \phi\right) \right)}
 \label{kkkk_1}
\eeq
where
\bea
f_1(x) = 1-\frac{x}{2} + \frac{x^2}{8} + ....,\qquad f_2 (x) = -x + \frac{x^2}{4} + ....
\eea
We see that the envelop of  $\Delta_{\infty, L} (\omega_m)$ varies at $|\omega_m| = O({\bar g})$, while
 the argument of $\cos[...]$ deviates from the low-frequency form $\beta_N \log {|\omega_m|/{\bar g}} + \phi$
 already at much smaller $|\omega_m| \sim {\bar g}/\beta_N$.
 \begin{figure}
 	\includegraphics[width=8cm]{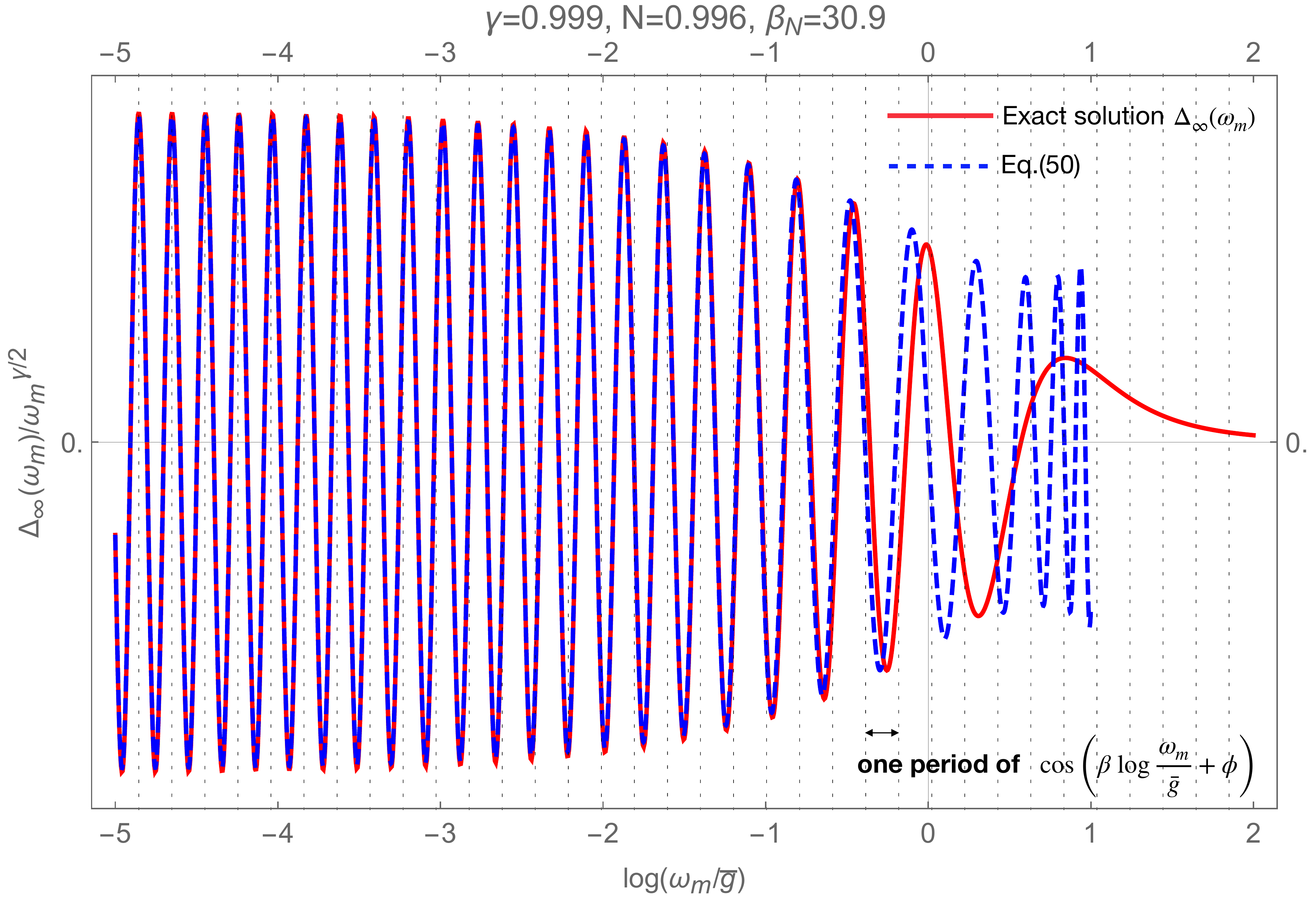}
 	\caption{
Comparison between the exact $\Delta_{\infty} (\omega_m)$ (red solid)  and $\Delta_{\infty,L} (\omega_m)$  from Eq. \eqref{kkkk_1} (blue dashed).
  The envelope of $\Delta_{\infty} (\omega_m)$ is well described by \eqref{kkkk_1} up to $\omega_m \sim {\bar g}$, and the period of oscillations is well described up to $\omega_m \sim {\bar g}/\beta^{1/3}_N$.
   Note that $\Delta_{\infty} (\omega_m)$ deviates from the low-frequency $\cos{(\beta_N \log {|\omega_m|/{\bar g}} + \phi)}$ form  beginning from much smaller $\omega_m \sim {\bar g}/\beta_N$.
   }\label{fig:compare_large_beta}
 \end{figure}
 In Fig.\ref{fig:compare_large_beta} we compare $\Delta_{\infty} (\omega_m)$ with Eq. (\ref{kkkk_1}). We see that the envelope of $\Delta_{\infty} (\omega_m)$ is well described by Eq. (\ref{kkkk_1}), while oscillations become non-logarithmic already at small $|\omega_m| \sim {\bar g}/\beta_N$ and are captured by Eq. (\ref{kkkk_1}) up to $\omega_m \leq \bar g/\beta^{1/3}_N$.

\subsubsection{Non-linear gap equation, $T=0$.}
\label{non_lin_gamma_to_one}

From a generic point of view, the behavior of $\Delta_n (\omega_m)$ for $\gamma \leq 1$
  qualitatively similar to that for smaller $\gamma$. Namely, $\Delta_n (\omega_m)$ form a discrete, infinite set. A function $\Delta_n (\omega_m)$ behaves as $1/|\omega_m|^\gamma$ at the highest frequencies,  oscillates $n$ times at smaller $\omega_m$, and at even smaller $\omega_m$ approaches a
 finite $\Delta_n (0)$.  The condensation energy $E_{c,n}$ is different for different $n$ and is the largest for
 $n=0$.

  On a more careful look, we find that new features in $\Delta_n (\omega_m)$ gradually develop as $\gamma$ approaches one.  To see this, consider the non-linear gap equation at $N \neq 1$:
  \beq
    \label{eq:gapeq_d_d}
    \Delta_{n} (\omega_m) = \frac{{\bar g}^\gamma}{2N}  \int d \omega'_m \frac{\Delta_{n} (\omega'_{m}) - N \Delta_{n} (\omega_m) \frac{\omega'_{m}}{\omega_m}}{\sqrt{\Delta^2_n (\omega'_m) + (\omega'_{m})^2}}
    ~\frac{1}{|\omega_m - \omega'_{m}|^\gamma}
    \eeq
   For $\gamma \to 1$ and $N \neq 1$, the dominant contribution to the r.h.s. of (\ref{eq:gapeq_d_d}) comes from
   $\omega'_m \approx \omega_m$. Keeping only this contribution, we obtain
   \beq
   \Delta^2_n (\omega_m) + \omega^2_m = \Delta^2 (0)
    \label{aa_1}
   \eeq
   where $\Delta (0) \approx {\bar g}^\gamma ((1-N)/(N (1-\gamma)) \sim \omega_{max}$.

\begin{figure}
	\includegraphics[width=8cm]{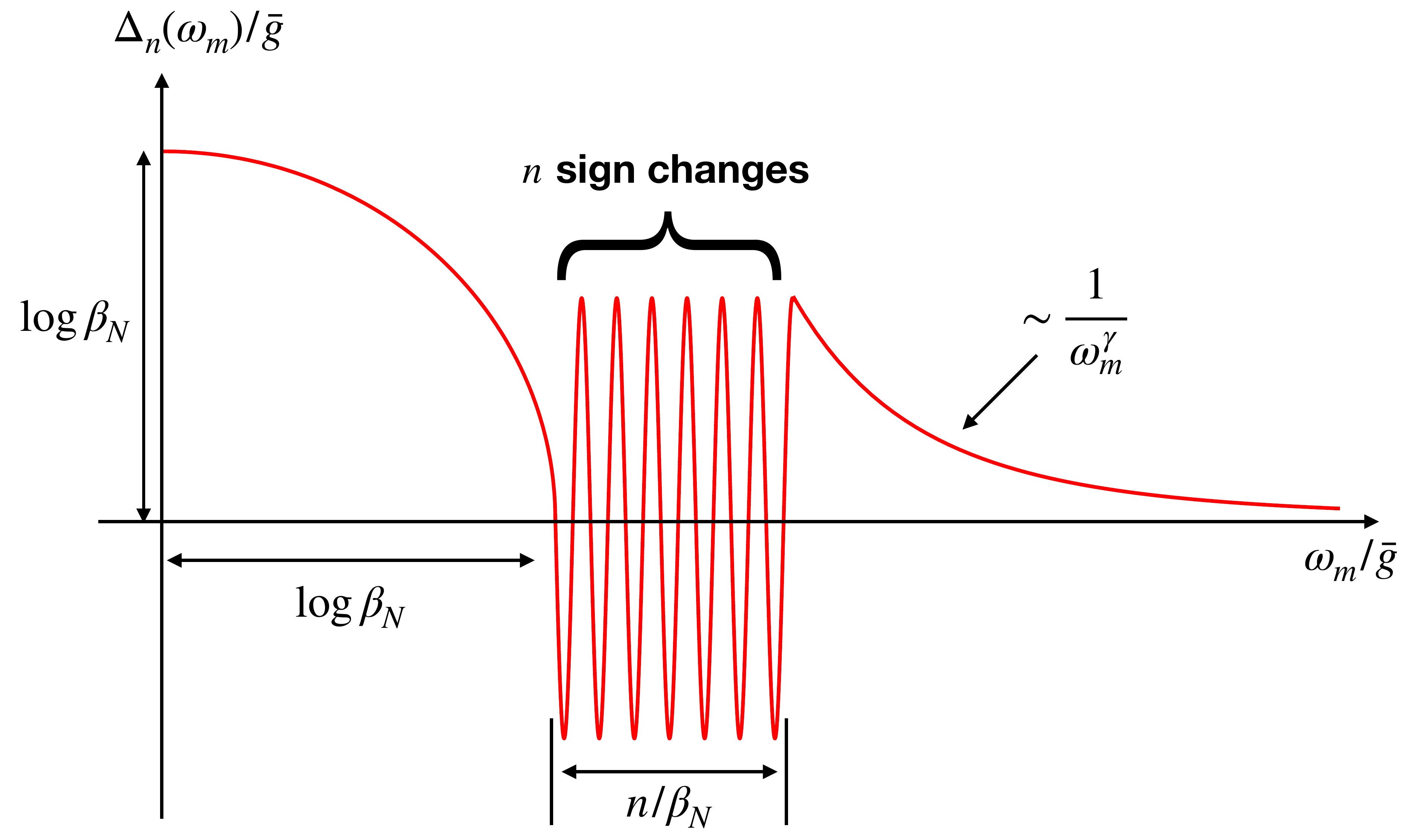}
	\caption{A sketch of the behavior of $\Delta_n(\omega_m)$ at $T=0$ and $\gamma\to1$.
   At $\omega_m =0$, $\Delta_n(0)\sim\bar g\log\beta_N$ is almost independent on $n$.
   $\Delta_n(\omega_m)$ is weakly frequency-dependent at $\omega_m < \Delta_n (0)$, then it oscillates $n$ times, and at even larger $\omega_m$ decays as $1/\omega_m^\gamma$. The width of the range where
    $\Delta_n(\omega_m)$ oscillates is of order $\bar g n/\beta_N$. At large $\beta_N$, this range is smaller than $\Delta_n (0)$ for nearly all $n$, except the very large ones.  }\label{fig:nonlinear_solution}
\end{figure}

\begin{figure}
	\includegraphics[width=8cm]{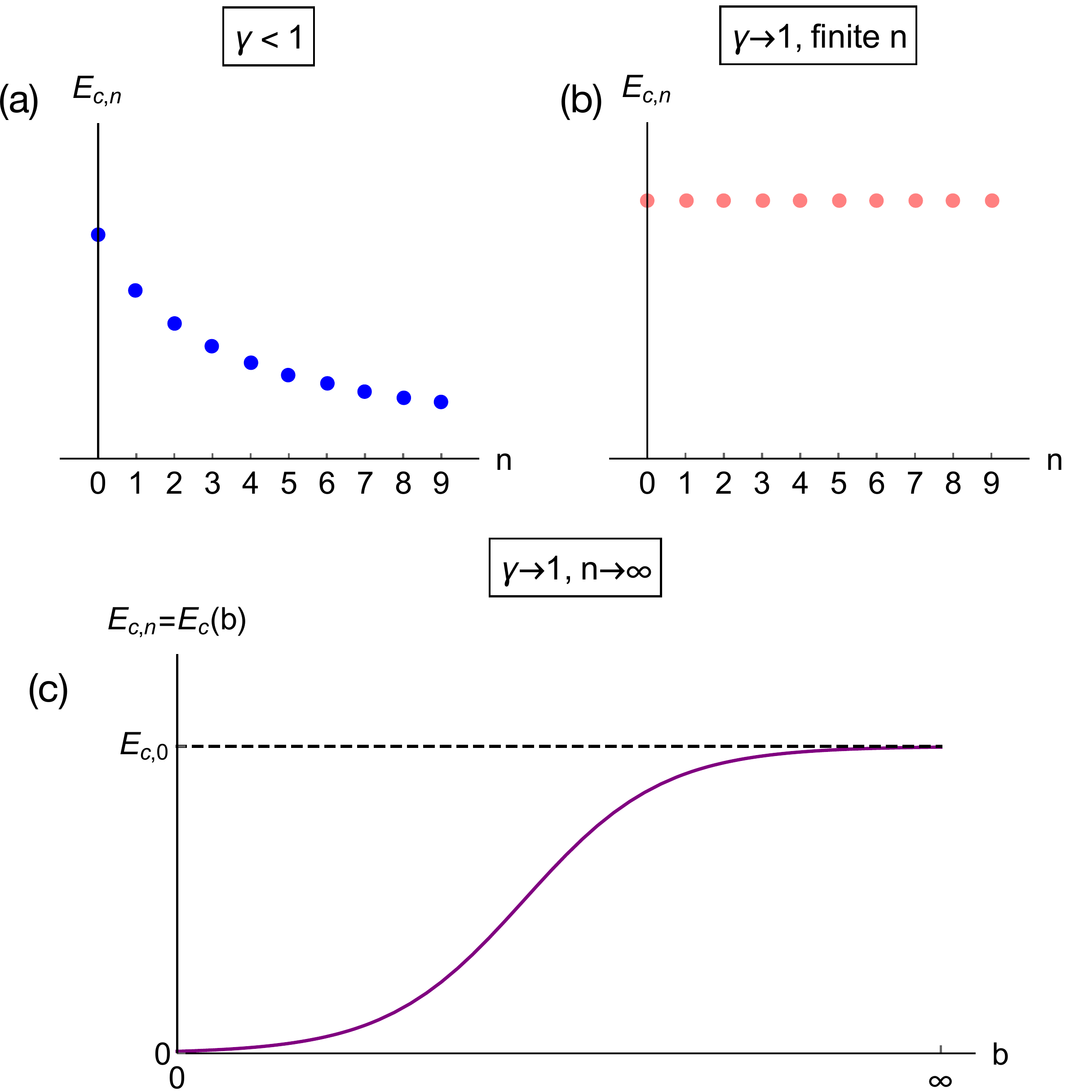}
	\caption{A sketch of the spectrum of the condensation energy for $T=0$ and $N<1$.
(a) For $\gamma<1$, $E_{c,n}$ is discrete and decreases with  $n$. The condensation energy $E_{c,0}$ for the  sign-preserving solution $\Delta_0 (\omega_m)$ is the largest.
 (b)  For $\gamma \to 1$, $\beta_N \to \infty$, and $E_{c,n}$  for all finite $n$ approach $E_{c,0}$, as the frequency range where
 $\Delta_n (\omega_m)$ differs from $\Delta_0 (\omega_m)$ scales as $1/\beta_N$ and vanishes at $\beta_N \to \infty$.
   (c) The condensation energy at $n\to\infty$ depends on the order in which the double limit $n\to\infty$ and $\beta_N\to\infty$ is taken and  becomes a continuous function $E_c (b)$ of
   $b \sim \beta_N/n$.  In the two limits, $E_c(0) =0$ and $E_c(\infty) = E_{c,0}$.
    }\label{fig:condensationEngery}
\end{figure}

   We see that at $\omega_m  \ll \Delta (0) $, $\Delta_n (\omega_m) \approx \Delta (0)$ is nearly independent on frequency and is also independent on $n$.
   The corrections to Eq. \ref{aa_1} do depend on $n$, but these corrections are small in $(1-\gamma)/(1-N) \sim {\bar g}/\omega_{max}$.  At $\omega_m \geq \Delta (0)$, $\Delta_n (\omega_m)$ oscillates $n$ times and then decreases as $1/|\omega_m|$.  Because in this range $\Delta_n (\omega_m) < \omega_m$, the oscillating term is the same as for
   $\Delta_{\infty, L} (\omega_m)$ in Eq. (\ref{kkkk_1}).  A simple analysis then shows that the relative width  of the frequency range for $n$ oscillations compare to $\Delta (0)$ is $n/n_{max}$, where $n_{max} \sim \beta_N$ up to a prefactor, which depends on  $\log \beta_N$.
    As long as $n < n_{max}$, this width is smaller that $\omega_{max}$, although the upper boundary of oscillations increases with $n$.  We illustrate this in Fig.\ref{fig:nonlinear_solution}
     As a result, the range, where $\Delta_n (\omega_m)$ oscillates, accounts only for a subleading contribution to the condensation energy $E_{c,n}$,  the leading one comes from frequencies $\omega_m \leq \omega_{max}$,
      where $\Delta_n \sim \omega_{max}$ is independent on $n$, up to corrections of order $1/\log{\beta_N} \sim (1-\gamma)/(1-N)$.
      At $\gamma \to 1$ and $N <1$,  $n_{max}$ tends to infinity and the ratio $E_{c,n}/E_{c,0}$ tends to one for all finite $n$. At $n \to \infty$, the result for the condensation energy depends on how the limit $n \to \infty$ and $n_{max} \to \infty$ is taken. If $b = n_{max}/n$ is large, $E_{c,n} \approx E_{c,0}$. In the opposite limit $b \ll 1$, oscillations start at  a frequency much smaller that $\omega_{max}$ and run up to $\omega_{max}$. In this case, the corrections to Eq. (\ref{aa_1}) are no longer small, and the analysis has to be modified. Obviously, at such large $n$, $\Delta_n (0)$ become smaller, and $E_{c,n}$ drops.
     The outcome of this consideration is that at $\gamma \to 1$, the spectrum of the condensation energy becomes
       continuous: $E_{c,n}$ for all finite $n$ becomes equal to $E_{c,0}$, while at $n \to \infty$, $E_{c,n} = E_{c} (b) = E_{c,\infty} f(b)$ becomes a continuous variable, ranging between $f(0) =0$ and $f(\infty) =1$.
       We illustrate this in Fig.\ref{fig:condensationEngery}

       There is a certain similarity between our case and  how a continuum spectrum  develops for lattice vibrations, when the system size becomes infinite and a momentum becomes a continuous variable.

       The transformation of the spectrum of $E_{c,n}$ from  a discrete one to continuous
        represents the major qualitative difference between $\gamma =1$ and  $\gamma <1$.
        For $\gamma <1$,  the set of $E_{c,n}$  is discrete and
         $E_{c,0}$ is the largest.  Because the condensation energy is proportional to the total number of particles, other $E_{c,n}$ are only relevant for spatially inhomogeneous fluctuations at a finite $T$.
         When the  spectrum of $E_{c,n}$ becomes a continuous one for spatially homogeneous $\Delta (\omega_m)$, fluctuations become one-dimensional and will likely restore $U(1)$ phase symmetry.  This issue requires further study because although the spectrum becomes continuous, $E_{c,0}$ by itself tends to infinity at $\gamma =1$.  We show in a subsequent publication that a continuous spectrum of condensation energies develops also in more physically transparent case of $N=1$ and $\gamma =2$. In this last case, $E_{c,0}$ remains finite.

\subsubsection{Linearized gap equation, finite $T$}

\begin{figure}
    \includegraphics[width=7cm]{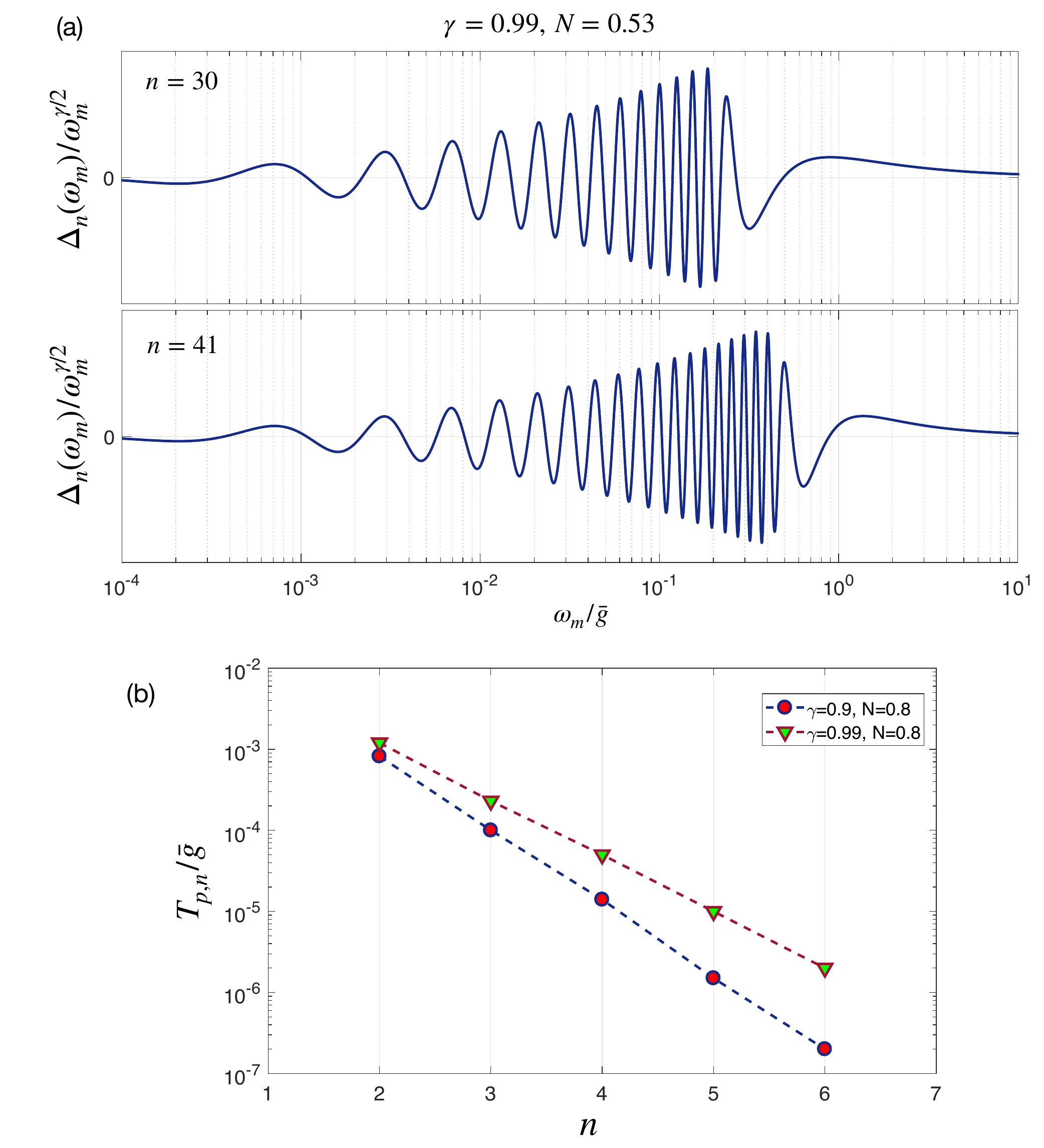}
    \caption{(a)$\Delta_n (\omega_m)$ at the onset temperatures $T_{p,n}$ for two different $n$, at a fixed $N$ and $\gamma\to 1$. The frequency, up to which $\Delta_n (\omega_m)$ oscillates, increases with $n$.
      This is consistent with the analysis at $T=0$. (b) $T_{p,n}$ as a function of $n$ for $\gamma=0.9$ and $0.99$ and $N=0.8$.  $T_{p,n}$ still
       displays an exponential dependence on $n$, but the slope is smaller than in Fig.\ref{fig:Tpng15}
        for  $N=1$.}\label{fig:Delta_gto1}
  \end{figure}

  We recall that for $N=1$, system behavior  evolves smoothly through $\gamma =1$. Namely, the onset temperature $T_{p,n}$ is of order $\Delta_n (0)$ at $T=0$, and both scale as ${\bar g} e^{-An}$, where $A \sim 1/\beta_{N=1}$ remains $O(1)$ for $\gamma =1$. An eigenfunction $\Delta_n (\omega_m)$ has the same structure as $\Delta_{\infty} (\omega_m)$  down to $\omega_m \sim T_{p,n}$, and saturates at smaller frequencies.
   For  $N <1$,   we find two  new features, both are consistent with the results at $T=0$. First,  the scale,  up to which $\Delta_n (\omega_m)$ oscillates, increases with $n$ (see Fig.\ref{fig:Delta_gto1}(a))
   Second,  $T_{p,n}$ at large $n$ still scales as ${\bar g} e^{-A n}$, but $A \propto 1/\beta_N$  increases as $\gamma \to 1$,  hence $T_{p,n}$ also increases (see Fig.\ref{fig:Delta_gto1}(b)).
   We show more numerical results later, in Sec. \ref{sec_c} and Appendix \ref{app_b}.

\subsubsection{Non-linear gap equation, finite $T$}

We didn't attempt to solve the non-linear  equation for $\Delta_n (\omega_m)$  at $T < T_{p,n}$.  Like for $N=1$, we expect, based on the analysis in Sec. \ref{non_lin_gamma_to_one},  that all $\Delta_n (\omega_m)$ with finite $n >0$ rapidly increase below $T_{p,n}$ and at $T \to 0$ merge with
$\Delta_0 (\omega_m)$, which, we recall, develops at a much larger $T_{p,0}$. We illustrate this in
Fig. \ref{fig:nonlinear}.

\subsection{Case $\gamma \geq 1$}

\subsubsection{Linearized gap equation, $T=0$}

For $\gamma >1$,  a simple analysis of the linearized gap equation (\ref{eq:gapeq_a}) shows that there is no  solution with $\Delta_n (\omega_m) \neq 0$.   Indeed,  for $N \neq 1$, the integral over $\omega'_m$ diverges at $\omega'_m = \omega_m$, leaving $\Delta (\omega_m) =0$  as the only option.

\subsubsection{Nonlinear gap equation, $T=0$}
\label{sec_a}
 The solution of the non-linear gap equation does not exist for $N >1$ and is singular for $N <1$. Namely, all $\Delta_n (\omega_m)$ with finite $n$ tend to infinity at any finite $\omega_m$, while the solutions with $n \to \infty$ form a continuous spectrum  of the condensation energies. The way to see this is to consider the $T=0$ case as the limit $T \to 0$. This is what we do in Sec. \ref{sec_b} below.

\subsubsection{Linearized gap equation, $T \neq 0$}
\label{sec_c}

At a finite $T$ the sum over $m' \neq m$ in (\ref{eq:gapeq_d_a}) does not diverge.  In this situation, it is natural to expect that  $\Delta_n (\omega_m)$  is non-zero and finite below a certain  $T_{p,n}$, which, we recall, remains finite for $\gamma =1$ and $N <1$.

  Like for $N=1$, the  calculations are more straightforward, when done for the  pairing vertex $\Phi (\omega_m)$,
   expressed via the normal state ${\tilde \Sigma}_{\text{norm}} (\omega_m)$. The gap function $\Delta (\omega_m) = \omega_{m}\Phi (\omega_m)/{\tilde \Sigma}_{\text{norm}} (\omega_m)$.
     We have from (\ref{eq:gapeq_a})
 \beq \label{eq:gapeq_a_1}
    \Phi (\omega_m) =
     \frac{{\bar g}^\gamma}{N}  \pi T \sum_{m' \neq m} \frac{\Phi (\omega_{m'})}{|{\tilde \Sigma}_{\text{norm}} (\omega_{m'})|}
    ~\frac{1}{|\omega_m - \omega_{m'}|^\gamma},
\eeq
 where ${\tilde \Sigma}_{\text{norm}} (\omega_{m'})$ is given by Eq. (\ref{ee_7}) and, we remind,
$K = ({\bar g}/(2\pi T))^\gamma$.
For $\gamma >1$,
 $A(m \to \infty) = 2\zeta (\gamma)$.
  Substituting the self-energy into the equation for $\Phi (\omega_m) = \Phi (m)$ and eliminating the term
   with $m=0$, we obtain
   \bea
 N \Phi (m>0) &=&  \sum_{n=1,n\neq m}^\infty \frac{\Phi (n)}{A (n) + \frac{2n+1}{K}} \frac{1}{|n-m|^\gamma} + \sum_{n=1}^\infty \frac{\Phi (n)}{A (n) + \frac{2n+1}{K}}\frac{1}{(n+m+1)^\gamma} + \nonumber \\
  && \frac{K}{N-K}  \sum_{n=1}^\infty \frac{\Phi (n)}{A (n) + \frac{2n+1}{K}} \left(\frac{1}{n^\gamma} + \frac{1}{(n+1)^\gamma}\right) \left(\frac{1}{m^\gamma} + \frac{1}{(m+1)^\gamma}\right)
   \label{ch_3}
 \eea
 Consider the limit of small $T$, when $K \gg 1$.
Like for $\gamma <1$ (Paper II), one solution of (\ref{ch_3}) exists for $N \approx K \gg 1$ .
 We express $N$ as $N = K + b_\gamma$, where $b_\gamma = O(1)$ and substitute into (\ref{ch_3}). The divergent $K$ cancels out, and in the remaining equation the kernel factorizes between internal $m'$ and external $m$.
   Solving the equation, we obtain
\bea
\Phi (m>0) &=&   C  \left(\frac{1}{m^\gamma} + \frac{1}{(m+1)^\gamma}\right), \nonumber \\
\Phi (0) &=&  \frac{1}{b_\gamma} \sum_{n=1}^{\infty} \frac{\Phi (n)}{A(n)} \left(\frac{1}{n^\gamma} + \frac{1}{(n+1)^\gamma}\right)
 \label{ch_5}
 \eea
 and
  \beq
 b_\gamma =  \sum_{n=1}^\infty \frac{1}{A (n)}
 \left(\frac{1}{n^\gamma} + \frac{1}{(n+1)^\gamma}\right)^2
  \label{ch_6}
 \eeq
 Both $\Phi (m)$ and $b_\gamma$ evolve smoothly through $\gamma =1$.  The pairing vertex $\Phi (m)$ and the gap function $\Delta  (m)$ do not have nodes and in our classification are $\Phi_{n=0} (m)$ and $\Delta_{n=0} (m)$. The corresponding $T_{p,0} \approx ({\bar g}/2\pi)/N^{1/\gamma}$. We discussed the $n=0$ solution for $N \gg 1$ in length in earlier papers~\cite{Abanov_19,*Wu_19_1,Chubukov_2020a}.  In short: for both $\gamma <1$ and $\gamma >1$,  $\Delta_{n=0} (m)$ displays a re-entrant behavior, i.e., it emerges at a finite $T_{p,0}$ and vanishes at $T=0$.   We verified that for $\gamma >1$ this behavior holds for all $N >1$.

\begin{figure}
  \includegraphics[width=7cm]{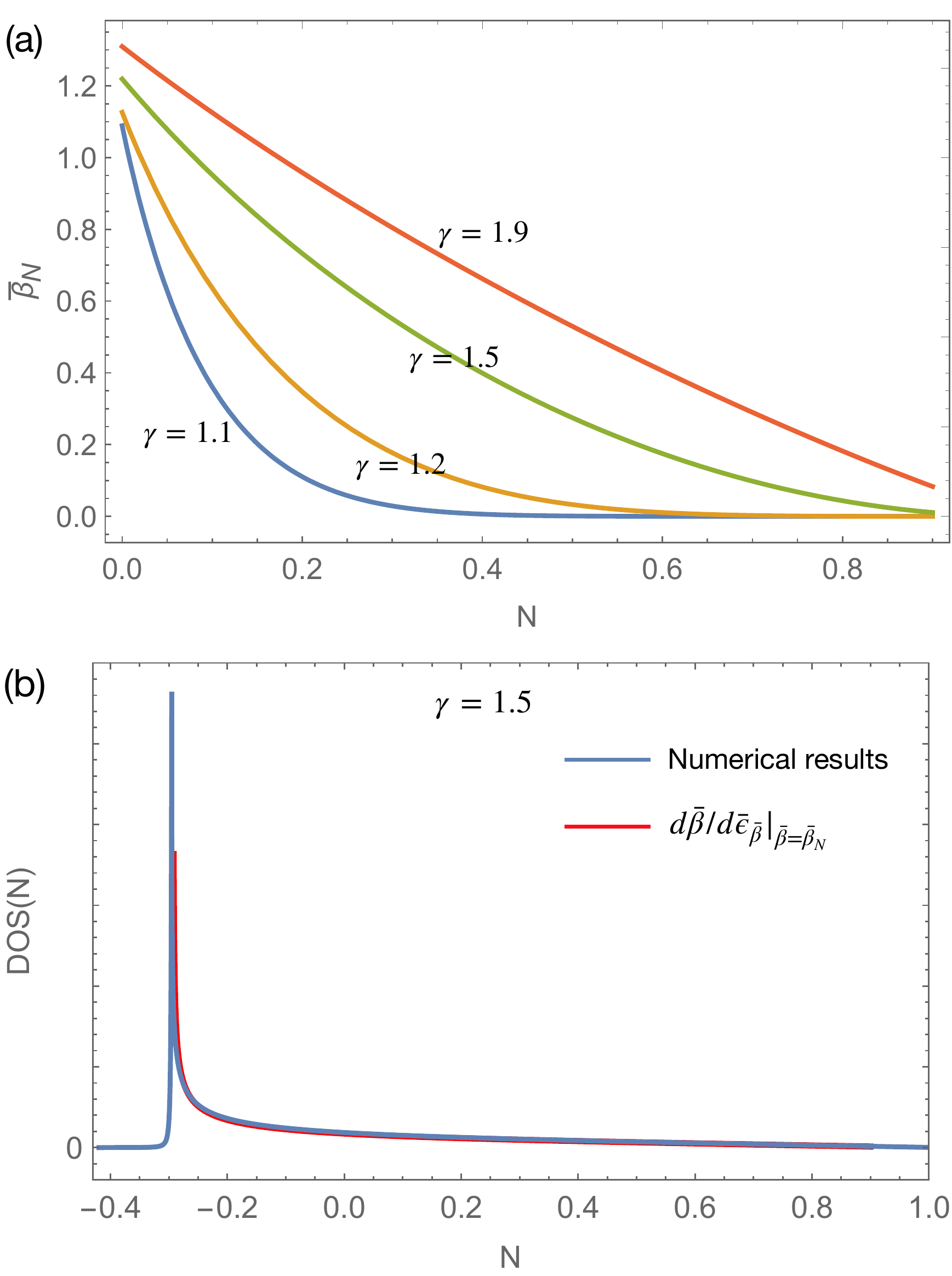}
  \caption{(a) ${\bar \beta}_N$ as a function of $N$ for several different $\gamma$, see Eq. (\ref{3_2}).
     (b) The comparison between DoE's, obtained by using \eqref{chh_11} and  by solving numerically the actual gap equation. We set $\gamma=1.5$. The two DoE's  are almost identical. The DoE  diverges at $N_{min}=-0.292893 = -1 + 1/\sqrt{2}$, and tends to zero as $N_{max} \to 1$.}\label{fig:betaNbar}
\end{figure}

We now turn to $N <1$, where, as we will see, system behavior differs qualitatively between $\gamma <1$ and $\gamma >1$.
 For $N <1$ and $K \to \infty$, we obtain from (\ref{ch_3}):
   \bea
 N \Phi (m>0) &=&  \sum_{n=1,n\neq m}^\infty \frac{\Phi (n)}{A (n)} \frac{1}{|n-m|^\gamma} + \sum_{n=1}^\infty \frac{\Phi (n)}{A (n)}\frac{1}{(n+m+1)^\gamma} - \nonumber \\
  &&  \sum_{n=1}^\infty \frac{\Phi (n)}{A (n)} \left(\frac{1}{n^\gamma} + \frac{1}{(n+1)^\gamma}\right) \left(\frac{1}{m^\gamma} + \frac{1}{(m+1)^\gamma}\right)
   \label{ch_3_a}
 \eea
  For $m,n \gg 1$, the last term becomes irrelevant, and Eq. (\ref{ch_3_a}) reduces to
   \beq
 N \Phi (m >0) =  \frac{1}{2 \zeta (\gamma)} \sum_{n=1,n\neq m}^\infty \left(\frac{\Phi (n)}{|n-m|^\gamma} + \frac{\Phi (n)}{(n+m)^\gamma} \right)
\label{ch_3_b}
 \eeq
We search for the solution in the form
\beq
\Phi (m>0) \propto \cos\left({\bar \beta}_N m + {\bar \phi}\right)
\label{3_1}
\eeq
 where $0 < {\bar \beta}_N < \pi$. This corresponds to
 \beq
\Delta (\omega_m) \propto \omega_m \cos\left(\frac{{\bar \beta}_N}{2\pi T} \omega_m  + {\bar \phi}\right)
\label{3_1a}
\eeq
 Substituting (\ref{3_1}) into (\ref{ch_3_b}), we find after simple algebra that Eq. (\ref{ch_3_b}) is satisfied if ${\bar \beta}_N$ satisfies ${\bar \epsilon}_{{\bar \beta}_N} = N$, where
\beq
{\bar \epsilon}_{{\bar \beta}} = \frac{1}{\zeta (\gamma)} \sum_{n=1}^\infty \frac{\cos({\bar \beta} n)}{n^\gamma} = \frac{1}{\zeta (\gamma)} {\text{Re}} \left[Li_{\gamma} \left(e^{-i{\bar \beta}}\right)\right]
 \label{3_2}
 \eeq
 where $Li$ is the polylogarithm function.
In Fig.\ref{fig:betaNbar}(a) we plot ${\bar \beta}_N$ as a function of $N$ for several $\gamma >1$.
The solution exists for $N$ between maximal $N_{max} \to 1$ and negative minimal $N_{min} = 2^{1-\gamma}-1$.
 At $N \to N_{max}$,
 ${\bar \beta}_N \to 0$
  at $N=N_{min}$,
 ${\bar \beta}_N = \pi$.
   For $\gamma \geq 1$, ${\bar \beta}_N \sim (1-N)^{1/(\gamma-1)}$ is small
  for all $N >0$. At $N=1$, ${\bar \beta}_N$ vanishes.  In this case, the solution (log-oscillating function)  has to be obtained as in Sec. \ref{sec:e}.

  Like we did for $\gamma <1$, we interpret ${\bar \epsilon}_{{\bar \beta}_N} =N$ as the dispersion relation and identify ${\bar \beta}_N$ with the effective momentum and $N$ with the effective energy.  Then one can define the density of eigenvalues (DoE) as
\beq
\nu (N) \propto \left.\frac{d{\bar \beta}}{d{\bar \epsilon}_{{\bar \beta} }} \right|_{{\bar \beta} = {\bar \beta}_N}.
\label{chh_11}
\eeq
  We plot this function in Fig.\ref{fig:betaNbar}(b)
   along with the DoE obtained numerically by solving the full
   Eq. (\ref{ch_3_a})
   as an eigenvalue/eigenfunction equation. We see that the analytic and numerical DoE are quite similar. Both show divergence at $\gamma-$dependent  $N_{min}$
   and vanish at $N_{max} =1$ as $\nu (N) \propto (1-N)^{\frac{2-\gamma}{1-\gamma}}$.  Note that the behavior
    of $\nu (N)$ near $N=1$ changes at $\gamma =2$.

We now use the form of $\Phi (m)$ to obtain $T_{p,n}$.   As before, we use the initially free parameter ${\bar \phi}$ to match with $\Phi (m)$ at $m = O(1)$, and match with the power-law form at $\Sigma (\omega_m) \sim \omega_m$, i.e., at $m \sim K  = ({\bar g}/2\pi T)^\gamma$.  In more precise form, we have $m/K \sim (1-N)$, where $(1-N)$ appears because the constant term in the self-energy $2 \pi T K \zeta (\gamma)$
 cancels out for $N=1$.    The matching condition is ${\bar \beta}_N ({\bar g}/2\pi T)^\gamma (1-N) = n \pi + O(1)$  Solving for $T = T_{p,n}$, we obtain
\beq
T_{p,n} \sim \frac{{\bar g}}{2\pi} \left(\frac{{\bar \beta}_N (1-N)}{n \pi}\right)^{1/\gamma}
\label{3_3}
\eeq
We see that the $n-$dependence of $T_{p,n}$ is now $1/n^{1/\gamma}$ rather than $e^{-An}$.  This implies that for a given $n$ and $N$,  $T_{p,n}$ rapidly increases as $\gamma$ crosses one.
 For $\gamma \geq 1$,
 \beq
 T_{p,n} \sim {\bar g}  (1-N)^{\frac{1}{\gamma-1}} \left(\frac{1}{n}\right)^{1/\gamma}
\label{3_3_a}
\eeq
 \begin{figure}
  \includegraphics[width=7cm]{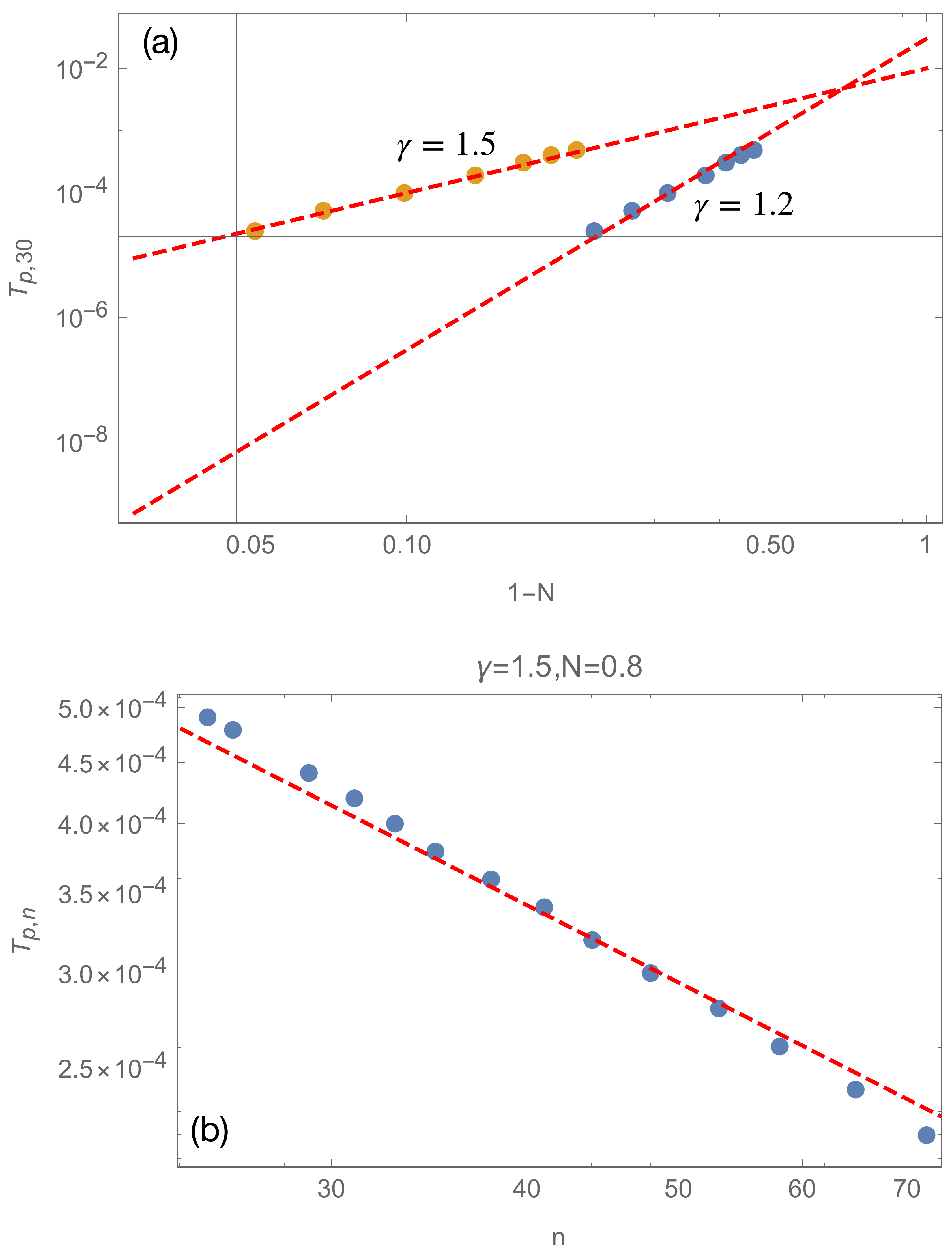}
  \caption{(a) The onset temperature $T_{p,n}$ for $n=30$ as a function of $1-N$ for $\gamma=1.2$ and $\gamma=1.5$. (b) The dependence of $T_{p,n}$ on $n$ for $\gamma=1.5$ and $N=0.8$.  In both panels,
  dots are numerical results and red dashed lines are obtained from \eqref{3_3_a}. }\label{fig:T_N_scaling}
 \end{figure}
 In Fig.\ref{fig:T_N_scaling}
 we plot numerical results for $T_{p,n}$ as a function of $N$ for a given $n$ for several $\gamma \geq 1$
  and as a function of $n$ for given $\gamma$ and $N$.  We see that the agreement is quite good.

\begin{figure*}
  \includegraphics[width=16cm]{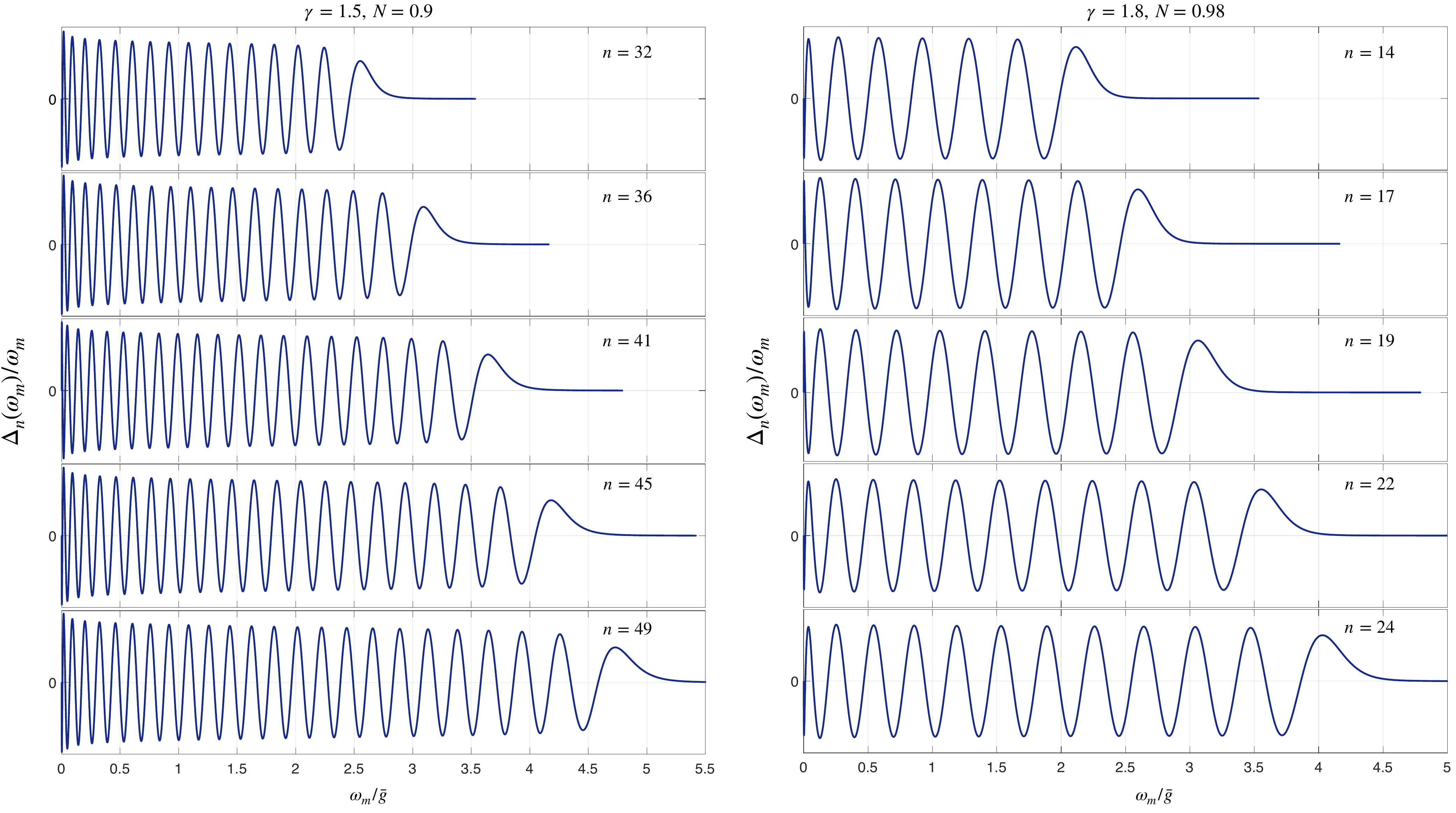}
  \caption{Numerical results for $\Delta_n(\omega_m)$ for different $n$ and two sets of $\gamma>1$ and $N <1$. Observe that (i) the period of oscillations of $\Delta_n(\omega_m)$  is set by $\omega_m$ instead of $\log {\omega_m}$, (ii) the envelop of $\Delta_n(\omega_m)$ is proportional to $1/\omega_m$, and (iii) for fixed $N$, the frequency, $\omega_{max}$, at which  oscillations end, increases with  $n$. }
\label{fig:engenfunctions_gamma>1}
\end{figure*}

We now look at the eigenfunctions $\Phi_n (m)$, or, equivalently, $\Delta_n (m)$.  We use $\Delta_n (m)$ for easier comparison with the results for $\gamma <1$.
 The eigenfunctions behave as $\Delta_n (m) \sim m \cos ({\bar \beta}_N m + {\bar \phi})$  up to $n-$dependent
 $m_{max} (n) \sim ({\bar g}/(2\pi T_{p,n}))^{\gamma}$. At larger $m$, oscillations end and each $\Delta_n (m)$ decays as $1/m^\gamma$.  Comparing with the form of $\Delta_n (m)$ for a generic $\gamma <1$,  we see two key differences.  First, for $\gamma >1$, the period of oscillations is set by $m$ rather than by $\log m$. Second,  for $\gamma <1$, $m_{max} \propto {\bar g}/T_{p,n}$,  hence the frequency, where oscillations end, $\omega_{max} =2 \pi T m_{max}$, is  $O({\bar g}$ for all $n$.   For $\gamma >1$,  $\omega_{max}  \sim {\bar g}  ({\bar g}/T_{p,n})^{\gamma -1}$ becomes n-dependent ($\omega_{max} = \omega_{max} (n)$), i.e., the larger is $n$, the larger is the range of frequencies where $\Delta_n (m)$ oscillates as $\cos{({\bar \beta}^{eff}_N
 (\omega_m/{\bar g}) + {\bar \phi})}$, where ${\bar \beta}^{eff}_N = {\bar \beta}_N {\bar g}/(2\pi T_{p,n}) \sim n^{1/\gamma}$.
  At $n \to \infty$, oscillations extend to $\omega_m \to \infty$.
   We earlier found the precursor to this behavior for $\gamma \to 1$.
  In Fig. \ref{fig:engenfunctions_gamma>1}
   we present numerical results for the eigenfunctions $\Delta_n (\omega_m)$ for two different $\gamma \geq 1$.
   We see that the eigenfunctions indeed oscillate with the period set by $\omega_m$  rather than $\log {\omega_m}$, and that as $n$ increases, oscillations extend to larger $\omega_{max}$.
 These numerical results confirm that there is indeed a qualitative change of  system behavior for $N <1$ between $\gamma <1$ and $\gamma >1$.  We also note  that the divergence of
  ${\bar \beta}^{eff}_N \propto n^{1/\gamma}$ at $n \to \infty$ is consistent with the divergence of $\beta_N$ as $\gamma \to 1$ from below.

The  crossover from log-oscillations of $\Delta_n (\omega_m)$  for $\gamma <1$ to oscillations with a period set by $\omega_m$ for $\gamma >1$ is sharp at $n \gg  1$, when $T_{c,n}$ is small and relevant Mastubara numbers are large. For smaller $n$, the crossover gets smoothen up.  In numerical calculations, there is an additional smothering due to sampling of a finite number of Matsubara points. In Appendix \ref{app_b} we
 show the numerical results for the crossover behavior and its dependence on the number of Matsubara points, sampled in numerical calculations.

\begin{figure}
	\includegraphics[width=7cm]{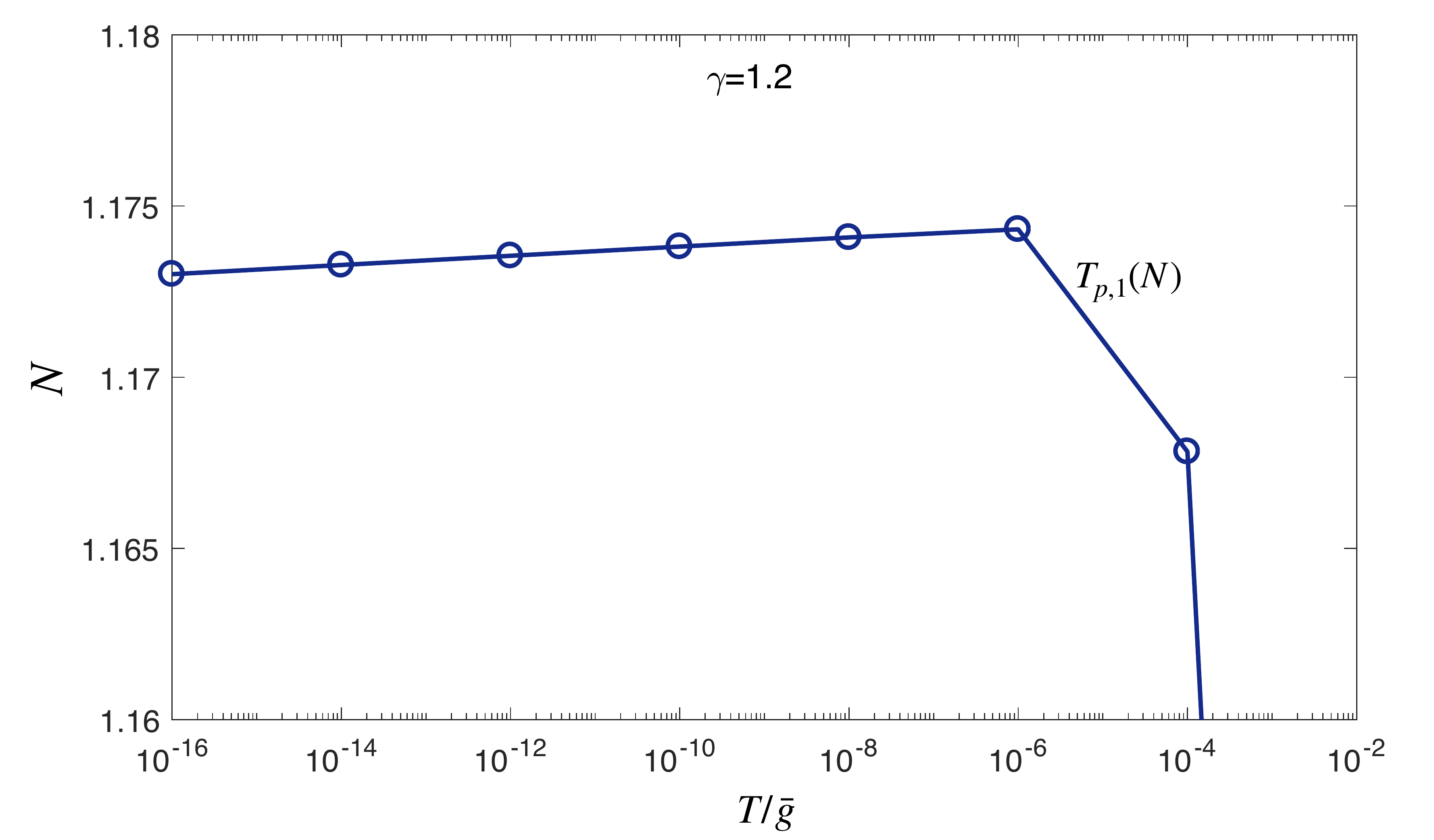}
	\caption{
Numerical results for $T_{p,n=1}(N)$ for $N$ near $1$ and $\gamma=1.2$. $T_{p,1}$ is finite in some range of  $N\geq 1$, but the dependence is non-monotonic, and eventually smaller $T_{p,1}$ correspond to $N$ closer to $N=1$, i.e., at $T \to 0$, the line $T_{p,1}$ approaches $N=1$.  To verify that $T_{p,1} (N) =0$  right at $N=1$, one needs to go to far smaller $T$ than in the figure, which is numerically quite challenging. }\label{fig:Tp1nearN1}
\end{figure}

Finally, in Fig.\ref{fig:Tp1nearN1}
 we show the dependence of $T_{p,n=1}$ on $N$ near $N =1$ (or, equivalently, the temperature dependence of the second
   eigenvalue of the gap equation).  The onset temperature $T_{p,1} (N)$ decreases as $N$ approaches one from below, but because  $T_{p,1} (N=1)$ is finite, it has to remain finite also for $N >1$.
We see that $T_{p,1} (N)$ continuous, as a function of $N$,  into the range $N >1$, but
  then reverses trend, such that smaller $T_{p,1}$ correspond to $N$ closer to $N=1$.
  This reentrant behavior is the consequence of the fact that at $T=0$ there is no solution of the linearized gap equation for any $N$, except for $N =1$.

\subsubsection{Non-linear gap equation, $T \neq 0$}
\label{sec_b}

\begin{figure}
   \includegraphics[width=7cm]{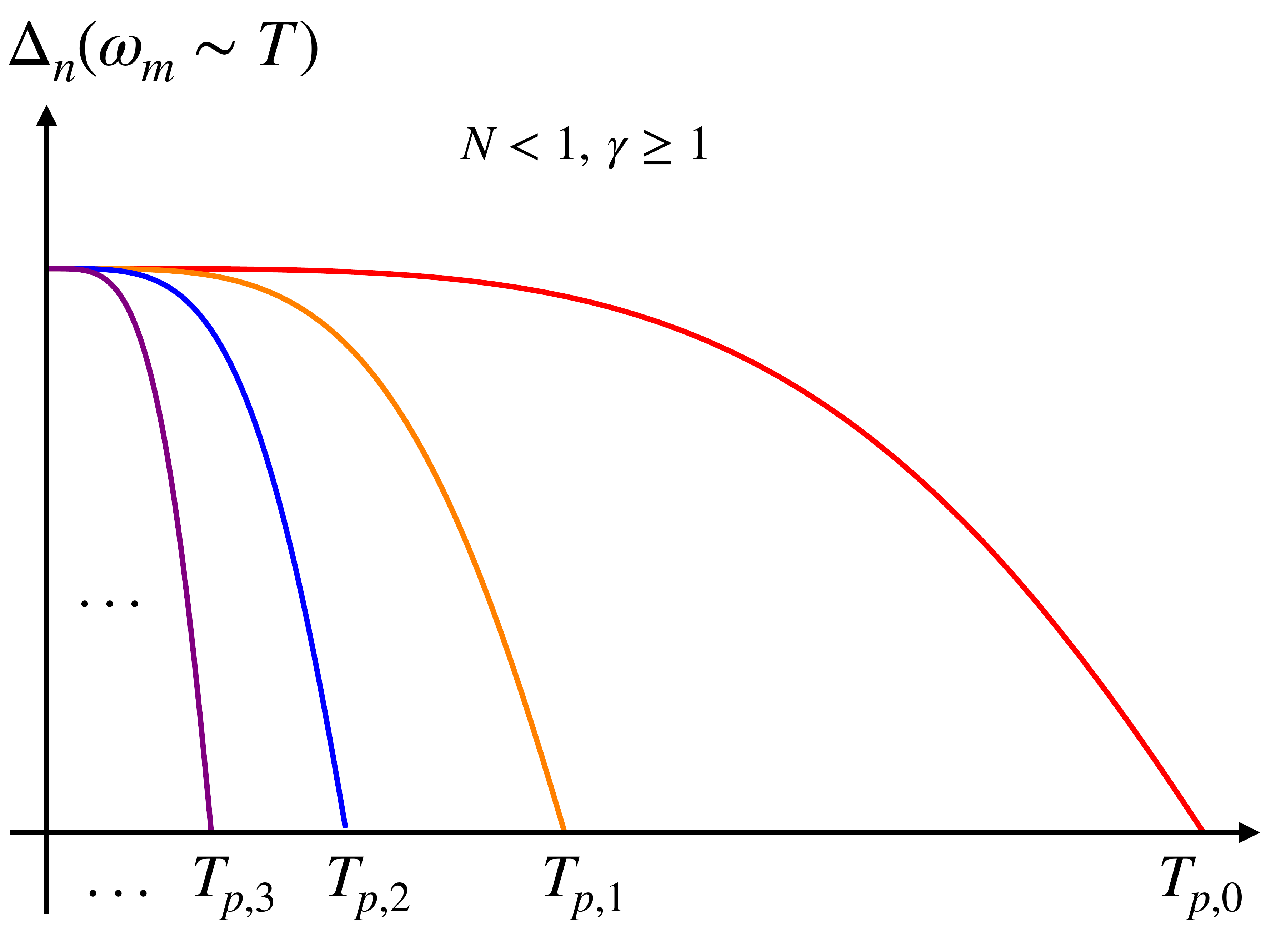}
   \caption{A sketch  of the temperature dependence of $\Delta_n$ for $\gamma>1$ and $N<1$.
     Different $\Delta_n (\omega_m)$ with finite $n$ appear at different $T_{p,n}$, but at $T\to0$ coincide with   $\Delta_0 (\omega_m)$.
       This holds for $n$ up to $n_{max} \sim ({\bar g}/T)^\gamma$.}\label{fig:nonlinear}
 \end{figure}

We analyze non-linear equations for the pairing vertex and the self-energy, Eqs. (\ref{eq:gapeq_a}),  at small but finite $T$.
 It is convenient to introduce dimensionless ${\bar \Phi} = (2\pi T)^{\gamma-1}\Phi /{\bar g}^\gamma$ and
 ${\bar {\tilde\Sigma}} = (2\pi T)^{\gamma-1}{\tilde \Sigma} /{\bar g}^\gamma$. In these variables, Eqs.
   (\ref{eq:gapeq_a}) become, for the $n-th$ solution
    \bea \label{eq:gapeq_a_3}
    {\bar \Phi}_n (m) &=&
     \frac{1}{2N} \sum_{m' \neq m} \frac{{\bar \Phi}_n (m')}{\sqrt{{\bar {\tilde \Sigma}}^2_n (m') +{\bar \Phi}^2_n (m')}}
    ~\frac{1}{|m - m'|^\gamma}, \nonumber \\
     {\bar {\tilde \Sigma}}_n (m) &=&  \left(\frac{2\pi T}{{\bar g}}\right)^{\gamma} (m+1/2)
   + \frac{1}{2}  \sum_{m' \neq m}  \frac{{\bar {\tilde \Sigma}}_n (m')}
   {\sqrt{{\bar {\tilde \Sigma}}^2_n (m') +{\bar \Phi}^2_n (m')}}
    ~\frac{1}{|m - m'|^\gamma}
\eea
Based on our earlier analysis of the case $\gamma \to 1$ from below, we expect  that at small $T < T_{p,n}$, $\Delta_n (m) = \pi T (2m+1)   {\bar \Phi}_n (m)/ {\bar {\tilde \Sigma}}_n (m)$
 is large and weakly dependent on $n$, up to large $n$.
   This holds if ${\bar \Phi}_n (m) \gg {\bar {\tilde \Sigma}}_n (m)$.  Using this inequality, we obtain from (\ref{eq:gapeq_a_3})
\beq
{\bar \Phi}_n (m) \approx \frac{\zeta(\gamma)}{N}, ~~{\bar {\tilde \Sigma}}_n (m) \approx \left(\frac{2\pi T}{\bar g}\right)^\gamma \frac{m+1/2}{1-N}
\label{3_5}
\eeq
 and, hence,
 \beq
 \Delta_n (m) = {\bar g} \left(\frac{\bar g}{2\pi T}\right)^{\gamma -1} \frac{1-N}{N}  + ...
 \label{3_7}
 \eeq
  where dots stand for subleading corrections, which depend on $n$ and $m$.
  We see that $\Delta_n (m)$ diverges as $1/T^{\gamma -1}$ and all $\Delta_n (m)$ merge into the same
 gap function at $T \to 0$.  This holds  for $n$ up to some $n_{max} (T)$, for which $T_{p,{n_c}} \sim T$.
  Using $T_{p,n} \propto 1/n^{1/\gamma}$, we obtain $n_{max} \sim ({\bar g}/T)^\gamma$.  As $T \to 0$, $n_{max} \to \infty$, hence  $\Delta_n (m)$ becomes independent on $n$ for all finite $n$, despite that $T_{p,n}$ are all different.  We note in this regard that at  $T \leq T_{p,n}$, $\Delta_n (m)$  is of order ${\bar g} n^{(\gamma-1)/\gamma}$, i.e., it increases below $T_{p,n}$ with a slope, which increases with  $n$. We illustrate this in Fig.\ref{fig:nonlinear}
     As the consequence, at $T \to 0$ , the condensation energy $E_{c,n}$  becomes equal for all finite $n$, as we
       anticipated in Sec. \ref{sec_a}.  The gap functions $\Delta_n (m)$ with  $n > n_{max}$ do not have a $T$ range to develop into Eq. (\ref{3_7}), and have smaller condensation energy at $T \to 0$.  The condensation energy for
    these solutions depends on $ b = n_{max}/n = ({\bar g}/(n^{1/\gamma} T))^{\gamma}$. At $T \to 0$,  $n_{max}$ tends to infinity,  and at $n \to \infty$, $b$ becomes a continuous variable.  The condensation energy
    $E_c (b) = E_c (\infty) {\tilde f} (b)$, where ${\tilde f} (0) =0$ and ${\tilde f} (\infty) =1$.   This
     is consistent with the results in Sec. \ref{non_lin_gamma_to_one} on  $T=0$ and $\gamma \to 1$.
      The behavior of the condensation energy is illustrated in Fig.\ref{fig:condensationEngery}
     At small $b$, we used $\Delta_n \propto 1/n^{1/\gamma}$ and  the expression for the condensation energy from Refs. \cite{Chubukov_2020a,haslinger,with_emil}
     and obtained $E_c \propto (b)^{2(2-\gamma)/\gamma}$.  For $\gamma \geq 1$, this reduces to $E_c \propto b^2$.

      We emphasize again  that this behavior is qualitatively different from the one in a non-critical BCS/Eliashberg superconductor, where there are at most a few different solutions of the gap equation for ant given $N$
       and from quantum-critical superconductivity for $\gamma <1$, where there exists an infinite set of gap functions for $N < N_{cr}$, but the spectrum of the condensation energy is discrete.  We also emphasize that this behavior does not extend to $N =1$, for which a discrete set of $\Delta_n (m)$ and $E_{c,n}$ survives for $\gamma >1$.   The difference with $N=1$ is obvious from Eq.
       (\ref{3_7}), which shows that  the divergent term cancels out  for $N =1$.

\section{Another extension of the $\gamma$ model}
\label{sec:tilde N}

We now propose another extension of the original model, which does not introduce
 divergencies.
For this we return to the original model with $N=1$ and  re-express  Eqs. (\ref{eq:gapeq_a})
 for the  pairing vertex and the self-energy by  pulling out the divergent terms from the r.h.s., like we
  did in Sec. \ref{sec:e}.
 We obtain
  \bea \label{eq:gapeq_aa}
    &&\Phi (\omega_m) \left(1- {\bar g}^\gamma \frac{\zeta (\gamma)}{(2\pi T)^{\gamma-1}}
    \frac{1}{\sqrt{\tilde{\Sigma}^2(\omega_m)+\Phi^2(\omega_m)}} \right) = \nonumber \\
    && {\bar g}^\gamma \pi T \sum_{m' \neq m} \left(\frac{\Phi (\omega_{m'})}{\sqrt{{\tilde \Sigma}^2 (\omega_{m'}) +\Phi^2 (\omega_{m'})}} - \frac{\Phi (\omega_{m})}{\sqrt{{\tilde \Sigma}^2 (\omega_{m}) +\Phi^2 (\omega_{m})}}\right)
    ~\frac{1}{|\omega_m - \omega_{m'}|^\gamma}, \nonumber \\
    && {\tilde \Sigma} (\omega_m)  \left(1- {\bar g}^\gamma \frac{\zeta (\gamma)}{(2\pi T)^{\gamma-1}} \frac{1}{\sqrt{\tilde{\Sigma}^2(\omega_m)+\Phi^2(\omega_m)}}\right) =\omega_m + \nonumber \\
     &&  {\bar g}^\gamma \pi T \sum_{m' \neq m}\left(  \frac{{\tilde \Sigma}(\omega_{m'})}{\sqrt{{\tilde \Sigma}^2 (\omega_{m'})  +\Phi^2 (\omega_{m'})}} -  \frac{{\tilde \Sigma}(\omega_{m})}{\sqrt{{\tilde \Sigma}^2 (\omega_{m})  +\Phi^2 (\omega_{m})}}\right)
    ~\frac{1}{|\omega_m - \omega_{m'}|^\gamma}
\eea
We then introduce
\bea
{\bar \Phi} (\omega_m) &=& \Phi (\omega_m) \left(1- {\bar g}^\gamma \frac{\zeta (\gamma)}{(2\pi T)^{\gamma-1}}\frac{1}{\sqrt{\tilde{\Sigma}^2(\omega_m)+\Phi^2(\omega_m)}}\right) \nonumber \\
{\bar {\tilde \Sigma}} (\omega_m) &=& {\tilde \Sigma} (\omega_m) \left(1- {\bar g}^\gamma \frac{\zeta (\gamma)}{(2\pi T)^{\gamma-1}}\frac{1}{\sqrt{\tilde{\Sigma}^2(\omega_m)+\Phi^2(\omega_m)}}\right)
\label{3_9}
\eea
Because $\Phi (\omega_m)/{\tilde \Sigma} (\omega_m) = {\bar \Phi} (\omega_m)/{\bar{\tilde \Sigma}} (\omega_m)$, Eqs. (\ref{eq:gapeq_aa}) can be re-expressed solely in terms of ${\bar \Phi} (\omega_m)$ and ${\bar{\tilde \Sigma}} (\omega_m)$:
  \bea \label{3_10}
   && {\bar \Phi} (\omega_m)=
     {\bar g}^\gamma \pi T \sum_{m' \neq m} \left(\frac{{\bar \Phi} (\omega_{m'})}{\sqrt{{\bar {\tilde \Sigma}}^2 (\omega_{m'}) +{\bar \Phi}^2 (\omega_{m'})}} - \frac{{\bar \Phi} (\omega_{m})}{\sqrt{{\bar {\tilde \Sigma}}^2 (\omega_{m}) +{\bar \Phi}^2 (\omega_{m})}}\right)
    ~\frac{1}{|\omega_m - \omega_{m'}|^\gamma}, \nonumber \\
    && {\bar {\tilde \Sigma}} (\omega_m)  = \omega_m \nonumber \\
     && +  {\bar g}^\gamma \pi T \sum_{m' \neq m}\left(  \frac{{\bar {\tilde \Sigma}}(\omega_{m'})}{\sqrt{{\bar {\tilde \Sigma}}^2 (\omega_{m'})  +{\bar \Phi}^2 (\omega_{m'})}} -  \frac{{\bar {\tilde \Sigma}}(\omega_{m})}{\sqrt{{\bar {\tilde \Sigma}}^2 (\omega_{m})  +{\bar \Phi}^2 (\omega_{m})}}\right)
    ~\frac{1}{|\omega_m - \omega_{m'}|^\gamma}
\eea
These equations are now free from singularities, even if we replace a summation over Mascara numbers by an integration over Matsubara frequencies.

We now extend the modified  Eliashberg equations (\ref{3_10}) in the same way as before, by multiplying the interaction in the particle-particle channel by a factor $1/M$:
  \beq \label{3_12}
    {\bar \Phi} (\omega_m)=
     \frac{{\bar g}^\gamma}{M}  \pi T \sum_{m' \neq m} \left(\frac{{\bar \Phi} (\omega_{m'})}{\sqrt{{\bar {\tilde \Sigma}}^2 (\omega_{m'}) +{\bar \Phi}^2 (\omega_{m'})}} - \frac{{\bar \Phi} (\omega_{m})}{\sqrt{{\bar {\tilde \Sigma}}^2 (\omega_{m}) +{\bar \Phi}^2 (\omega_{m})}}\right)
    ~\frac{1}{|\omega_m - \omega_{m'}|^\gamma}
\eeq
The gap function $\Delta (\omega_m)$ is expressed via ${\bar \Phi} (\omega_m)$ and ${\bar {\tilde \Sigma}} (\omega_m)$ in the same way as via the original $\Phi (\omega_m)$ and ${\tilde \Sigma} (\omega_m)$:
 $\Delta (\omega_m) = \omega_m {\bar \Phi} (\omega_m)/{\bar {\tilde \Sigma}} (\omega_m)$. The gap equation becomes
\beq
\Delta (\omega_m) =\frac{{\bar g}^\gamma}{M}  \pi T \sum_{m' \neq m} ~\frac{1}{|\omega_m - \omega_{m'}|^\gamma} \left(\frac{\Delta (\omega_{m'}) - M \frac{\Delta (\omega_{m})}{\omega_m} \omega_{m'}} {\sqrt{\Delta^2 (\omega_{m'}) + \omega^2_{m'}}} - \frac{\Delta (\omega_{m}) (1-M)}{\sqrt{\Delta^2 (\omega_{m}) + \omega^2_{m}}}\right)
\label{3_12a}
\eeq

\subsection{Linearized gap equation, $T=0$}

For infinitesimally  small ${\bar \Phi} (\omega_m)$, the self-energy coincides with that in the normal state.  Converting $\pi T \sum_{m'}$ into $\int d \omega'_m/2$ and evaluating the frequency integral in (\ref{3_10}), we obtain at $T=0$,
\beq
{\bar {\tilde \Sigma}} (\omega_m)  = \omega_m B(|\omega_m|)
 \label{3_14}
\eeq
where
\beq
B(|\omega_m|) = 1 - \left(\frac{\bar g}{|\omega_m|}\right)^\gamma \frac{1}{\gamma -1}
\label{3_15}
\eeq
 Substituting ${\bar {\tilde \Sigma}} (\omega_m)$ into the equation for ${\bar \Phi} (\omega_m)$, we obtain
 \beq
 {\bar \Phi}_\infty (\omega_m) = \frac{{\bar g}^\gamma}{2M} \int d \omega'_m \left(\frac{{\bar \Phi}_\infty (\omega'_m)}{|\omega'_m| B(|\omega'_m|)|} - \frac{{\bar \Phi}_\infty (\omega_m)}{|\omega_m| B(|\omega_m|)|}\right) \frac{1}{|\omega'_m-\omega_m|^\gamma}
 \label{3_16}
 \eeq
 we label infinitesimally small ${\bar \Phi} (\omega_m)$ as ${\bar \Phi}_\infty (\omega_m)$, like in earlier analysis, anticipating that the non-linear  equation for the pairing vertex will have a discrete set of solutions ${\bar \Phi}_n (\omega_m)$.
 At small $\omega_m$,  the bare $\omega_m$ term in ${\bar {\tilde \Sigma}} (\omega_m)$ can be neglected, and (\ref{3_16}) reduces to
 \beq
 {\bar \Phi}_\infty (\omega_m) = \frac{\gamma -1}{2M}
 \int d \omega'_m \frac{{\bar \Phi}_\infty (\omega_m) |\omega_m|^{\gamma -1}-{\bar \Phi}_\infty (\omega'_m)|\omega'_m|^{\gamma-1}} {|\omega'_m-\omega_m|^\gamma}
 \label{3_17}
 \eeq
 This equation is similar, but not equivalent, to Eq. (\ref{eq:gapeq_0})  for the pairing vertex  for $\gamma <1$. In both cases, the kernel is marginal, and we search for the solution in the form
  ${\bar \Phi} (\omega_m) \propto |\omega_m|^{-\gamma/2 + b}$.  Like before, a normalizable solution
   exists when $b = \pm i \beta_M$ is purely imaginary.  Substituting power-law form with the complex exponent,
    we find that
    (\ref{3_17})
    is satisfied if $\epsilon_{i\beta_M} = M$, where $\epsilon_{i\beta}$ is exactly the {\it same} function as in Eq. (\ref{su_15_2}).
    In this respect, the extension to $M \neq 1$ for $\gamma>1$ is quite similar to the extension to $N \neq 1$ for $\gamma <1$.  The similarity is particularly transparent for the linearized  equation for
      $D(\omega_m) = \Delta (\omega_m)/\omega_m$. From (\ref{3_17}) we obtain
    \beq
 D_\infty(\omega_m) \omega_m \left(1 + {\bar \lambda} \left(\frac{\bar g}{|\omega_m|}\right)^\gamma \right) = \frac{{\bar g}^\gamma}{2M} \int d \omega'_m \frac{D_\infty(\omega'_m)-D_\infty (\omega_m)}{|\omega_m-\omega'_m|^\gamma} {\text{sign}} \omega'_m,
 \label{3_19}
 \eeq
  where ${\bar \lambda} = (1/M -1)/(\gamma -1)$.   This equation is identical to Eq. (\ref{3_11}) once we replace $N$ by $M$.

\begin{figure}
  \includegraphics[width=7cm]{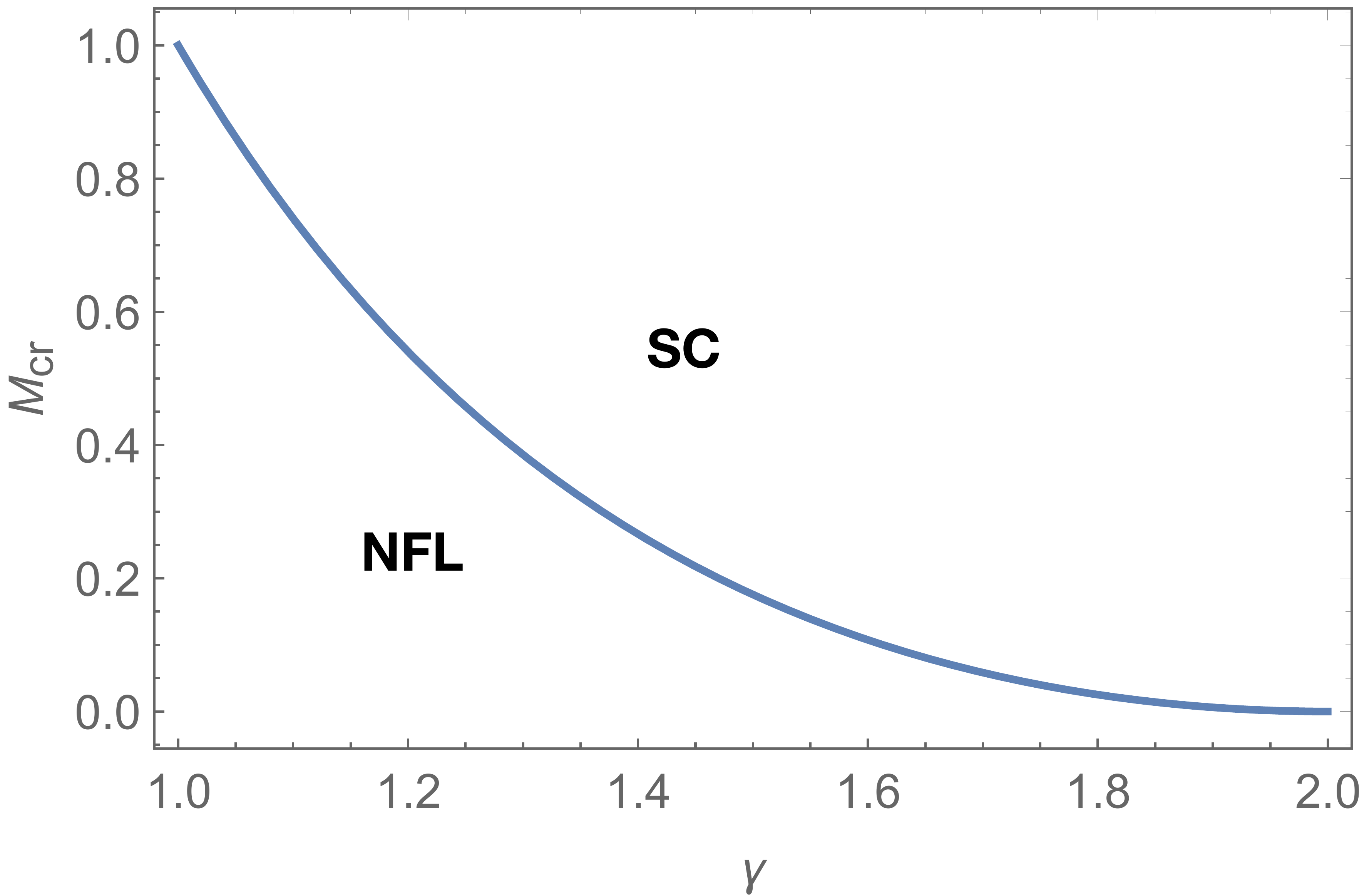}
  \caption{Critical $M_{cr}$ as a function of $\gamma>1$. A normalizable solution of the gap equation exists for $M > M_{cr}$. }\label{fig:Mcr}
\end{figure}

Because $d\epsilon_{i\beta}/d \beta $ is  positive for $\gamma >1$, a normalizable
      solution of the gap equation    exists for
      $M > M_{cr}$, where $M_{cr}$  satisfies $\epsilon_0 = M_{cr}$.   We plot $M_{cr}$ as a function of $\gamma$ in Fig.\ref{fig:Mcr}.  We see that $M_{cr}$  gradually decreases as $\gamma$ increases, from $M_{cr} = 1$ at $\gamma =1$ to
     $M_{cr} =0$ at $\gamma =2$.

\subsection{Linearized gap equation at a finite $T$}

At a finite $T$, we obtain from (\ref{3_12a}) for infinitesimally small $\Delta_{\infty} (\omega_m)$ = $\Delta_{\infty} (m)$ and $m >0$
\beq
S_m \left(|2m+1| - 2 K (\zeta(\gamma) - H(m, \gamma))\right) = \frac{K}{M} \sum_{m' \neq m} \frac{S_{m'}-S_m}{|m'-m|^\gamma}
\label{3_20}
\eeq
where, $S_m = \Delta (m)/|2m+1|$,  $K = ({\bar g}/(2\pi T))^\gamma$, and $H(m, \gamma) = \sum_{1}^m 1/p^\gamma$ is the
the Harmonic number.  At small $T$, (\ref{3_20}) simplifies to
 \beq
S_m = \frac{1}{2M (H(m, \gamma)-\zeta(\gamma))} \sum_{m' \neq m} \frac{S_{m}-S_{m'}}{|m'-m|^\gamma}
\label{3_21}
\eeq
The solution of this equation exists at a particular $T$, which determines the onset temperature for the pairing.

\begin{figure}
  \includegraphics[width=8cm]{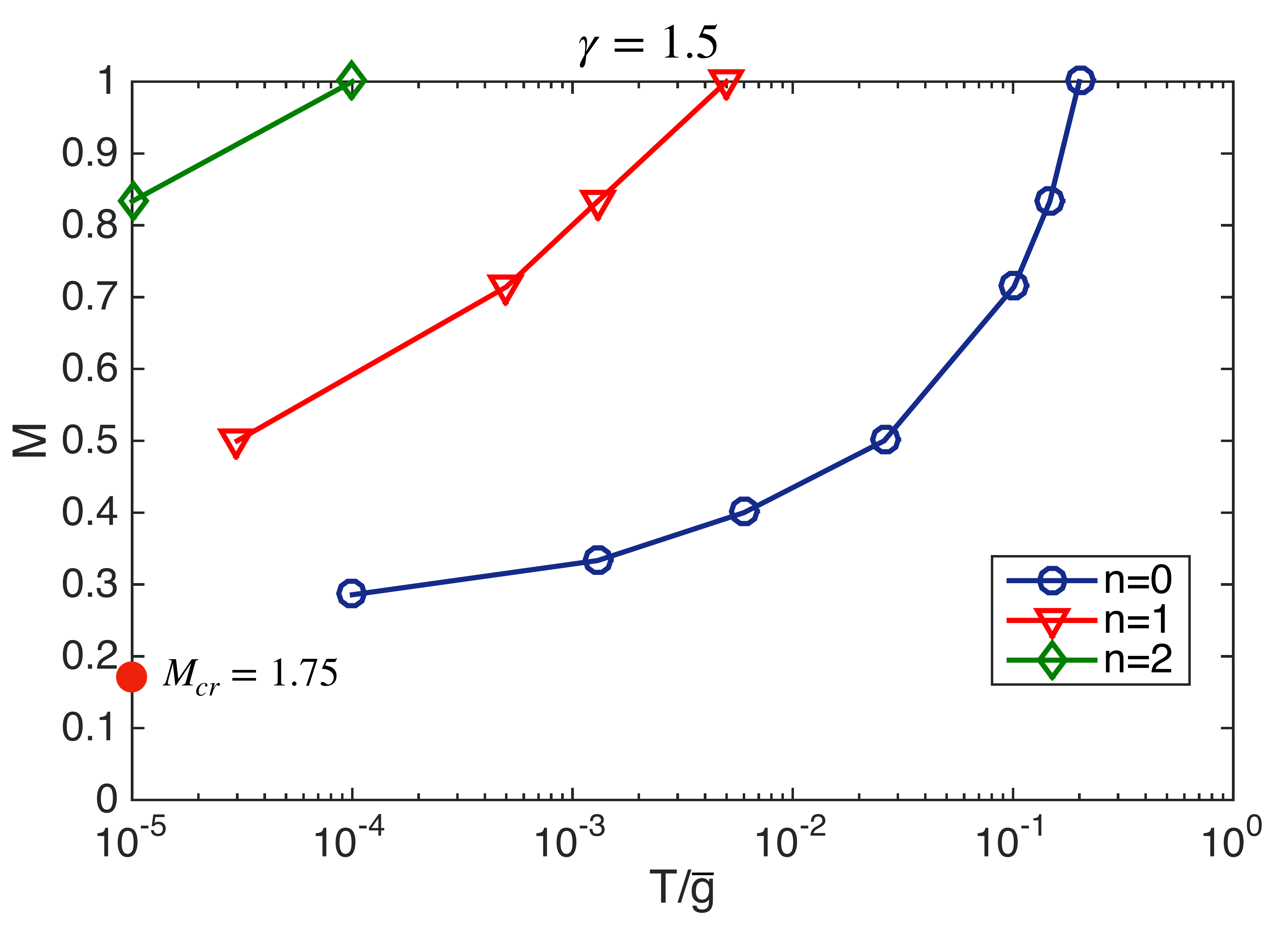}
  \caption{The onset temperatures $T_{p,n}(M)$, obtained by solving \eqref{3_21} for a particular $\gamma=1.5$. Solutions exists at a discrete set of temperatures for any
    $M>M_{cr}=1.75$. The lines  $T_{p,n} (M)$ all terminate at $M_cr$  at $T =0$. To verify this numerically for $n \geq 1$, one needs to go to very low $T$. }
    \label{fig:MvsT}
\end{figure}

We show results of the numerical solution of Eq. (\ref{3_20}) in Fig.\ref{fig:MvsT}.  Like for $\gamma <1$, we find that there exists a discrete set of onset temperatures  $T_{p,n}$, and an eigenfunction
  $\Delta_n (m)$ changes sign $n$ times as a function of Matsubara number $m$.  Different $T_{p,n} (M)$ all approach $M=M_{cr}$ at $T=0$, although for larger $n$ one has to go to very low $T$ to see this.
     Such behavior  is similar to the one for $\gamma <1$, the only distinction is that now there is no special  behavior of $T_{p,0}$ because ${\bar {\tilde \Sigma}} (\omega_m)$  does not vanish at the first two Matsubara frequencies $\omega_m = \pm \pi T$.

Because the behavior of the gap function is the same in  all models with $M > M_{cr}$,
  including the original model  with $M=1$, the extension to $M \neq 1$ allows one to extract this behavior by focusing at either $M \geq M_{cr}$, where the analytical analysis is simplified because relevant  frequencies are  small, or at $M >1$, where  $T_{p,n}$ are larger and can be detected more easily in numerical studies.

 \section{Conclusions}
\label{sec:conclusions}
In this paper, we continued our analysis of the interplay between the pairing and the non-Fermi liquid behavior in a metal for a set of quantum-critical systems, which at low-energies  are described  by a model
of fermions with an effective dynamical electron-electron interaction
 $V(\Omega_m) \propto 1/|\Omega_m|^\gamma$   between fermions at the Fermi surface (the $\gamma$-model).
  We analyzed both the original model and its extension, in which we introduce an extra parameter $N$ to account for non-equal
   interactions in the particle-hole and particle-particle channel.
  In the two previous papers we considered the case $0 < \gamma <1$ and argued that (i) at $T=0$, there exists an infinite discrete set of topologically different gap functions, $\Delta_n (\omega_m)$, all with the same  spatial symmetry, and (ii) each $\Delta_n$
  evolves with temperature and terminates at a particular $T_{p,n}$.  In this paper we analyze how the system behavior changes between  $\gamma <1$ and $\gamma >1$, both at $T=0$ and a finite $T$.
    We show that the limit $\gamma \to 1$ is singular due to infra-red  divergence of $\int d \omega_m V(\Omega_m)$, and the system behavior is highly sensitive to how this limit is taken.  We
  showed that in the original model with $N =1$  the divergencies cancel out in the gap equation, and the gap functions $\Delta_n (\omega_m)$  smoothly evolve through $\gamma=1$ both at $T=0$ and a finite $T$.  However, for $N \neq 1$, the evolution through $\gamma =1$ is not smooth, and qualitatively new behavior emerges for $\gamma \geq 1$. Namely, there still exists a discrete set of
  $T_{p,n}$, below which $\Delta_n (\omega_m)$ appears, but (i) the functional forms of $T_{p,n}$ and $\Delta_n (\omega_m)$ change qualitatively, and (ii)  at $T \to 0$ all $\Delta_n (\omega_m)$  with $n < n_{max} \sim ({\bar g}/T)^\gamma$ tend to the same  gap function.  At $T \to 0$, $n_{max}$ tends to infinity, and the spectrum of the  condensation energy $E_{c,n}$ becomes  a continuous one.
   This opens up the new channel of one-dimensional gap fluctuations.  We also discussed  another extension of the $\gamma$-model for $\gamma >1$, to $M \neq 1$,  for which the extended model is free from singularities, and displays the same behavior as the original model with $M=1$.   This allows one to
       better understand the physics of the original model by  zooming into ranges of $M$ where either analytical or numerical analysis is simplified.

  In the next paper in the series, Paper IV, we consider the original $\gamma$-model in the range $1< \gamma <2$ in more detail. We argue that dynamical vortices appear one-by-one in the upper half-plane of frequency as $\gamma$ increases between one and two, and the new  physics emerges at a finite $T$.
 In Paper V we show that for $\gamma =2$, the number of these vortices becomes infinite, and
   the new physics extends down to $T=0$.

\acknowledgments
  We thank   I. Aleiner, B. Altshuler, E. Berg, R. Combescot, D. Chowdhury, L. Classen,  K. Efetov, R. Fernandes,  A. Finkelstein, E. Fradkin, A. Georges, S. Hartnol, S. Karchu, S. Kivelson, I. Klebanov, A. Klein, R. Laughlin, S-S. Lee, G. Lonzarich, D. Maslov, F. Marsiglio, M. Metlitski, W. Metzner, A. Millis, D. Mozyrsky,  C. Pepin, V. Pokrovsky,  N. Prokofiev,  S. Raghu,  S. Sachdev,  T. Senthil, D. Scalapino, Y. Schattner, J. Schmalian, D. Son, G. Tarnopolsky, A-M Tremblay, A. Tsvelik,  G. Torroba,  E. Yuzbashyan,  J. Zaanen, and particularly Y. Wang,  for useful discussions.   The work by  AVC was supported by the NSF DMR-1834856.

\appendix

\section{The exact solution of the linearized gap equation for $\gamma <1$.}
\label{app:exact}

In Paper I  we obtained the exact solution of the linearized equation for $T=0$, $\gamma <1$ and any $N$.
The solution has the form:
\begin{equation}\label{eq:lenearDeltaExact}
\Delta_\infty (\omega_m ) \propto |\omega_m|^{-\gamma } \tilde{f}(\log |\omega_m/\omega_{0} |^{\gamma })
\end{equation}
where $\omega_0 = {\bar g}/(1-\gamma)^{1/\gamma}$ and
\begin{equation}\label{eq:fTildeB}
 \tilde{f}(x)=\int_{-\infty }^{\infty }\tilde{b}(\beta )e^{-i\beta x}d\beta,
\end{equation}
 where
\begin{equation}\label{eq:btilda}
\tilde{b}(\beta )=\frac{\sinh (\pi \beta_{N})}{\sqrt{\cosh (\pi (\beta -\beta_{N})\cosh ( \pi (\beta +\beta _{N}))}}e^{-iI(\beta )}.
\end{equation}
Here, $\beta_N$ is the solution of $\epsilon_{i\beta_N} = N$,  $\epsilon_{i\beta}$ is given by \eqref{su_15_2}, and
\begin{equation}\label{eq:I1}
I(\beta )=\frac{1}{2}\int_{-\infty }^{\infty } \log |1-\frac{1}{N}\epsilon_{ i\beta '}|\left[\tanh (\pi (\beta '-\beta ))-\tanh (\pi \beta ') \right]d\beta '
\end{equation}
Notice that $I(\beta )$ is real and antisymmetric.

In the limit $\gamma \to 1$,
$$
\epsilon_{i\beta }=N_{cr}+(1-\gamma )R(\beta ) ,
$$
 where $R(\beta)$ is given by (\ref{an_1}). Then
\begin{equation}
 1-\frac{\epsilon_{i\beta }}{N}\approx (1-\gamma )\left(\frac{N-N_{cr}}{1-\gamma }-R(\beta ) \right)=(1-\gamma )\left(R(\beta_{N})-R(\beta ) \right),
\end{equation}
and the function $I(\beta )$ in \eqref{eq:I1} becomes
\beq
I(\beta )  =-\beta \log (1-\gamma )+J(\beta )
\eeq
 where
\beq
J(\beta )\equiv \frac{1}{2}\int_{-\infty }^{\infty }\log |R(\beta_{N})-R(\beta ')|\left[\tanh (\pi (\beta '-\beta ))-\tanh (\pi \beta ') \right]d\beta '
\eeq
 does not depend on $\gamma $. We see that the function $\tilde{f}(x)$ in (\ref{eq:fTildeB})
is in fact the function of $x-\log (1-\gamma )$. Substituting into (\ref{eq:lenearDeltaExact})
we find that at $\gamma \to 1$,  the argument of $\tilde{f}$ is $\log{|\omega_m|/\omega_0} -\log{(1-\gamma)} =
\log{|\omega_m|/{\bar g}}$, i.e.,  relevant scale for $\Delta_\infty (\omega_m)$ is ${\bar g}$ rather than the divergent  $\omega_0$:
\begin{equation}\nonumber
\Delta_{\infty } (\omega_m)=|\omega_m|^{-1}  \tilde{f}(\log |\omega_m|/\bar{g}).
\end{equation}

\section{Numerical results for the crossover from logarithmic to power-law oscillations}
\label{app_b}

\begin{figure*}
  \includegraphics[width=16cm]{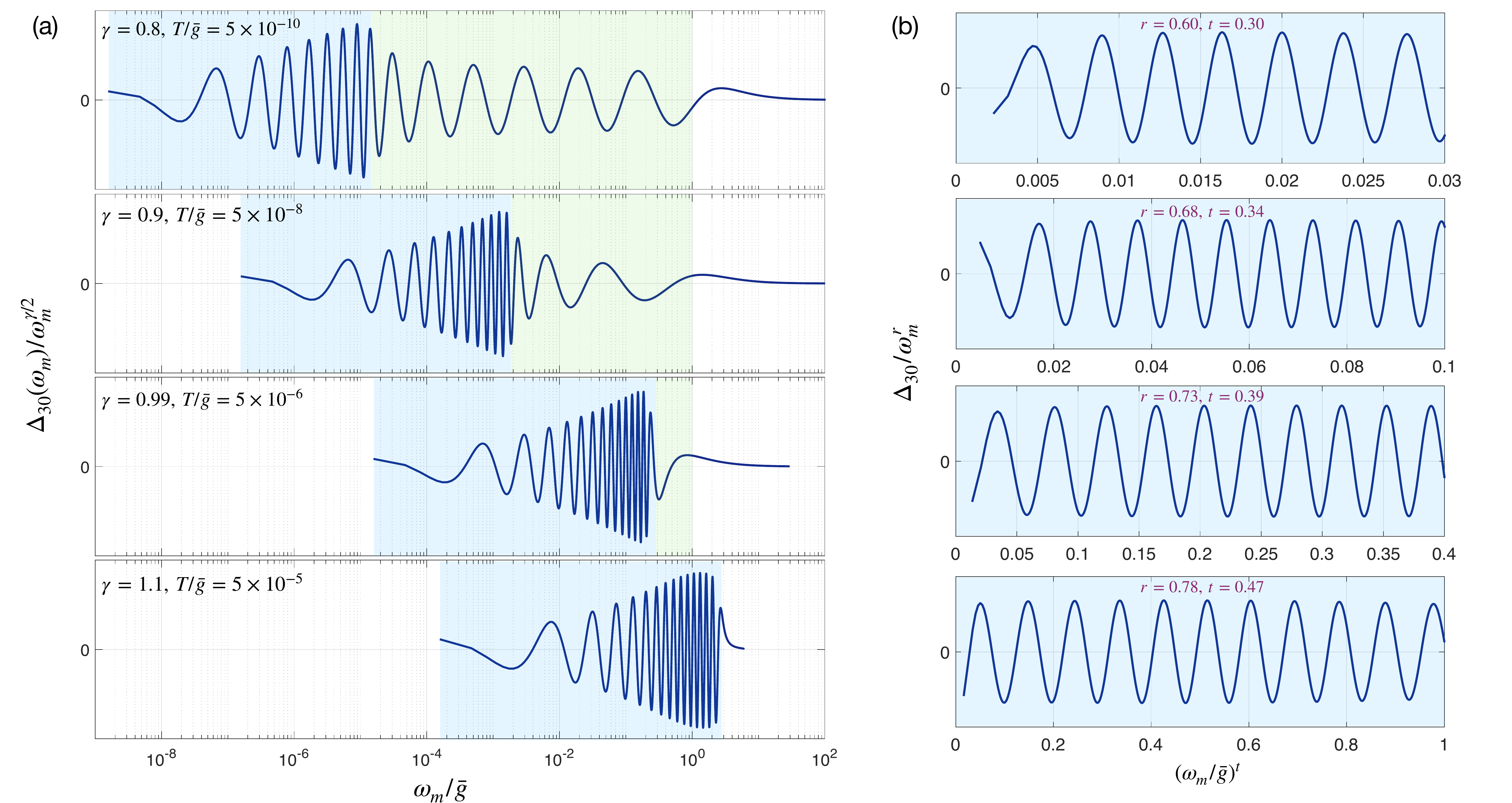}
  \caption{ Left panel. Evolution of $\Delta_n (\omega_m)$ with $\gamma$ around $\gamma=1$. We set $n =30$.
   As $\gamma$ increases towards $1$, the $|\omega_m|^{\gamma/2} \cos{(\beta_N \log (\omega_m/\bar g)^\gamma)}$ form of the gap function (the green region) progressively get replaced by $|\omega_m|^{r} \cos{(\omega_m/\bar g)^t}$ form (the blue region).  Right panel: the plot of $\Delta_n (\omega_m)/|\omega_m|^{r}$ vs $(\omega_m/\bar g)^t$.  The exponents $r$ and $t$ are presented in Fig. \ref{fig:rt}.  Different $T= T_{p,n}$ for the same $n$ and $\gamma$ correspond to different $N$.}
\label{fig:Delta_aboutg1}
\end{figure*}
\begin{figure}
  \includegraphics[width=.5\columnwidth]{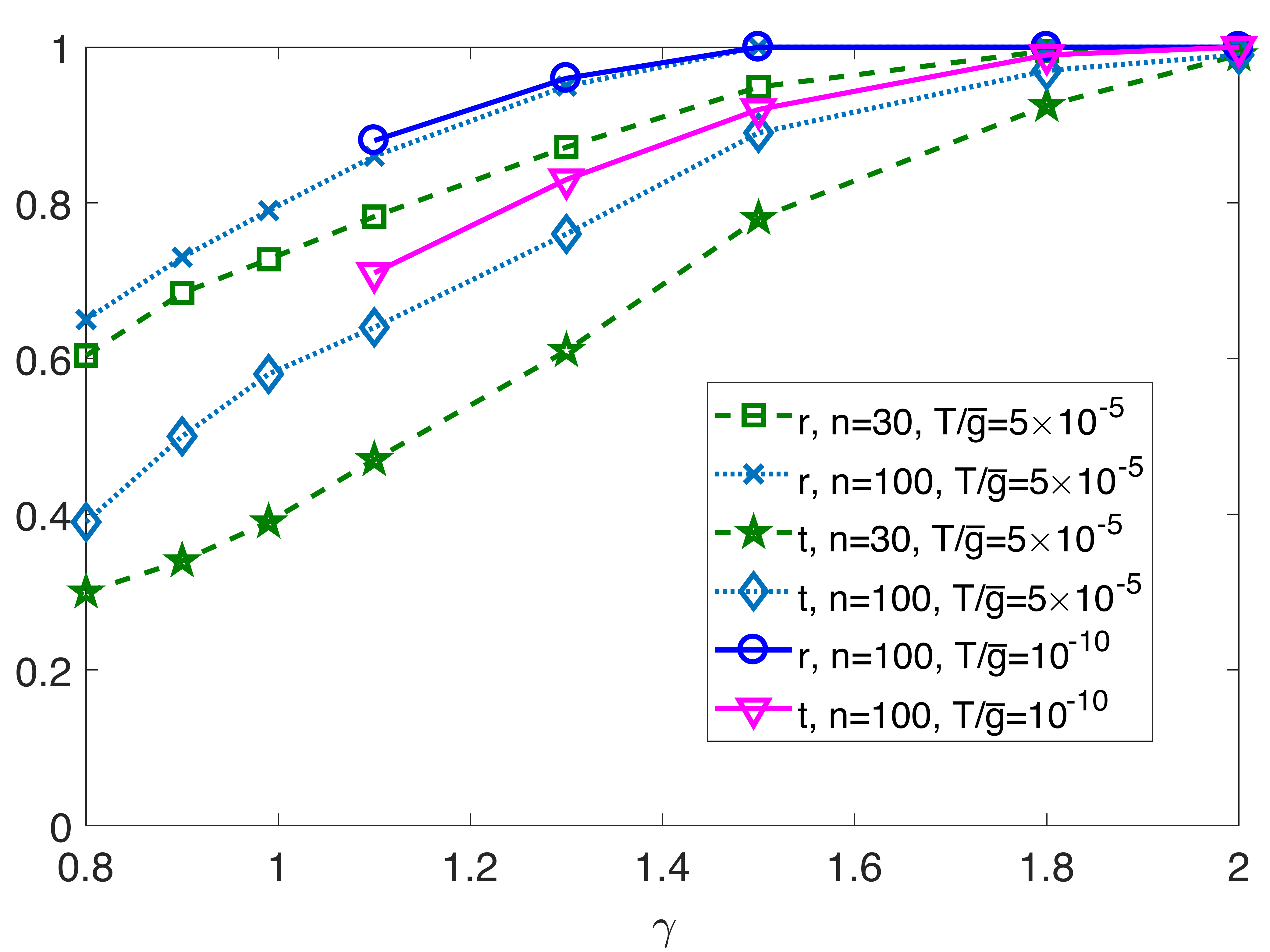}
  \caption{The exponents $r$ and $t$, defined in the text and in the caption to Fig. \ref{fig:Delta_aboutg1}, vs $\gamma $ for $n=30$ and $100$. As $n$ increases, both $r$ and $t$ get closer to the analytical result $r=t=1$, valid for $n \gg 1$. }
\label{fig:rt}
\end{figure}

In this Appendix, we present
 the results of a detailed numerical analysis of the  crossover from log-oscillations of $\Delta_n (\omega_m)$  for $\gamma <1$ to oscillations with a period set by $\omega_m$ for $\gamma >1$.
   Like we said in Sec. \ref{sec_c}, the
   transformation at $\gamma =1$ is sharp at $n \gg 1$,  when $T_{c,n}$ is small and relevant Matsubara numbers
   are large. For smaller $n$, the crossover gets smoothen up.  In numerical calculations, there is an additional smothering due to sampling of a finite number of Matsubara points.

  We show the results in Figs. \ref{fig:Delta_aboutg1} and  \ref{fig:rt}.   We see from Fig. \ref{fig:Delta_aboutg1}
 that as $\gamma$ approaches one, log-oscillations of $\Delta_n (\omega_m)$  at a given $n$ progressively get replaced by power-law oscillations.  The period of power-law oscillations and the envelope are best fitted by $(\omega_m/{\bar g})^t$ and $(\omega_m/{\bar g})^r$, respectively.  For an infinite number of sampling points, we expect a sharp crossover at  $\gamma =1$  and $n \to \infty$ between logarithmic and power-law oscillations.

In Fig.\ref{fig:rt} we show the results for the exponents $r$ and $t$, extracted from Fig. \ref{fig:Delta_aboutg1}.  For a given $\gamma$, the values of $r$ and $t$ vary with  $n$ and temperature.
 Analytically, we obtained in Sec. \ref{sec_c} $r=t=1$ for large $n$, when $T_{p,n} \ll {\bar g}$.
 We see that $r$ and $t$  are different from one for a given $n$, but both tend to $1$ when $n$ becomes large enough and $T\to0$. This result confirms our analysis in Eq.\eqref{3_1} for $\gamma\geq1$.
  Overall, the numerical results clearly show the main result of our analysis --- the transformation of log-oscillations for $\gamma <1$ into power-law oscillations for $\gamma >1$.

\bibliography{gamma_around_1}

\end{document}